\documentclass[reprint,aps,prc,amsmath,amssymb,nofootinbib,superscriptaddress]{revtex4-1}

\usepackage{braket} 
\usepackage[dvipsnames]{xcolor}  
\usepackage{float}
\usepackage[T1]{fontenc}
\usepackage[utf8]{inputenc}
\usepackage{graphicx,color,rotating,pifont}
\usepackage{amsmath,amssymb}
\usepackage{mathrsfs}
\usepackage{mathtools}
\usepackage{subfigure}
\usepackage{siunitx}
 
\usepackage{bm}
\usepackage{ae}
\usepackage{dcolumn}
\usepackage{txfonts}
\usepackage{tensor}
\usepackage{braket}
\usepackage{booktabs}
\usepackage{xspace}
\usepackage{soul}
\usepackage{multirow}
\usepackage{tikz}
\usepackage{pgffor} 
\setstcolor{red}
\setul{}{.2ex} 
 
\usepackage[normalem]{ulem}

\usepackage{hyperref}
\usepackage{cleveref}

\newcommand{\QRPA}{\ensuremath{\mathrm{QRPA}}}
\newcommand{\HFB}{\ensuremath{\mathrm{HFB}}}

\newcommand{\beq}{\begin{equation}}
\newcommand{\eeq}{\end{equation}}
\newcommand{\beqn}{\begin{eqnarray}}
\newcommand{\eeqn}{\end{eqnarray}}
\newcommand{\bsub}{\begin{subequations}}
\newcommand{\esub}{\end{subequations}}
\newcommand{\bpm}{\begin{pmatrix}}
\newcommand{\epm}{\end{pmatrix}}

\DeclareSIUnit{\fm}{\femto\meter}
\DeclareSIUnit{\MeVc}{\MeV\per\text{\ensuremath{c}}}

\allowdisplaybreaks[3]

\newcommand{\ed}[1]{{\color{red}#1}} 
\newcommand{\eda}[1]{{\color{black}#1}}

\begin{document} 
\title{Symmetry restoration in the axially deformed proton-neutron quasiparticle random phase approximation for nuclear $\beta$ decay: The effect of angular-momentum projection}

 \author{R. N. Chen}
 \email{chenrn6@mail2.sysu.edu.cn}
  \affiliation{School of Physics and Astronomy, Sun Yat-sen University, Zhuhai 519082, P.R. China}   
 \affiliation{Guangdong Provincial Key Laboratory of Quantum Metrology and Sensing, Sun Yat-Sen University, Zhuhai 519082, P.R. China }

  \author{Y. N. Zhang}  
  \email{zhangyn226@mail.sysu.edu.cn}
  \affiliation{ Sino-French Institute of Nuclear Engineering and Technology, Sun Yat-Sen University, Zhuhai, 519082 Guangdong, P.R. China}   
  
  \author{J. M. Yao}    
  \email{yaojm8@sysu.edu.cn}
  \affiliation{School of Physics and Astronomy, Sun Yat-sen University, Zhuhai 519082, P.R. China}   
 \affiliation{Guangdong Provincial Key Laboratory of Quantum Metrology and Sensing, Sun Yat-Sen University, Zhuhai 519082, P.R. China }

\author{J. Engel}  
\email{engelj@physics.unc.edu}
\affiliation{Department of Physics and Astronomy, University of North Carolina, Chapel Hill, North Carolina 27516-3255, USA}

\begin{abstract}

We examine the effects of symmetry restoration on nuclear $\beta$ decay within the axially  deformed proton–neutron quasiparticle random phase approximation (QRPA). We employ the proton–neutron finite-amplitude method (pnFAM) to compute transition amplitudes, and perform angular-momentum projection both after variation and after the QRPA to restore rotational symmetry. Exact projection reduces the calculated $\beta$ decay half-lives from those that use the needle approximation by up to 60\%, and even more when taking the effects of projection on the ground-state energy into account.

\end{abstract}
 
\maketitle

\section{Introduction}
\label{sec:introduction}
Accurate values for nuclear $\beta$ decay rates are crucial for understanding the stability of atomic nuclei, simulating nucleosynthesis~\cite{Langanke:2003RMP,Fischer:2024PPNP,Suzuki:2022PPNP}, and testing the Standard Model~\cite{Severijns:2006RMP,Towner_2010,Hayen:2024,Herczeg:2001PPNP,Severijns:2006,Otten:2008RPP,Falkowski:2021JHEP}.
One example of nucleosynthesis is the rapid neutron-capture process ($r$-process)~\cite{Burbidge:1957RMP,Cowan:1991PR,Qian:2007PR}, which is responsible for creating about half of the elements heavier than iron and occurs on a timescale that depends significantly on $\beta$ decay rates~\cite{Moller:2003PRC}.
These rates for many neutron-rich nuclei along the $r$-process path have not yet been measured, however, and must therefore be computed within nuclear models~\cite{Pereira:2009PRC,Nishimura:2011PRL,Moller:1997}. Reliable theoretical predictions of nuclear $\beta$ decay rates are thus essential. 

In the past decade, {\em ab initio} methods, in which both strong and weak interactions are consistently derived from chiral effective field theory~\cite{Weinberg:1991}, have led to major progress~\cite{Ekstrom:2014PRL,Gysbers:2019,Stroberg:2021,King:2023,King:2024,Li:2025,Beaujeault:2022ayi,Zaragoza:2024jvk}. 
Nevertheless, applying these methods to medium-mass, open-shell, deformed nuclei, especially those far from the $\beta$  stability line, remains challenging. To date, nuclear $\beta$ decay has primarily been investigated within valence-space shell models~\cite{Langanke:2003RMP,Suzuki:2022PPNP} and the proton-neutron quasiparticle random-phase approximation (pnQRPA)~\cite{Schuck:2021,Delion:1997vr}.
The shell model, widely used for $\beta$ decay in light nuclei or those near magic numbers~\cite{Caurier:2004gf,Martinez-Pinedo:1999:PRL,Suzuki:2022PPNP,Zhi:2013hg}, defines effective-interaction matrix elements within a truncated valence space, with parameters fitted to reproduce selected nuclear structure data~\cite{Caurier:2005RMP}.
However, the dimensionality of the configuration space increases rapidly with the number of valence nucleons, making it difficult to apply the shell model to open-shell heavy nuclei far from magic numbers. 
By contrast, pnQRPA methods, based on realistic nuclear forces~\cite{Fang:2013,Ni:2014}, non-relativistic energy-density functionals (EDFs)~\cite{Engel:1999,Niu:2015,Gambacurta:2020,Bai:2022pys,Suhonen:1988umy}, or relativistic EDFs~\cite{Paar:2004,Niksic:2005,Marketin:2007,Niu:2013PLB,Liang:2010dy,Marketin:2015gya}, are generally formulated in large single-particle spaces, with the configuration space truncated at the level of two-quasiparticle or two-quasihole excitations on top of an unspecified QRPA vacuum. 
Recent studies indicate that this framework can be improved by incorporating particle–vibration coupling~\cite{Niu:2015,Liu:2024PRC} or by including two-particle–two-hole (2p-2h) excitations~\cite{Gambacurta:2020}.

Although the QRPA is typically formulated as a matrix-diagonalization problem, large single-particle spaces make its application to open-shell deformed nuclei computationally challenging. The need to compute numerous two-body matrix elements from EDF-based residual interactions, combined with the diagonalization of the QRPA matrix, makes the process time-consuming—though such calculations have been performed in select cases for both charge-conserving~\cite{Arteaga:2008,Peru:2008,Yoshida:2008,Losa:2010,Terasaki:2010,Repko:2018gcn,Kvasil:2019giv} and charge-changing~\cite{Terasaki:2016} transitions.
To mitigate the computational burden, Nakatsukasa et al.~\cite{Nakatsukasa:2007qj} introduced the finite amplitude method (FAM), which computes the fields induced by an external one-body operator and solves the corresponding linear response equations iteratively.  This technique has since been widely adopted in QRPA calculations \cite{Inakura:2009PRC,Inakura:2010zz,Stoitsov:2011zz,Avogadro:2011PRC,Niksic:2013ega,Liang:2013pda,Hinohara:2013qda,Liang_2014,Oishi:2015lph,Stoitsov:2011zz,Inakura:2009tqn,Shafer:2016etk}. In particular, the extension of the QRPA+FAM (QFAM) to the proton-neutron channel known as the proton-neutron finite amplitude method (pnFAM)—has enabled systematic studies of $\beta$ decay half-lives across a wide range of nuclei when combined with modern Skyrme EDFs in axially deformed systems~\cite{Mustonen:2014bya,Mustonen:2015sfa,Ney:2020mnx}. It is worth noting that the QFAM has also been implemented in studies of electromagnetic strength distributions in doubly open-shell nuclei, both at zero temperature and at finite temperature, starting from chiral Hamiltonians~\cite{Beaujeault:2022ayi,Zaragoza:2024jvk}.

In the QRPA studies of deformed nuclei, excited states are built from quasiparticle excitations of deformed Hartree-Fock-Bogoliubov (HFB) states.
As a consequence, the QRPA wave functions do not have well-defined values for conserved quantities such as total angular momentum and particle number.
To compare with laboratory-frame observables, one must project intrinsic-frame states onto states with good quantum numbers.
A symmetry-projected random phase approximation (RPA) with variation after projection (VAP) was derived in Ref.\cite{Federschmidt:1985}.
This approach, however, has only been applied in a schematic R(8) model; in most studies of well-deformed nuclei, angular momentum projection (AMP) is implemented approximately, in the so-called needle approximation~\cite{Krumlinde:1984hbw,Arteaga:2008,Yousef:2009,Ravlic:2024hpi,Ring:1980}.
This approximation, which assumes that rotated and unrotated wave functions are fully orthogonal regardless of the (non-zero) rotation angle.  It is a reasonable approximation for large deformation, but is expected to fail for small deformation, where  
rotational symmetry must be restored through exact AMP. \footnote{In the spherical limit, however, the needle approximation is accidentally exact, even though even-even ground states are unchanged under rotation.} 
In recent years, the need to go beyond the needle approximation has led to the use of exact AMP with projection after variation (PAV) in axially deformed RPA calculations~\cite{Erler:2014,Porro:2024} of multipole strength functions. 
The projection was shown to significantly modify both monopole and quadrupole transition strengths in $^{24}$Mg~\cite{Porro:2024}.
However, the method has never been applied to open-shell nuclei or to charge-changing processes such as $\beta$ decay.

In this paper, we extend the axially-deformed pnFAM developed in Ref.~\cite{Mustonen:2014bya} for $\beta$ decay by incorporating exact AMP within a PAV scheme. This extension restores the rotational symmetry broken by deformed HFB solutions and, in some respects, is less involved than for charge-conserving processes, as it avoids complications from spurious rotational modes. We demonstrate the impact of symmetry restoration by computing Gamow-Teller (GT) strength functions and $\beta$ decay rates for several neutron-rich iron isotopes.   It is worth pointing out that this work constitutes a nontrivial generalization of Ref.~\cite{Porro:2024} to charge-changing processes, in which  pairing correlations between nucleons are included and the QRPA equations are solved within the pnFAM framework.

The paper is organized as follows: Section~\ref{sec:framework} outlines the axially  deformed pnFAM formalism and presents general expressions for computing transition strengths with AMP in the PAV scheme. Section~\ref{sec:Results and discussion} presents results for the Fermi and GT strength functions, along with the corresponding $\beta$-decay rates, both with and without exact AMP. Finally, Section~\ref{sec:summary} offers concluding remarks and directions for future research.
 
\section{Framework}
\label{sec:framework} 
 
\subsection{The pnFAM with axial symmetry}
The pnFAM was developed in Ref.~\cite{Mustonen:2014bya}, which provides a detailed introduction. For completeness, we present a brief outline of the method below.

The pnFAM solves equations for the linear response of a nuclear quasiparticle vacuum perturbed by a weak charge-changing external field $\hat{F}(t)$: 
\beq 
\label{eq:F_perturbation}
\hat F(t) = \eta \left( \hat F e^{-i\omega t} + \hat F^\dagger e^{i\omega t} \right) \,,
\eeq 
where  $\omega$ is the complex frequency of the external field, the parameter $\eta$ is a small real number that controls the size of the perturbation, and $\hat{F}$ is a one-body operator of the form,
\beq 
\label{eq:F_sps}
\hat F = \sum_{p,n} f_{p n}c^\dagger_{p} c_{n}.
\eeq 
Here $(c^\dagger, c)$ are single-particle creation and annihilation operators for states in a harmonic-oscillator basis. 
The indices $p, n$ label states of protons and neutrons, and the $f_{p n}$ are single particle matrix elements of the operator $\hat F$.

To represent the unperturbed ground state, we take the HFB state $\ket{\Phi_0}$, a vacuum with respect to quasiparticle operators $(\beta^{\dagger}, \beta)$, which  are connected with the single particle operators $(c^\dagger, c)$ via the Bogoliubov transformation~\cite{Ring:1980},
 \beqn 
  \begin{pmatrix}
\beta \\
\beta^\dagger
\end{pmatrix}
= \mathcal{W}^\dagger
 \begin{pmatrix}
c \\
c^\dagger
\end{pmatrix}, \quad 
  \mathcal{W}^\dagger =
 \begin{pmatrix}
U^\dagger & V^\dagger \\
V^T & U^T 
\end{pmatrix} \,. 
\eeqn 
In the quasiparticle basis, the one body operator $\hat{F}$ can be rewritten as 
\begin{eqnarray}
    \hat{F} = \frac{1}{2} \begin{pmatrix}
        \beta^{\dagger} & \beta
    \end{pmatrix} \mathcal{F} \begin{pmatrix}
        \beta\\\beta^{\dagger} 
    \end{pmatrix}+F^{0} \,,
\end{eqnarray}
 where $F^{0}$ is a zero-body term and it vanishes for charge-changing operators. The matrix $\mathcal{F}$ has the form, 
 \begin{eqnarray}
 \label{eq:F_qp}
     \mathcal{F} = \begin{pmatrix}
         F^{11} & F^{20}\\
         F^{02} & -(F^{11})^{T}
     \end{pmatrix} \,. 
 \end{eqnarray}
 The matrix elements of $F^{20}$ and $F^{02}$ in the quasiparticle basis are related to those of $f$ in Eq.(\ref{eq:F_sps}) by, 
 \begin{eqnarray}
       F^{20}
       = \Bigg(U^{(\pi)\dagger} fV^{(\nu)*}\Bigg),  \quad  F^{02}=-\Bigg(V^{(\pi)T} f U^{(\nu)}\Bigg)  \,. 
 \end{eqnarray}
 Here, $ {U}^{(\nu)}$ and $ {U}^{(\pi)}$ denote the matrix U of neutrons and protons, respectively, as well as $ {V}^{(\nu)}$ and $ {V}^{(\pi)}$.
 
 The external field $\hat{F}(t)$ induces an oscillation of the generalized density $\mathcal{R}$ ~\cite{Niksic:2013ega},
 \begin{eqnarray}
   \mathcal{R}(t) =  \mathcal{R}_0+ \eta [\delta \mathcal{R}(\omega)e^{-i\omega t}+\delta \mathcal{R}^{\dagger}(\omega)e^{i\omega t}] \,, 
\end{eqnarray}
where the generalized density in the single-particle basis is 
\beqn 
\mathcal{R}(t) = \begin{pmatrix} \rho (t)& \kappa(t) \\
-\kappa^*(t) & 1 - \rho^* (t)
\end{pmatrix}  \,,
\eeqn 
$\mathcal{R}_0$ is the static HFB version of this density, and 
\beqn 
    \delta \mathcal{R}(\omega) 
    = 
    \begin{pmatrix}
        \delta \rho(\omega) & \delta \kappa(\omega) \\
        -\delta \bar{\kappa}^*(\omega) & -\delta \rho^*(\omega)
    \end{pmatrix}
\equiv 
 \mathcal{W}\begin{pmatrix}
         0 &  \mathcal{X}(\hat F ; \omega) \\
        \mathcal{Y}(\hat F ; \omega) & 0
    \end{pmatrix}\mathcal{W}^\dagger 
\eeqn 
From here one obtains the variation of the one-body densities,
\bsub\beqn 
\delta \rho( \omega) & =U^{(\pi)} \mathcal{X} (\hat F; \omega)V^{(\nu)T}+V^{(\pi)*} \mathcal{Y}(\hat F ; \omega) U^{(\nu)\dagger} \,, \\
\delta \kappa(  \omega) & =U^{(\pi)} \mathcal{X}(\hat F ; \omega) U^{(\nu)T}+V^{(\pi)*} \mathcal{Y}(\hat F ; \omega) V^{(\nu)\dagger} \,, \\
\delta \bar{\kappa}^{*}(  \omega) & =-U^{(\pi)*} \mathcal{Y}(\hat F ; \omega) U^{(\nu)\dagger}-V^{(\pi)} \mathcal{X}(\hat F ; \omega) V^{(\nu)T}\,. 
\eeqn 
\esub
 The matrix elements $\mathcal{X}_{pn}(\hat{F};\omega)$ and $\mathcal{Y}_{pn}(\hat{F};\omega)$, defined in the quasiparticle basis, correspond to the amplitudes for changing neutrons into protons. 
 Refs.~\cite{Nakatsukasa:2007qj,Avogadro:2011PRC} show that the variation of the densities can be expressed in terms of a time-dependent perturbation of the neutron quasiparticle-annihilation operators, 
  \beq 
 \beta_{n}(t) =  \Bigg[\beta_{n}  + \delta\beta_{n}(t)\Bigg]e^{i\mathcal{E}_{n} t} \,, 
  \eeq 
 where $\mathcal{E}_n$ is the $n^\text{th}$ neutron quasiparticle energy. The second term, arising from the perturbation in Eq.~(\ref{eq:F_perturbation}), can be expanded in terms of proton quasiparticle creation operators: 
   \begin{equation}
   \label{eq:qp_operator_change}
       \delta \beta_n(t) = \eta\sum_{p} \beta^{\dagger}_{p}\Bigg[\mathcal{X}_{pn}(\hat{F};\omega)e^{-i\omega t }+\mathcal{Y}_{pn}(\hat{F};\omega)e^{i\omega t }\Bigg] \,. 
   \end{equation}

The variation of the generalized density induces a corresponding variation in the HFB Hamiltonian~\cite{Niksic:2013ega},
 \begin{eqnarray}
      \mathcal{H}(t) =  \mathcal{H}_0+ \eta[\delta \mathcal{H}(\omega)e^{-i\omega t}+\delta \mathcal{H}^{\dagger}(\omega)e^{i\omega t}] \,. 
 \end{eqnarray}
 Here, $\mathcal{H}_0$ is the HFB Hamiltonian, and $\mathcal{H}(t)$, expressed in the single-particle basis, is determined by the functional derivative of the nuclear EDF, $E[\mathcal{R}]$, with respect to the generalized density~\cite{Ring:1980}:
\beqn 
\mathcal{H}=\frac{\delta E[\mathcal{R}]}{\delta \mathcal{R}}=\left(\begin{array}{cc}
h & \Delta \\
-\Delta^{*} & -h^{*}
\end{array}\right) \,, 
\eeqn
where $h$ and $\Delta$ are the single-particle Hamiltonian and pairing field, respectively. 
In actual calculations, it is convenient to transform the variation of the Hamiltonian, $\delta \mathcal{H}(\omega)$, from the single-particle basis to the quasiparticle basis, through the relation,
 \beqn 
 \label{eq:change_in_H}
 \delta \mathcal{H}(\omega)
   &=&   \begin{pmatrix}
U^{(\pi)\dagger} & V^{(\pi)\dagger} \\
V^{(\pi)T} & U^{(\pi)T }
\end{pmatrix}
\begin{pmatrix}
\delta h(\omega) & \delta \Delta(\omega) \\
-\delta \bar{\Delta}^*(\omega) & -\delta h^T (\omega)
\end{pmatrix}
\begin{pmatrix}
U^{(\nu)} & V^{(\nu)*} \\
V^{(\nu)} & U^{(\nu)*} 
\end{pmatrix}  \nonumber\\
  &\equiv&  
  \begin{pmatrix}
      \delta H^{11}(\omega) & \delta H^{20}(\omega) \\
      -\delta H^{02}(\omega) & -(\delta H^{11})^T(\omega)
  \end{pmatrix}  \,,
\eeqn
from which one finds 
\beqn 
\delta H_{}^{20}(\omega) 
&\equiv & U^{(\pi)\dagger} \delta h(\omega) V^{(\nu)*}-V^{(\pi)\dagger} \delta  \Bar{\Delta}^{ *} (\omega)V^{(\nu)*}\nonumber\\&& +U^{(\pi)\dagger} \delta \Delta^{} (\omega)U^{(\nu)*}   
-V^{(\pi)\dagger} \delta h^{T} (\omega)U^{(\nu)*},   \\
\delta H_{}^{02}(\omega) &\equiv & -V^{(\pi)T} \delta h(\omega) U^{(\nu)}
+U^{(\pi)T} \delta \Bar{\Delta}^{ *}(\omega) U^{(\nu)}\nonumber\\&& 
-V^{(\pi)T} \delta \Delta^{} (\omega)V^{(\nu)} +U^{(\pi)T} \delta h^{T} (\omega)V^{(\nu)} \,.   
\eeqn 
Expressions for the variations in the $h$ and $\Delta$ can be obtained by writing the derivative of $\mathcal{H}$ in the form,
\begin{eqnarray}
    \delta \mathcal{H}(\omega) =\lim_{\eta\to 0} \frac{1}{\eta} [\mathcal{H}(\mathcal{R}_0 + \eta \delta \mathcal{R}(\omega) )- \mathcal{H}(\mathcal{R}_0)] \,,
\end{eqnarray}
leading to
\begin{subequations}
\begin{eqnarray}
    \delta h(\omega) &=& \lim_{\eta \to 0} \frac{1}{\eta} [h(\rho_0 +   \eta \delta \rho) - h(\rho_0)],  \\
   \delta \bar{\Delta}(\omega) &=& \lim_{\eta \to 0} \frac{1}{\eta} [\Delta(\kappa_0 + \eta   \delta \bar{\kappa} ) - \Delta(\kappa_0)], \\
  \delta\Delta^{ }(\omega) &=& \lim_{\eta \to 0} \frac{1}{\eta} [\Delta(\kappa_0 +  \eta  \delta \kappa ) - \Delta (\kappa_0)] \,. 
\end{eqnarray}
\end{subequations}
In these expressions, $\rho_0$ and $\kappa_0$ are  the density matrix and pairing tensor of the unperturbed HFB state,
\beq 
\rho_0 = V^\ast V^T, \quad \kappa_0 = V^\ast U^T. 
\eeq 

Now, the generalized density $\mathcal{R}(t)$  evolves according to the time-dependent HFB equation~\cite{Ring:1980},
   \begin{equation}
   \label{eq:TDHFB}
        i \partial_t \mathcal{R}(t)  
      =  \Bigg[ \mathcal{H}(t)+   \mathcal{F}(t),  \mathcal{R}(t)\Bigg], 
   \end{equation}
 from which, one can obtain the equation of motion for the neutron or proton quasiparticle-annihilation operators: 
 \beqn 
 \label{eq:TDHFB_qp_op}
      i \partial_t \delta\beta_{k}(t)  
      &=& \mathcal{E}_k \delta\beta_{k}(t) + [\mathcal{H}_0, \delta\beta_{k}(t)]
      +[\delta \mathcal{H} (t)+\mathcal{F}(t), \beta_k]. \nonumber
      \\
 \eeqn
Substituting Eqs.\ (\ref{eq:F_qp}), (\ref{eq:qp_operator_change}), and (\ref{eq:change_in_H}) into Eq.\ (\ref{eq:TDHFB_qp_op}), one obtains the pnFAM amplitude equations, 
\begin{subequations}
\label{eq:FAM_equations_XY}
\begin{eqnarray}
       (\mathcal{E}_p+\mathcal{E}_n-\omega)\mathcal{X}^{(K^{\pi})}_{pn}(\hat{F};\omega)+\delta H^{20}_{pn}(\omega) &=& -F^{20}_{pn}, \\
       (\mathcal{E}_{p}+\mathcal{E}_{n}+\omega) \mathcal{Y}^{(K^{\pi})}_{pn}(\hat{F};\omega)+\delta H^{02}_{pn}(\omega)&=&-F^{02}_{pn} 
\end{eqnarray}
\end{subequations}
Here, the $\mathcal{E}_{n(p)}$ are the neutron (proton)  quasiparticle energies.  
When the HFB state is axially deformed, single-quasiparticle excitations do not have a well-defined total angular momentum.  
Instead, each is labeled by $K^\pi_i$, where $K_i$ is the projection of the angular momentum of the $i$-th single-quasiparticle state onto the HFB symmetry axis, and $\pi$ denotes the state's parity.  A quasiparticle pair $(p, n)$ obeys the selection rules $K_p + K_n = K$ and $\pi_p \pi_n = \pi$, where $K$ is the projection of the total angular momentum of the pair onto the symmetry axis. 
In the following, $K^\pi$ will be used to label the amplitudes associated with different decay modes.  
We assume that the operator $\hat{F}$ has the same $K$ and $\pi$ as the quasiparticle pairs.
  
We solve the pnFAM equations iteratively for each frequency $\omega$.  
The strength distribution of the transition induced by the operator $\hat F$ is then given by the imaginary part of the response function,
   \begin{equation}
   \label{eq:strength}
       \frac{d B(\hat{F},\omega)}{d\omega} = -\frac{1}{\pi} {\rm Im}[S(\hat F, K^\pi; \omega)] \,,
   \end{equation} 
where $S(\hat F, K^\pi; \omega)$ at each $\omega$ is determined by the amplitudes $\mathcal{X}^{({K^{\pi}})}_{pn}(\hat F;\omega )$ and   $\mathcal{Y}^{({K^{\pi}})}_{pn}(\hat F;\omega )$, and  the matrix elements of the external field~\cite{Avogadro:2011PRC}:
\beqn
\label{eq:strength_sum}
   S(\hat F,K^{\pi};\omega)  
    & =& \sum_{pn} \Bigg(F^{20*}_{pn} \mathcal{X}^{(K^{\pi})}_{pn}
    (\hat F;\omega )+ F^{02*}_{pn} \mathcal{Y}^{(K^{\pi})}_{pn}(\hat F;\omega ) \Bigg) \,. \nonumber\\
\eeqn 
The response function has the spectral representation~\cite{Nakatsukasa:2007qj}
\beqn 
\label{strength-function}
   &&  S(\hat F, K^{\pi};\omega)  \nonumber\\
   &=& -\sum_{N}\left(\frac{|\bra{N, K^\pi}\hat{F}\ket{\QRPA}|^2}{\Omega_{N}-\omega}-\frac{|\bra{N, K^\pi}\hat{F}^{\dagger}\ket{\QRPA}|^2}{\Omega_{N}+\omega}\right) \,, \nonumber\\
\eeqn
where $\ket{\QRPA}$ and $\ket{N, K^\pi}$ are the wave functions of the ground state and the $N$-th excited state, respectively, and the QRPA relates these states but doesn't define any of them independently.  
The response function clearly has peaks at the excitation energies $\Omega_N$.  
From (\ref{eq:strength}) and (\ref{strength-function}), with the definition of the frequency of the external field $\omega=\Omega + i\gamma$, one finds the spectral representation of the strength function~\cite{Mustonen:2014bya},
\begin{eqnarray}
\label{eq:strength_function}
    \frac{dB(\hat{F},\omega)}{d\omega}
    &=&\frac{\gamma}{\pi} \sum_{N} \Bigg[ \frac{|\bra{N, K^\pi}\hat{F}\ket{\QRPA}|^2}{(\Omega-\Omega_N)^2+ \gamma^2} \nonumber\\
    &&-   \frac{|\bra{N, K^\pi}\hat{F}^{\dagger}\ket{\QRPA}|^2}{(\Omega+\Omega_N)^2+ \gamma^2}   \Bigg], 
\end{eqnarray}
where $\bra{N, K^\pi}\hat{F}\ket{\QRPA}$  is the transition matrix element from the QRPA ground state to the $N^\text{th}$ excited state $\ket{N, K^\pi}$. 
Both states are related to the axially deformed HFB solution, which generally breaks the conservation of total angular momentum and particle number spontaneously~\cite{Ring:1980}.

\begin{figure}[]
     \centering
     \includegraphics[width=\columnwidth]{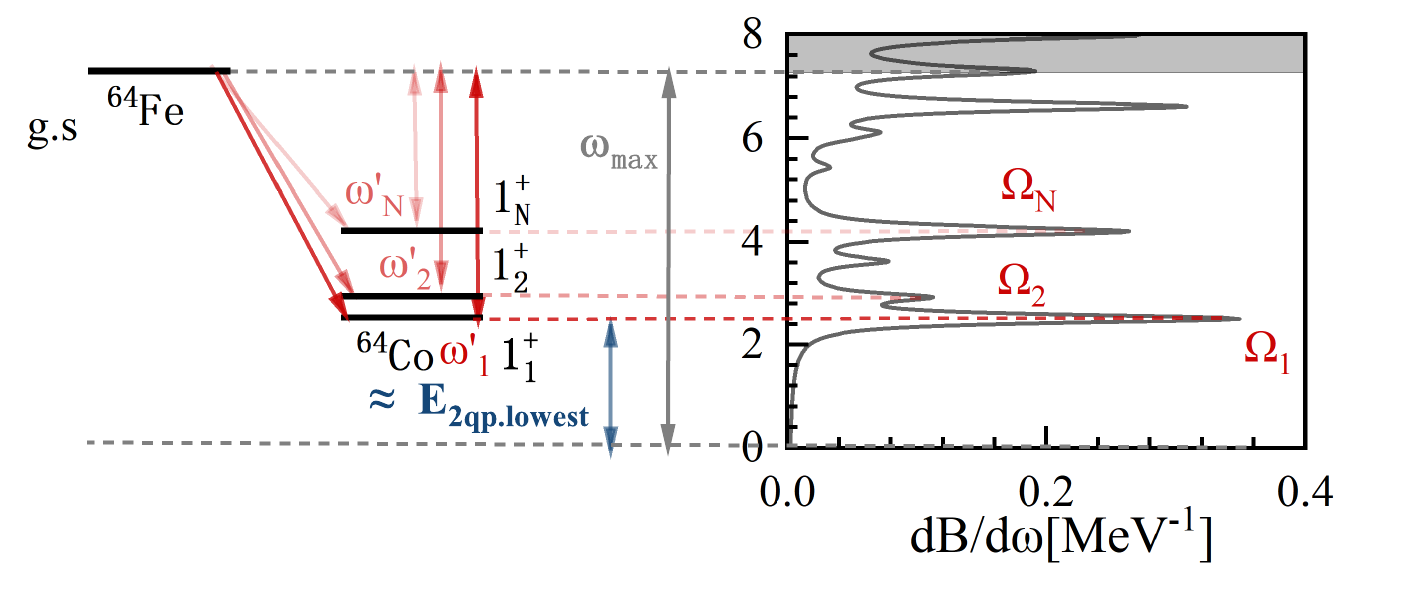}
     \caption{A schematic illustration of GT transitions from the ground state ($0^+_1$) of \nuclide[64]{Fe} to a set of final states ($1^+_N$ ) with $K=0$ in the daughter nucleus \nuclide[64]{Co}. 
     The corresponding portion of the GT strength function $dB(\hat{F},\omega)/d\omega$ from Eq.\ (\ref{eq:strength_function}) as a function of the energy is also shown. The heights of the peaks are for illustration only.    The shaded area contains transitions forbidden by energy conservation in $\beta$ decay. 
     See main text for details.}
     \label{fig:excitation-energy-schematic}
     \end{figure}

In the following calculations of strength functions and $\beta$-decay rates, we will frequently use the variable $\omega_N'$, which denotes the energy difference between the initial and final states. To illustrate the relationship between $\Omega_N$ and $\omega'_N$, we present a schematic diagram in Fig.~\ref{fig:excitation-energy-schematic}, showing GT transitions from the ground state $\ket{0^+_i}$ of \nuclide[64]{Fe} to the $N^\text{th}$ $1^+$ states $\ket{N,1^+}$ of the daughter nucleus. 
In addition, we show the one-to-one correspondence between the transition energies $\omega'_N$ and the locations of the peaks in the strength function $\dfrac{dB(\hat{F}, \omega)}{d\omega}$, the peak frequencies of which are given by $\Omega_N = \omega_{\rm max} - \omega'_N$.  
The maximum energy $\omega_{\rm max}$ is approximately~\cite{Engel:1999}
    \begin{equation}
        \omega_{\rm max} =Q_{\beta^-}+\mathcal{E}_{\rm 2qp   , lowest} =\lambda_{n}-\lambda_{p}+\Delta M_{n-H} \,,
    \end{equation}
where $\lambda_p$ and $\lambda_n$ are the proton and neutron Fermi energies from the HFB solution, $\Delta M_{n-H} \equiv m_n-m_p-m_e=0.782$ MeV is the mass difference between the neutron and hydrogen atom, the energy of the lowest excited state is taken to be the sum of the lowest proton and neutron quasiparticle energies, and
\beqn 
Q_{\beta^-} =  \Delta M_{n-H} + E_i - E_f \,, 
\eeqn 
where $E_i$ and $E_f$ are the energies of the initial and final nuclear states, respectively.

According to Eq.\ (\ref{eq:strength_function}), each transition matrix element can be written as,
\beqn
\label{eq:FAM4XY}
\Big|\bra{N, K^\pi}\hat{F}\ket{\QRPA}\Big|^2
&= & \lim_{\omega \to \Omega_N} \pi \gamma \frac{dB(\hat{F},\omega)}{d\omega} 
\eeqn
The derivation appears in  Appendix \ref{eq:appendC}.

\subsection{Transition matrix elements in the PAV-pnFAM}
\label{sec:trans}
In the axially  deformed pnQRPA, the wave function of the $N^\text{th}$ excited state in a given channel, $\ket{N, K^{\pi}}$, has the form~\cite{Ring:1980}
\begin{equation}
    \ket{N, K^{\pi}} = \hat{Q}^{\dagger}_{N, K^\pi} \ket{\QRPA} \,,
\end{equation}  
and its eigenenergy is $E_N=\Omega_N$ (we omit the $K$ and $\pi$ labels and take $\hbar=1$ here for convenience). 
The phonon creation operator $\hat{Q}^{\dagger}_{N, K^{\pi}}$ can be written in terms of transition amplitudes as~\cite{Arteaga:2008,Martini:2014,Hinohara:2022uip}
\begin{equation}
\hat{Q}^{\dagger}_{N, K^{\pi}}
=\sum_{pn}\Bigg[X^{(N, K^{\pi})}_{pn}\beta^\dagger_{p}\beta^\dagger_{n}
- Y^{(N, K^\pi)}_{pn}\beta_{  n}\beta_{  p}\Bigg] \,,
\end{equation} 
and
 \begin{eqnarray}
\hat{Q}_{N, K^{\pi}}
=\sum_{pn}
\Bigg[X^{(N, K^{\pi})\ast}_{pn}\beta_{n }\beta_{p }
- Y^{(N, K^\pi)\ast}_{pn}\beta^\dagger_{ p}\beta^\dagger_{  n}\Bigg] \,.
\end{eqnarray}
The QRPA transition amplitudes $X^{(N,K^{\pi})}_{pn}$ and $Y^{(N,K^{\pi})}_{pn}$ in $\hat{Q}$ are related to the pnFAM amplitudes, which are functions of $\omega$, by~\cite{Litvinova:2007gg,Hinohara:2013qda}
\bsub
\label{eq:transition_amplitude_pnFAM_pnQRPA}
\beqn 
X^{(N, K^\pi)}_{pn}
&=&\frac{1}{\bra{N, K^{\pi}}\hat{F}\ket{\QRPA}}\frac{1}{2\pi i} \oint_{C_N} \mathcal{X}^{(K^\pi)}_{{pn}}(\hat{F};\omega)d\omega \\
Y^{(N, K^\pi)}_{pn}
&=&\frac{1}{\bra{N, K^{\pi}}\hat{F}\ket{\QRPA}}\frac{1}{2\pi i} \oint_{C_N} \mathcal{Y}^{(K^\pi)}_{{pn}}(\hat{F};\omega)d\omega  \,,
\eeqn 
\esub
where $\bra{N, K^{\pi}}\hat{F}\ket{\QRPA}$ is obtained from Eq.(\ref{eq:FAM4XY}). As pointed out in Ref. \cite{Hinohara:2013qda}, the above relations are valid up to an unknown phase that cannot be uniquely determined. We note that this phase does not affect the physical observables discussed in the present work.
Instead of being derived from contour integrals, the QRPA transition amplitudes $X^{(N,K^{\pi})}_{pn}$ and $Y^{(N,K^{\pi})}_{pn}$ are obtained directly using the following relations~\cite{Liu:2024PRC}  
\bsub
\begin{eqnarray}
\label{eq:X_QRPA_aprooximation_X}
   X^{(N, K^{\pi})}_{pn}
   &=&\lim_{\omega \to\Omega_N}\frac{ (\omega-\Omega_N ){\rm Im}[\mathcal{X}^{(K^{\pi})}_{pn}(\hat{F};\omega)]}{\bra{N, K^{\pi}}\hat{F}\ket{\QRPA}}  \,,\\  
\label{eq:X_QRPA_aprooximation_Y}
   Y^{(N, K^{\pi})}_{pn}
   &=&\lim_{\omega \to\Omega_N}\frac{ (\omega-\Omega_N ){\rm Im}[\mathcal{Y}^{(K^{\pi})}_{pn}(\hat{F};\omega)]}{\bra{N, K^{\pi}}\hat{F}\ket{\QRPA}} \,. 
\end{eqnarray}
\esub 

With the $X^{(N, K^\pi)}_{pn}$ and $Y^{(N, K^\pi)}_{pn}$ obtained from Eqs. (\ref{eq:X_QRPA_aprooximation_X}) and (\ref{eq:X_QRPA_aprooximation_Y}), one constructs the $\hat {Q}_{N, K^\pi}$ operators and uses projection to restore the rotational symmetry that they violate. 
In the pnQRPA with AMP, the wave functions for the initial state and final state in the $\beta^-$ decay of an even-even nucleus are approximated as follows, 
\bsub
\label{eq:projected_wfs}
\beqn
\ket{  0^+_i  } &=& \frac{1}{\sqrt{\mathcal{N}_i}} \hat{P}^{J_i=0}_{00}    \ket{\rm QRPA},  \\
\ket{N, J^\pi M K} &=& \frac{1}{\sqrt{\mathcal{N}^{(N)}_f}} \hat P^{J}_{MK} \ket{N, K^{\pi}} \,, 
\eeqn 
\esub
where the corresponding normalization factors are given by 
\bsub\beqn 
\label{eq:normalization_factor_initial_state}
\mathcal{N}_i &=&\bra{\QRPA}    \hat{P}^{J_i=0}_{00}    \ket{\QRPA}, \\
\label{eq:normalization_factor_final_state}
\mathcal{N}^{(N)}_f &=& \bra{\QRPA}    \hat{Q}_{N, K^\pi}  \hat{P}^{J}_{KK}       \hat{Q}^{\dagger}_{N,K^\pi} - \hat{Q}^{\dagger}_{N, K^\pi}  \hat{P}^{J}_{K K} \hat{Q}_{N, K^\pi}  \ket{\QRPA}  \,.\nonumber\\
\eeqn
\esub
More detailed expressions appear in the Appendix \ref{eq:appendA}. 

With the wave functions (\ref{eq:projected_wfs}) of initial and final states projected onto the right angular momentum, the transition matrix element of the tensor operator $\hat{T}_{\lambda\mu}$ is given by
\begin{align}
 \label{eq:projected_ME}
M_{\text{PAV}}^{(N)} 
&= \langle N, J^\pi MK | \hat{T}_{\lambda\mu} | J^\pi_i = 0^+, M_i = 0, K_i = 0 \rangle \nonumber\\
&= \frac{1}{\sqrt{\mathcal{N}_i \mathcal{N}_f^{(N)}}}
\langle \text{QRPA} | \hat{Q}_{N,K^{\pi}} \hat{P}^J_{KM} \hat{T}_{\lambda\mu} \hat{P}^{0}_{00} | \text{QRPA} \rangle.
\end{align} 
The detailed expression is derived in Appendices~\ref{eq:appendB} and~\ref{eq:appendC}. 

In the above expressions, we have replaced the QRPA ground-state wave function with the HFB wave function, in close analogy to the quasi-boson approximation frequently employed in QRPA studies without symmetry projections~\cite{Ring:1980}. With this approximation, the transition matrix element is evaluated using Eq.~(\ref{eq:projected-ME-appendix}). It is worth emphasizing that, due to the presence of the projection operator, this approximation is not as well defined as in the standard QRPA framework, as discussed in Refs.~\cite{Erler:2014,Porro:2024}. It may lead to violations of sum rules, which will be investigated subsequently.

\subsection{$\beta$-decay half-lives}
\label{sec:beta}
The half-life of nuclear $\beta^-$ decay is related to the decay rate by the expression,
\begin{equation}
T_{1/2}=\frac{\ln2}{\sum_{F} \lambda^{\rm PAV}_F}, 
\end{equation}
where the summation runs over both Fermi and GT transitions. 
The corresponding transition operators are
\beqn
\hat F_{\rm Fermi} = \sum_i g_V \tau_- (i), \quad \hat F_{{\rm GT},K} = \sum_i g_A  \mathbf{ \hat \sigma}_K(i)\tau_-(i)  \,,
\eeqn 
where $\tau_-(i)$ changes the $i$-th neutron to a proton and we have now written the $z$ components $K$ of the GT operator explicitly. 
Following Ref.\cite{Mustonen:2014bya}, we use $g_V = 1$ and a quenched value for the axial-vector coupling constant, $g_A = 1.0$, instead of its bare value.
For each type of transition, the decay rate $\lambda_F$ is proportional to the sum of the individual transition strengths $B^{F}_N$ to all energetically allowed states in the daughter nucleus, each weighted by the corresponding phase space integrals~\cite{Mustonen:2014bya}. 
With AMP in the PAV scheme, the decay rate is given by
 \begin{eqnarray}
  \label{eq:transition-strength}
     \lambda^{\rm PAV}_{\hat F} 
     &=&   \frac{\ln2} {\kappa} \sum_{N,K}f(\Omega_N) B^{\hat F}_{K}(\Omega_N)\nonumber\\\nonumber\\
     &=&  
    \frac{\ln2}{\kappa}\frac{1}{2\pi i} \sum_{N,K} \oint_{C_{N}} d \omega f_{ }(\omega) S_{\rm PAV} \left(\hat F, K^{\pi};\omega \right) 
 \end{eqnarray} 
where the constant is $\kappa = 6147.0 \pm 2.4$ seconds (s) and the sum $K$ is over the Cartesian components of the GT operator.  
The sum for the Fermi operator has just a single term, with $K=0$. The projected response function $S_{\rm PAV} \left(\hat F, K^{\pi}; \omega \right)$ is defined in exactly the same way as $S\left(\hat F, K^{\pi}; \omega \right)$ in Eq.~(\ref{strength-function}), except that the transition matrix elements are replaced by their projected versions.

The phase space factor  $f(\Omega_N)$ in Eq. (\ref{eq:transition-strength}) encodes the details of final-state lepton kinematics,
\begin{equation}
\label{eq:phase_space_factor}
    f(\Omega_{N})=\int^{\omega'_N}_{1} d E_e p_e E_e(\omega'_N-E_e)^2   F_1(Z,E_e)L_1 \,,
\end{equation}
where $Z$ is the charge of the daughter nucleus, and  $\Omega_N=\omega_{\rm max}-\omega'_N$ with $\omega^\prime_N$ defined, e.g., in Fig.~\ref{fig:excitation-energy-schematic}.  The Coulomb correction $F_1(Z,E_e)$ can be derived analytically under the approximation that the electron scatters off a point charge $Z$ with non-relativistic kinematics~\cite{Bahcal:1966,Doi:1985,Mustonen:2014bya},
\beqn
\label{eq:Coulfn}
    F_{\ell_e}(Z, E_e) & =&\left[\ell_e\left(2 \ell_e-1\right)!!\right]^2 4^{\ell_e}(2 p_e R_A)^{2\left(\gamma_{\ell_e}-\ell_e\right)} \nonumber\\
& &\times \exp (\pi \eta) \frac{\left|\Gamma\left(\gamma_{\ell_e}+i \eta\right)\right|^2}{\left[\Gamma\left(2 \gamma_{\ell_e}+1\right)\right]^2} \,.  
\eeqn
Here, $\Gamma$ is the Gamma function, $\eta=Z\alpha E_e/p_e$ with the fine structure constant $\alpha\simeq 1/137$, and $\gamma_\ell =\sqrt{\ell^2 - (Z\alpha)^2}$. 
The quantities $\ell_e$, $E_e$, and $p_e\equiv \sqrt{E^2_e-1}$ are the orbital angular momentum, energy, and momentum of the emitted electron, respectively.  All energies are normalized to the electron mass.  We approximate the Coulomb function $L_1$ in Eq. (\ref{eq:phase_space_factor}) by $L_1\simeq(1+\sqrt{1-(Z\alpha )^2})/2$, and take the nuclear radius to be $R_A=1.2A^{1/3}$ fm.  We neglect screening by atomic electrons~\cite{Bahcal:1966}.

Although we could use the first line of Eq.~(\ref{eq:transition-strength}) to obtain the decay rate, we choose instead to construct the strength function and do the contour integral in the second line.  To evaluate $f(\omega)$ from Eq. (\ref{eq:transition-strength}) in the complex plane, we replace it with a high-order polynomial,
\begin{equation} 
    f_{\rm poly}(\omega)=\sum^{N}_{k=0} a_k\left( \frac{\omega_{\rm max} -\omega}{m_e c^2}\right)^k \,,
\end{equation}
with the parameters $a_k$ fitted to the phase-space integral on the real axis~\cite{Mustonen:2014bya}.  
The circle $C$ in Eq. (\ref{eq:transition-strength}) should contain all the poles of the function $S_{\rm PAV}(\hat F,K^{\pi};\omega)$ below $Q_{\beta^-}$. 
One can choose the contour to be,
\begin{equation}
    \omega(t)= \frac{\omega_{\rm max}}{2}(1+e^{it}) \,,
\end{equation}
where the phase parameter $t$ runs from 0 to $2\pi$.  The contour integrals are performed under the requirement that each circular contour encloses only a single pole. Accordingly, the radius of the $i$-th contour is chosen to be smaller than $(E_{i+1}-E_{i-1})/2$, where $E_{i-1}$ denotes the excitation energy of the $(i-1)$-th excited state. For the numerical evaluation of the contour integrals, we employ 60 mesh points, following Ref.~\cite{Ney:2020mnx}.

 \section{Results and discussion}
\label{sec:Results and discussion}
 
\begin{figure}[tb]
    \centering 
    \includegraphics[width=8cm]{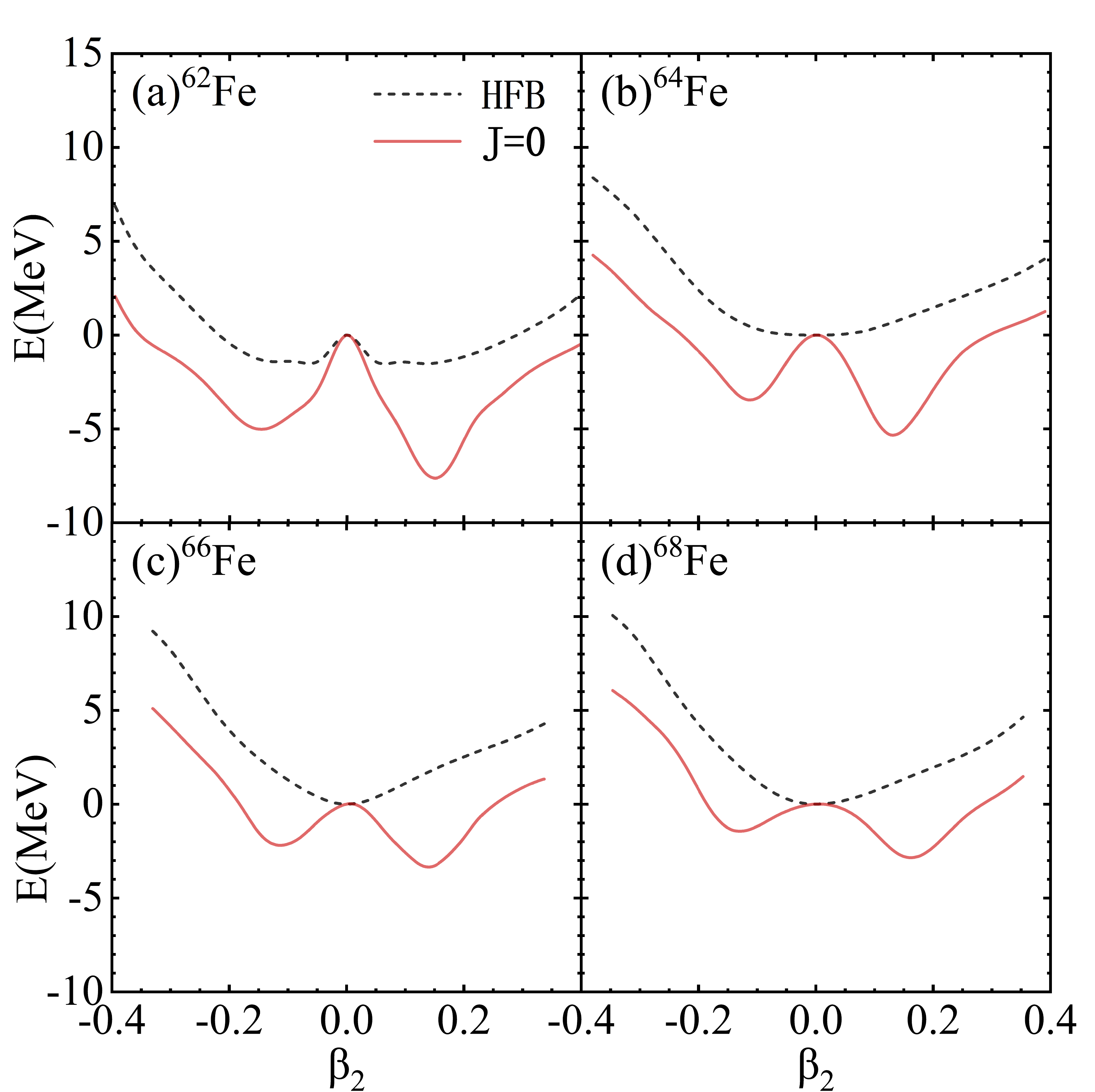}  
    \caption{The  energies of  HFB  states and those with projection onto angular momentum $J=0$ as a function of the quadrupole deformation $\beta_2$. All energies are normalized to those of the spherical states.}
    \label{fig:PEC-64Fe}
\end{figure}

The axially-deformed HFB wave functions $(U, V)$ are obtained from the code HFBTHO~\cite{STOITSOV200543,STOITSOV20131592,PEREZ2017363}, which is formulated in a harmonic oscillator (HO) basis, with an oscillator frequency $\omega_{\rm HO}$ of $\hbar\omega_{\rm HO} = 1.2 \times 41 A^{-1/3}$ MeV.   
Following Ref.~\cite{Mustonen:2015sfa}, we use the Skyrme EDF SKO'\cite{Reinhard:1999ut} throughout this paper.  
We generate the pairing correlations between nucleons with a density dependent pairing interaction of the form 
\beq 
\label{eq:pairing}
V_{\mathrm{pp}}=\left(V_0 \hat{\Pi}_{T=0}+V_1 \hat{\Pi}_{T=1}\right)\left(1-\alpha \frac{\rho_{t}(\mathbf{r})}{\rho_c}\right) \delta(\mathbf{r}) 
\eeq 
where $\rho_c=0.16$ fm$^{-3}$ is the saturation density, $\rho_{t}(\mathbf{r})$ is the total nucleon density, and $\alpha=0.5$, so that we use ``mixed pairing.'' 
The operator $\hat{\Pi}_{T}$ projects states onto those with well-defined isospin. 
Following Ref.~\cite{Mustonen:2015sfa}, we use $V_1=-264.216$ MeV fm$^3$ and limit the quasiparticle states to those with 
$E < E_{\rm cut}=60$ MeV, both in the HFB and in the sum over quasiparticle states $(p, n)$ in Eq.\ (\ref{eq:strength_sum}) for the response function.  
Figure~\ref{fig:PEC-64Fe} shows the energy surfaces of the nuclei \nuclide[62-68]{Fe}, obtained from the HFB calculation alone and from the HFB calculation supplemented by AMP onto the state with total angular momentum $J=0$. The Gauss-Legendre quadrature is employed with 12 mesh points for the Euler angle within the interval 
$[0,\pi]$. This calculation can be simplified further if one follows the so-called $M$-point Gauss-Legendre quadrature rule found in Ref.\cite{Bally:2020wkb}.
The mean-field energy surfaces for all the isotopes are rather soft around their energy minima.  In \nuclide[66,68]{Fe}, the lowest energy minima are located at spherical shapes in the HFB calculation.  
Projection shifts the lowest minima to prolate shapes with $\beta \simeq 0.15$, and makes them much more pronounced.  
In \nuclide[64]{Fe}, the $J=0$ minimum is located at $\beta_2\simeq 0.12$, which, while significant, is smaller than the value $\beta_2=0.29$ obtained from the measured $B(E2: 2^+\to 0^+)$~\cite{NNDC} and the rigid-rotor approximation~\cite{Ring:1980}.  

\begin{figure}[bt]
     \centering
     \includegraphics[width=8.4cm]{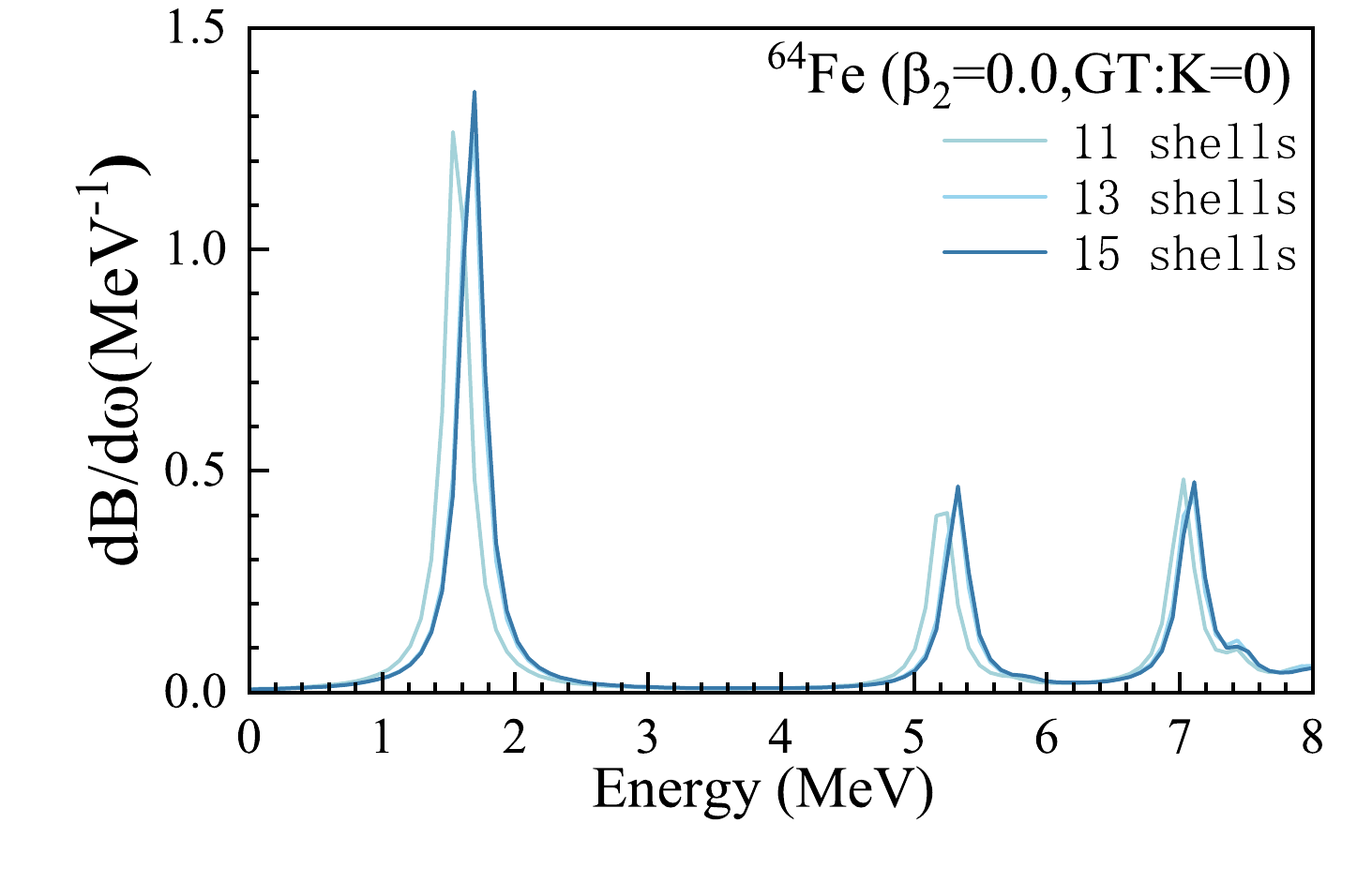}
     \caption{GT transition strength for $^{64}$Fe calculated in the pnFAM, starting from the spherical HFB state. Results are displayed for model spaces with increasingly large harmonic-oscillator shell numbers.
     }
     \label{fig:converge_check_for_64Fe}
 \end{figure}

For the pnFAM calculation, we take the time-odd parameters and isoscalar pairing strength ($V_0=-173.176$ MeV fm$^3$) from the {\tt 1A} parameter set of Ref.~\cite{Mustonen:2015sfa}.
With these choices, we examine the convergence of the GT transition strength in $^{64}$Fe with respect to the number  $N_{\rm sh}$ of major oscillator shells. 
The results of a calculation based on the spherical HFB state appear in Fig.\ref{fig:converge_check_for_64Fe}. 
To display the results, we have used a smearing parameter of $\gamma = 0.1$ MeV and an energy step size of $\Delta E = 0.02$ MeV, which is considerably smaller than in previous studies~\cite{Mustonen:2014bya,Liu:2024PRC,Ney:2020mnx,Mustonen:2015sfa}. 
This choice enables precise identification of both the positions and magnitudes of individual peaks in the strength function. 
We find that $N_{\rm sh} = 13$ major shells is enough for reasonably converged results.

 \begin{figure}[tb]
    \centering
\includegraphics[width=8cm]{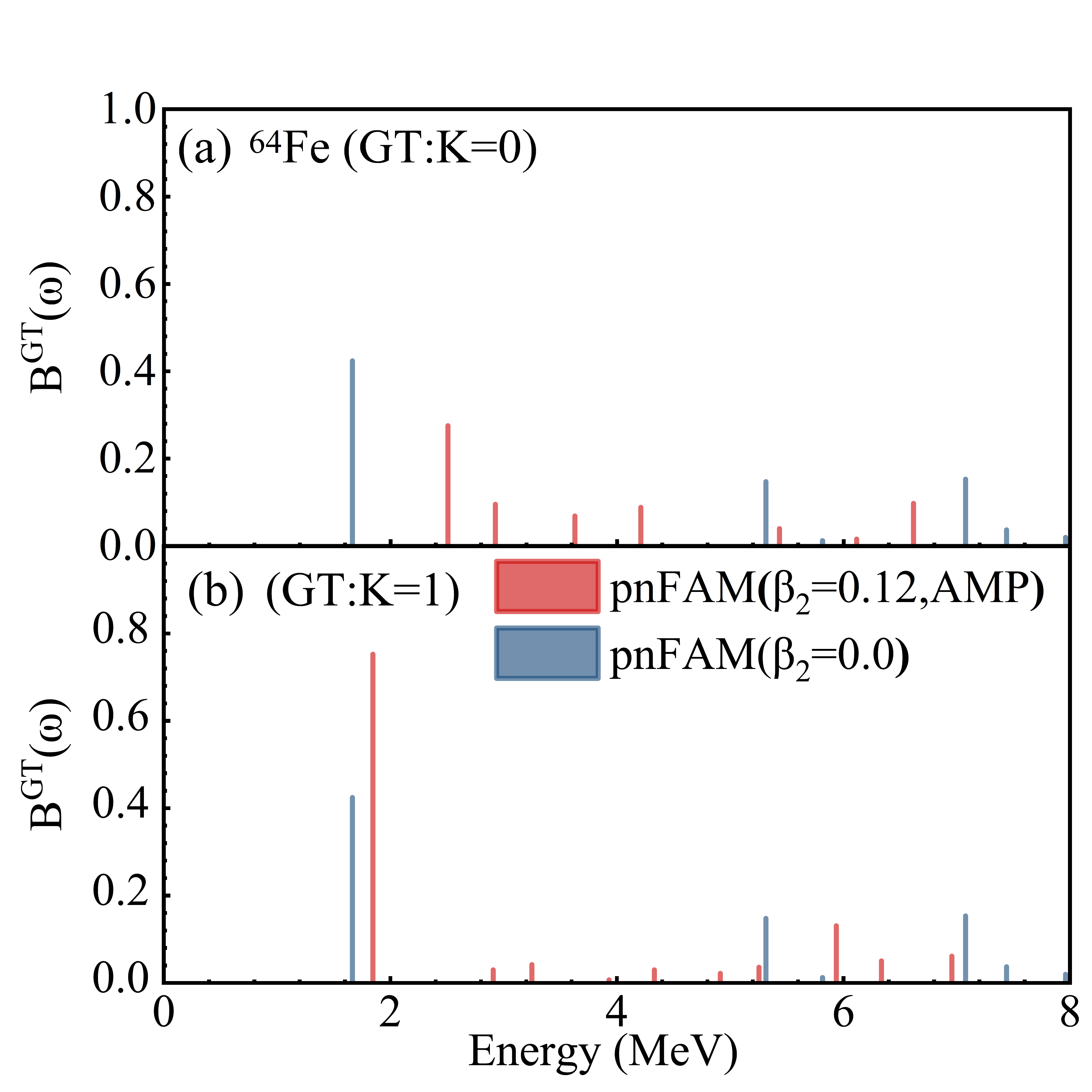}
\caption{The distribution of Gamow-Teller transition strength $B^{\rm GT}_{K}(\Omega_N)=|\bra{N, 1^+(K)}\hat\sigma\tau_-\ket{0^+_1}|^2$ in \nuclide[64]{Fe} as a function of energy, from calculations in the pnFAM and pnFAM+AMP, starting with HFB states with $\beta_2=0$ (a) and $\beta_2=0.12$ (b).
}  
  \label{fig:Fe64-spherical-prolate}
\end{figure}

 \begin{figure}[tb]
    \centering
  \includegraphics[width=8cm]{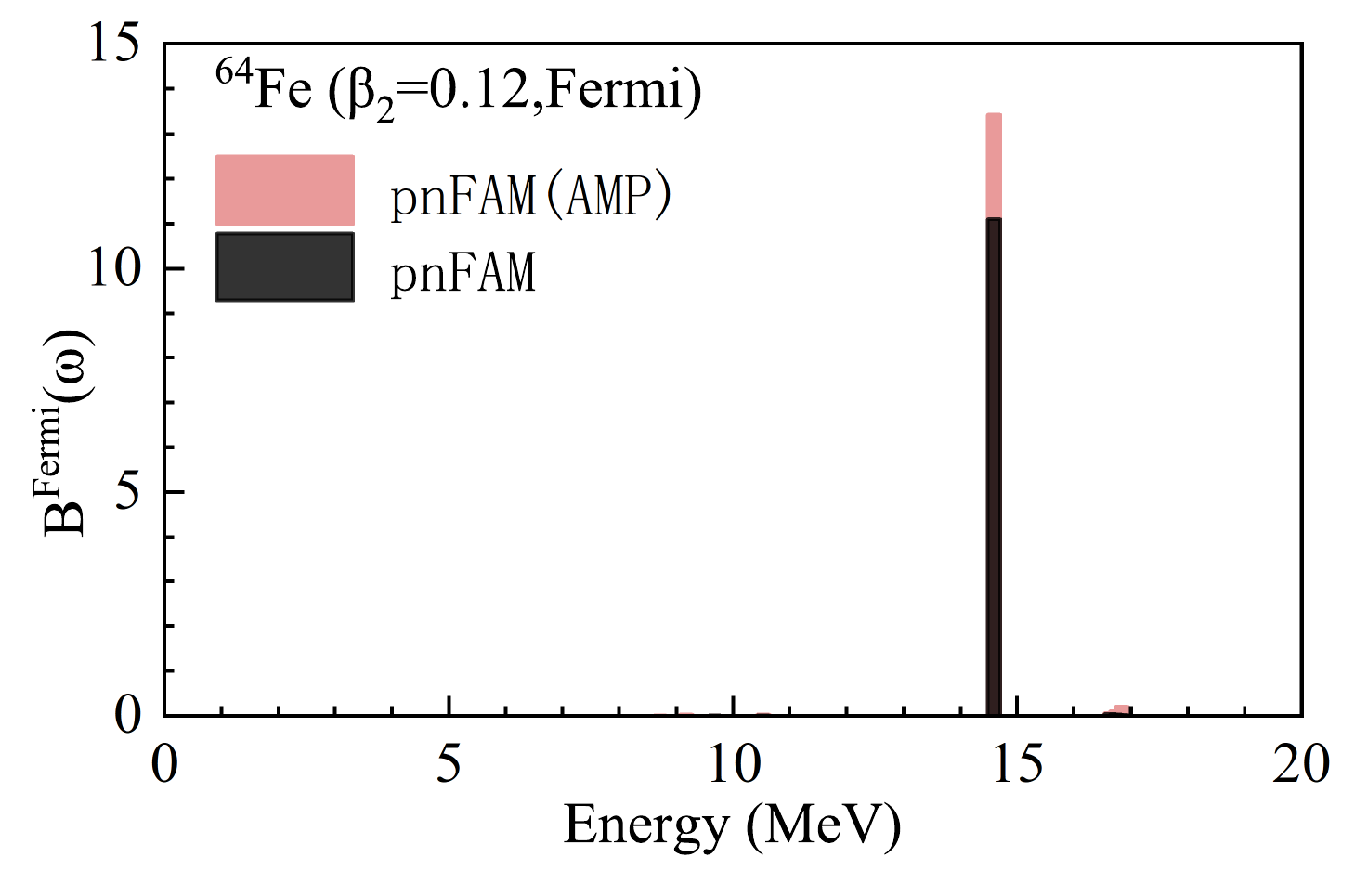}
  \caption{The distribution of Fermi transition strength $B^{\rm Fermi}_K(\Omega_N) =|\bra{N, 0^+(0)}\tau_-\ket{0^+_i}|^2$ in \nuclide[64]{Fe} as a function of energy, from calculations in pnFAM, with and without exact AMP, starting with the HFB state with $\beta_2=0.12$.}  
  \label{fig:Fe64-Fermi-AMP-PNP}
\end{figure}

\begin{table}[bt]
     \centering
   \tabcolsep=8pt
    \caption{The summed transition strengths in the Fermi and GT channels for \nuclide[64]{Fe}, from pnFAM calculations with and without exact AMP.}
    \begin{tabular}{ccccc}
     \hline\hline
       &  &     $S_-$ &  $S_+$ & $S_-  - S_+$   \\
          \hline 
    Fermi  &  pnFAM &  11.99&  0.01 &  11.98  \\
      &   pnFAM (AMP) &    13.42  & 0.73 & 12.69 \\ 
          \hline 
    GT  &  pnFAM & 36.33 &  0.34 & 35.99   \\
      &   pnFAM (AMP) &   36.74  &  0.42 &   36.32\\ 
        \hline\hline
     \end{tabular} 
  \label{tab:sum-rule-AMP-Fe64}
 \end{table}

To obtain most of the results to follow, and in particular the $\beta$-decay half-lives, we start from the  deformation-constrained HFB solution corresponding to the energy minimum for $J=0$, and use the code developed in Ref.~ \cite{Mustonen:2014bya} 
to solve the pnFAM equations for the amplitudes $\mathcal{X}^{(K^\pi)}_{pn}(\hat F; \omega)$ and $\mathcal{Y}^{(K^\pi)}_{pn}(\hat F; \omega)$. 
We then implement AMP through the procedure, described in Sections \ref{sec:trans} and \ref{sec:beta}.
That procedure is an approximation to VAP+AMP that was employed in a different context in Ref.~\cite{Erler:2014}.
It is worth noting that the energy-minimum state with $J=0$ is usually not the minimum of the underlying HFB energy surface. In such cases, the QRPA matrix may no longer be positive definite~\cite{Ring:1980}, rendering the QRPA calculation potentially ill defined. To explore whether this issue occurs in our QFAM calculations based on a reference state outside the energy minimum, we have computed the GT transition strengths in $^{62}$Fe using mean-field reference states with quadrupole deformations ranging from $\beta_2 = 0$ to $0.44$.  We obtained reasonable GT strength distributions for all deformations considered, indicating that stability issues are not problematic in the present study. This conclusion is consistent with that in the previous QFAM calculations of the monopole transition strength distribution in $^{32}$Mg \cite{Mercier:2022}. 

Figure~\ref{fig:Fe64-spherical-prolate} compares the GT transition strengths to states with $K=0$ and 1,
\beq
B^{\rm GT}_K(\Omega_N) = |\bra{N, J=1^{+} (K)}| \hat F_{\rm GT} |\ket{0^+}|^2 \,, 
\eeq  
from the pnFAM (with the usual needle approximation to projection) and pnFAM+AMP calculations based, respectively, on spherical and prolate HFB states. 
The strength distributions become more fragmented after including deformation. The differences have two main sources.
First, the change in the reference state from spherical to prolate shifts the locations of the excited states in energy, and second, the restoration of rotational symmetry in the deformed states modifies the transition matrix elements. 
As we will see later, these two effects significantly reduce the $\beta$-decay half-life. 

Table~\ref{tab:sum-rule-AMP-Fe64} contains the total Fermi and Gamow–Teller ($\mathrm{GT}$) $\beta^{\pm}$ transition strengths to states with excitation energies up to 50 MeV.  Without AMP, the Ikeda sum rule~\cite{Ikeda64} is perfectly satisfied. When AMP is included, the sum rule is violated by approximately 6\% and 1\% for the Fermi and GT operators, respectively. \eda{The violation of the Ikeda sum rule in the QRPA calculations with AMP is mainly the consequence of particle-number nonconservation in the projected reference state. Moreover, the presence of the normalization factor 
${\cal N}^{N}_f$ in the denominator of the transition matrix elements (\ref{eq:projected_ME}) also leads to a violation of the sum rule. One possible remedy is to remove this normalization factor for the projected intermediate states, thereby restoring the Ikeda sum rule. We note that the violation of the sum rule for Fermi is larger than that for GT transitions. It can be understood as follows. The Fermi operator is a pure isospin operator and couples almost exclusively to the isobaric analog state (IAS). Consequently,   the Fermi sum rule is highly sensitive to a small violation of particle-number conservation and to the normalization factor, so that any AMP-induced enhancement of the IAS strength directly leads to an overestimation of the Fermi sum rule. See Fig.~\ref{fig:Fe64-Fermi-AMP-PNP}. In contrast, the GT operator involves both spin and isospin and its strength is fragmented over many final states, making the GT sum rule much less sensitive to these effects.   We expect that the inclusion of particle-number projection, in addition to AMP, would further reduce the deviations of both Fermi and GT sum rules. We plan to investigate this improvement in future work.
}

\begin{figure}[tb]
    \centering
    \includegraphics[width=8.3cm]{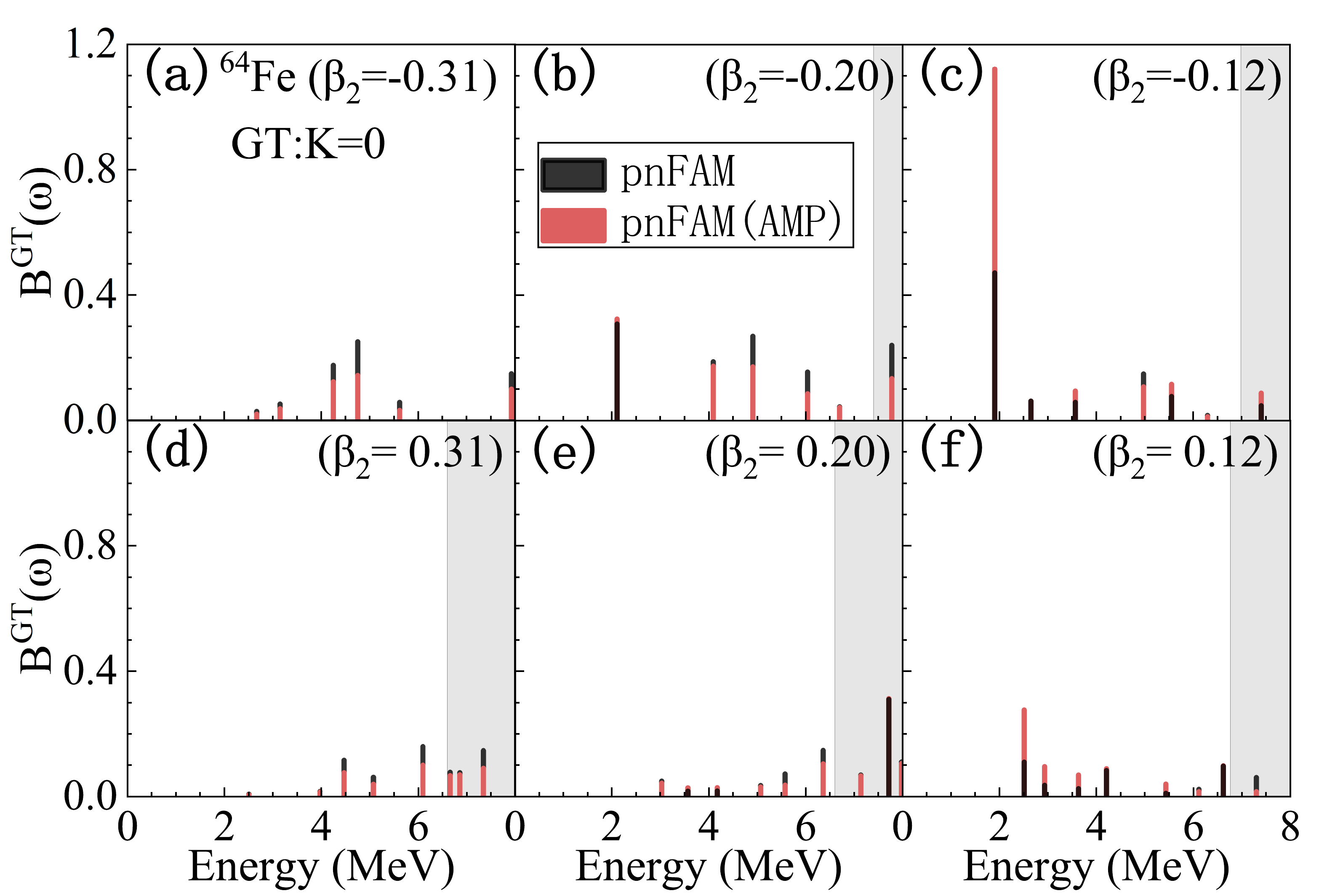}
  \caption{Transition strength $B^{\rm GT}_K$ for the $K=0$ mode of the GT channel in $^{64}$Fe, calculated in the pnFAM, with and without exact AMP, starting from HFB states with various values of $\beta_2$. The transitions in the shaded area are above the threshold for $\beta^-$ decay.  
  }
    \label{fig:shapes-Fe64-AMP-GT0}
\end{figure}

\begin{figure}[tb]
    \centering
    \includegraphics[width=8.3cm]{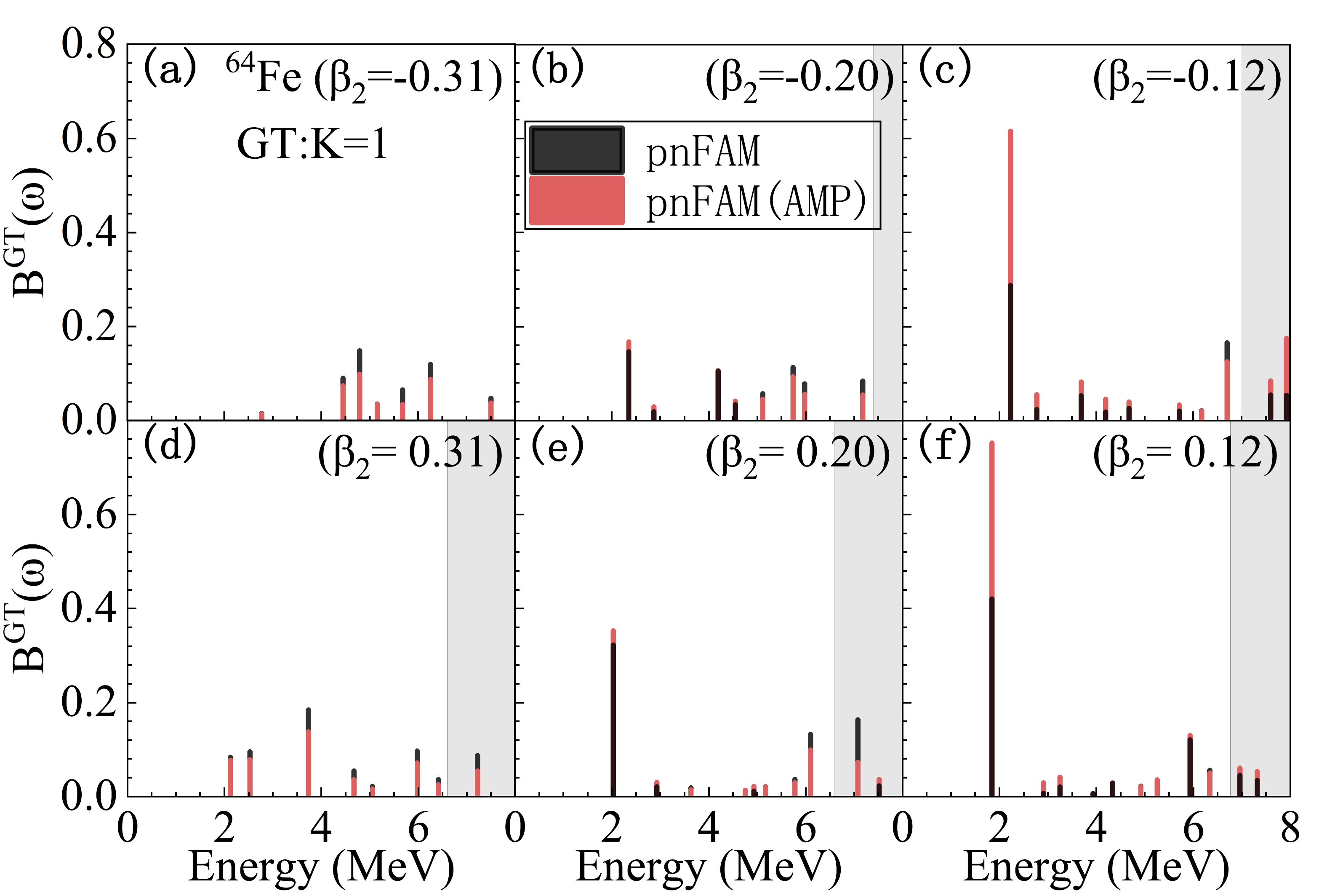}
    \caption{The same as Fig.~ \ref{fig:shapes-Fe64-AMP-GT0} , but for the $K=1$ mode in the  GT  channel.}
    \label{fig:shapes-Fe64-AMP-GT1}
\end{figure}

\begin{figure}[tb]
    \centering
    \includegraphics[width=7cm]{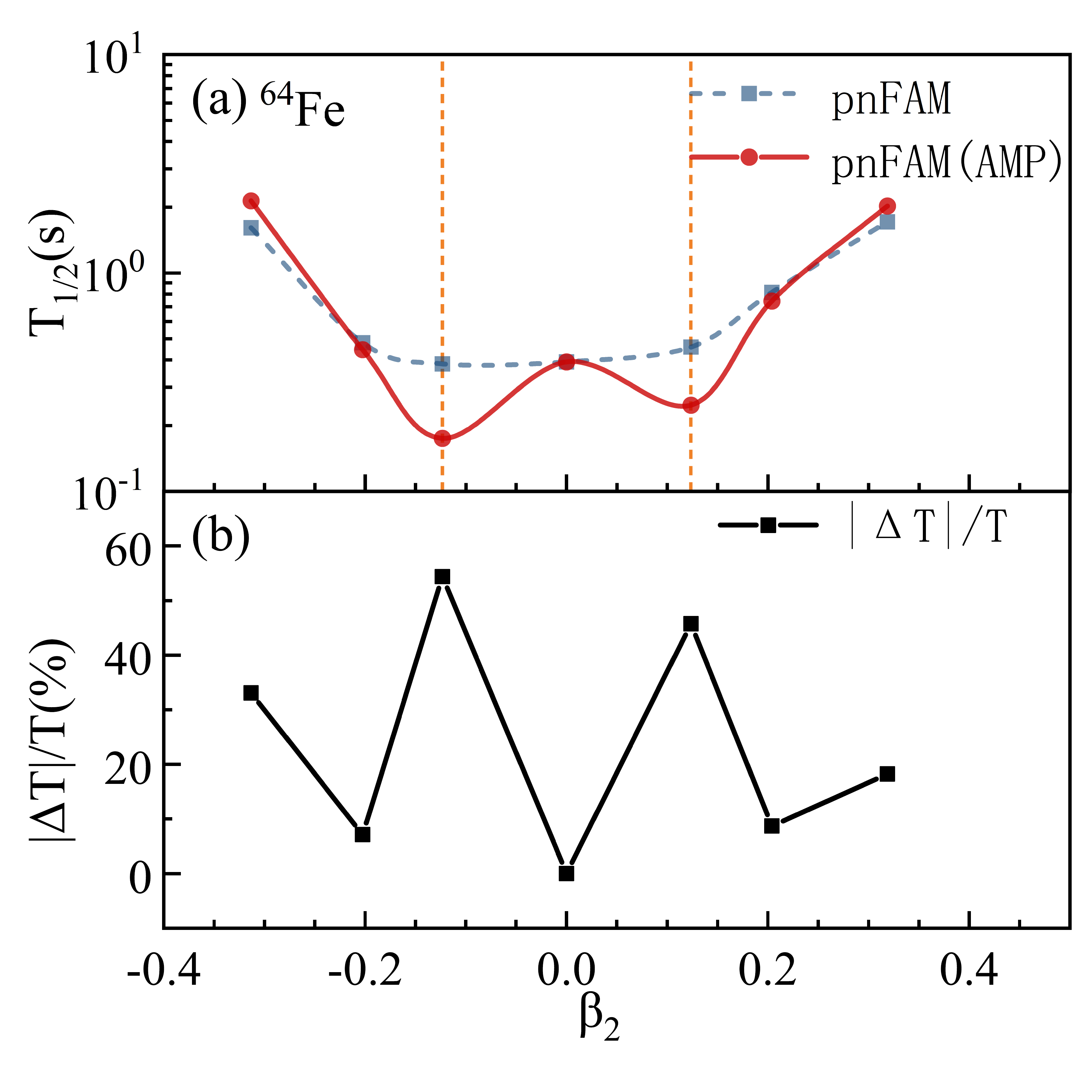}
    \caption{(a) The  $\beta$ decay half-life $T_{1/2}$ of \nuclide[64]{Fe} from  the pnFAM calculations with and without exact AMP, as a function of the quadrupole deformation $\beta_2$ of the intrinsic HFB state. 
    In the calculation without exact AMP, the needle approximation~\cite{Ring:1980} is used. The locations of the two energy minima are marked with dashed lines.  (b) The effect of AMP on  the $\beta$ decay half-life $T_{1/2}$ as a function of $\beta_2$. }
    \label{fig:Fe64-def-vs-half-lives-AMP}
\end{figure}

\begin{figure}[tb]
    \centering
    \includegraphics[width=8cm]{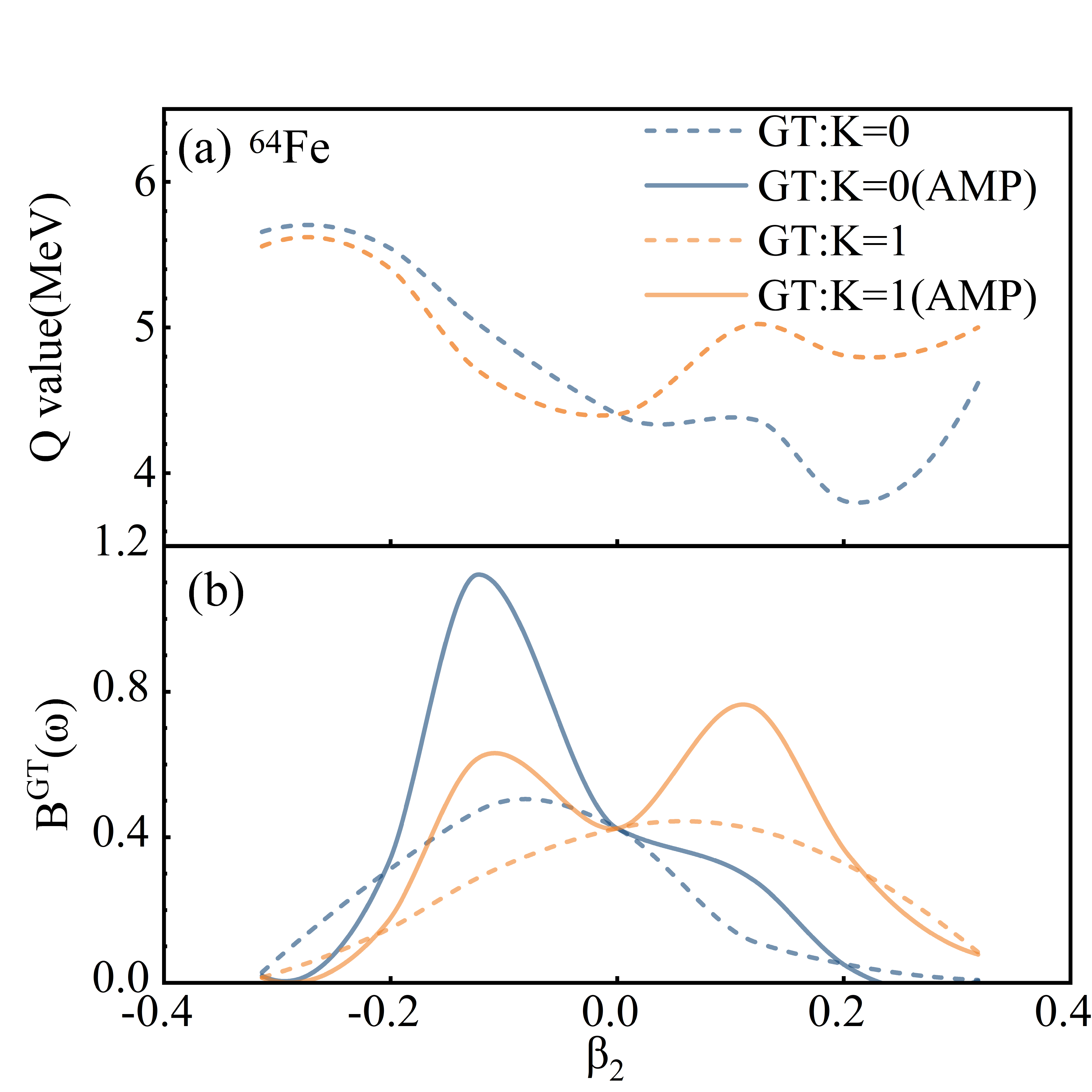}
    \caption{(a) Calculated (a) $Q$ values and (b) Gamow–Teller strengths $B^{\mathrm{GT}}(\omega)$ for the first strength peak in $^{64}$Fe, obtained with and without exact AMP, as functions of the quadrupole deformation $\beta_2$. The results correspond to those shown in Figs.~\ref{fig:shapes-Fe64-AMP-GT0} and \ref{fig:shapes-Fe64-AMP-GT1}, and are displayed separately for the $K=0$ and $K=1$ components.  }
    \label{fig:Q-value-BGT}
\end{figure}

 \begin{figure}[tb]
    \centering
    \includegraphics[width=8cm]{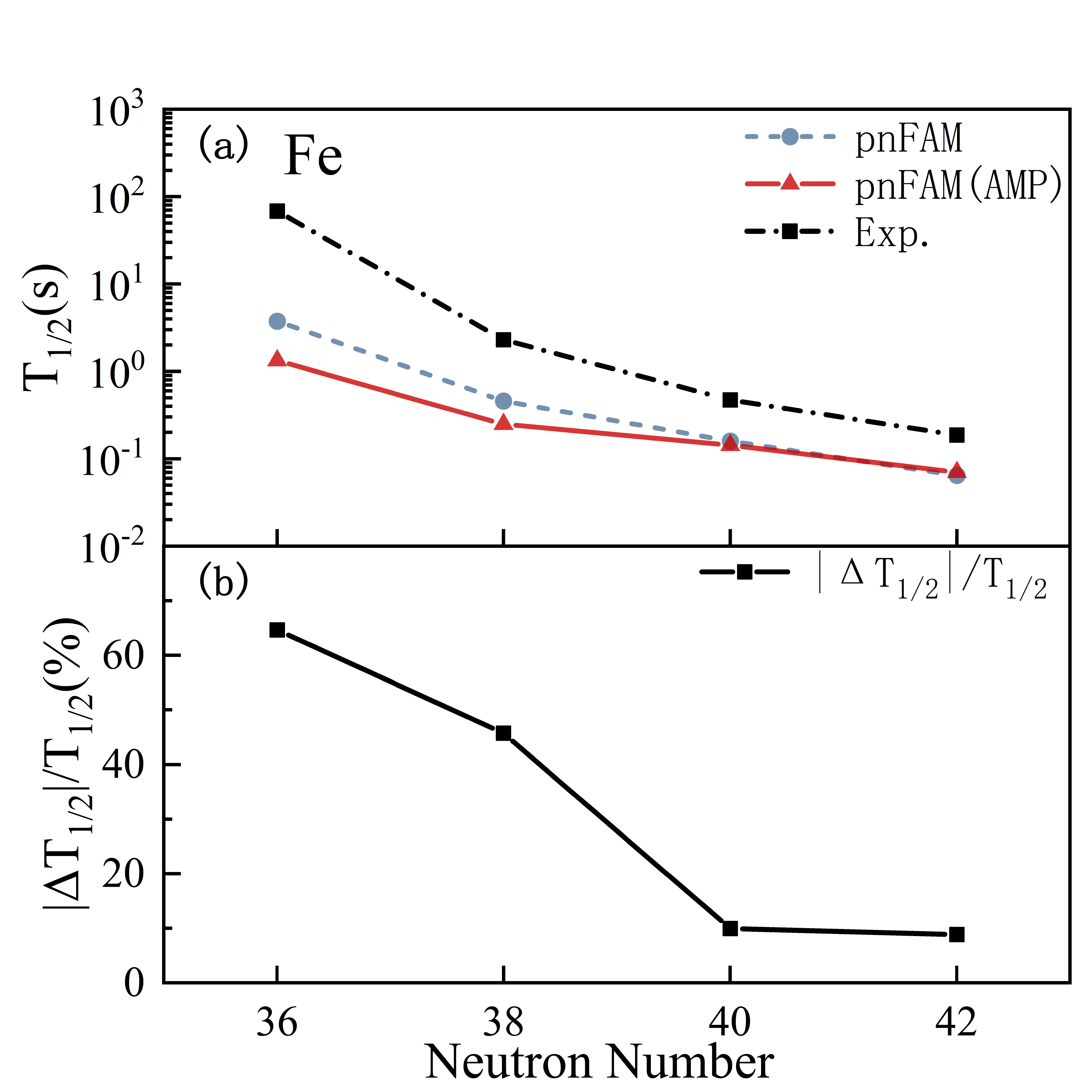}
    \caption{Comparison of calculated and experimental $\beta$-decay half-lives of Fe isotopes as a function of neutron number. (a) $\beta$ decay half-lives $T_{1/2}$ from the pnFAM (blue circles), pnFAM+AMP (red triangles) and experiment (black squares) on a logarithmic scale. (b) Relative deviation $|\Delta T_{1/2}|/T_{1/2}$ between the pnFAM+AMP results and pnFAM, expressed as a percentage.
}
    \label{fig:Fe-isotopes-half-lives-AMP}
\end{figure}

Figures~\ref{fig:shapes-Fe64-AMP-GT0} and \ref{fig:shapes-Fe64-AMP-GT1} show the effect of deformation on the GT transition matrix elements with AMP, for $K = 0$ and $K = 1$, respectively. 
The underlying HFB states span a range of quadrupole deformations $\beta_2$, enabling a comparative analysis of deformation effects. 
For the weakly-deformed configurations (prolate and oblate with $|\beta_2|=0.12$), exact AMP markedly increases the height of the first peak. 
As $|\beta_2|$ increases, the transition matrix elements gradually decrease, and the AMP effect evolves from enhancement to quenching. This trend demonstrates that quadrupole deformation can substantially alter the $\beta$-decay half-life, consistent with previous findings~\cite{Boillos:2015, Sarriguren:2017, Yoshida:2023, Ravlic:2024hpi}. Here, we explicitly quantify this deformation effect by restoring rotational symmetry. Figure~\ref{fig:Fe64-def-vs-half-lives-AMP} shows that the $\beta$-decay half-life of \nuclide[64]{Fe} decreases with increasing $|\beta_2|$ up to about $0.12$, beyond which it rises sharply at larger deformations.
In addition to deformation itself, exact AMP enhances or quenches strength by as much as 60\%. 
As expected, the AMP effect is strongest in weakly deformed states, generally reducing the half-life. 
This finding is consistent with earlier studies~\cite{Yousef:2009, Ravlic:2024hpi}, which showed that the needle approximation breaks down for weak deformation. 

To see more clearly how the $\beta$-decay half-life depends on quadrupole deformation $\beta_2$, we display in Fig.~\ref{fig:Q-value-BGT} the dependence of the $Q$ value and the Gamow–Teller (GT) strength in the first peak for $^{64}$Fe, both with and without exact AMP. 
For the weakly deformed configurations with $\beta_2 = \pm 0.12$, the Gamow–Teller (GT) strengths are markedly enhanced by AMP and become significantly larger than those of the spherical state. In addition, the $Q$ values increase with $|\beta_2|$, leading to a larger phase-space factor. Together, these effects reduce the $\beta$-decay half-life relative to that of the spherical configuration. Conversely, at stronger deformations, the GT strength is substantially quenched—despite the relatively large $Q$ value—resulting in a longer half-life.
It is particularly interesting that $K^\pi=0^+$ transitions mainly determine the $\beta$-decay half-life of oblate states, whereas the $K^\pi=1^+$ transitions are most important in prolate states. 
This behavior contrasts with that of the charge-conserving electric dipole (E1) strength, which is dominated by the $K^\pi=0^-$ mode in prolate deformed nuclei~\cite{Arteaga:2008}. 
With $\beta_2 = 0.12$, which corresponds to the energy minimum for projected states with $J=0$, exact AMP decreases the half-life of \nuclide[64]{Fe} from 0.46 s to 0.25 s with AMP. 
Both values are significantly smaller than the experimental half-life of 2.30 s~\cite{NNDC}.

We next extend the above analysis to other Fe isotopes. 
Figure~\ref{fig:Fe-isotopes-half-lives-AMP} compares the calculated $\beta$-decay half-lives of $\nuclide[62-68]{Fe}$ obtained from the pnFAM calculations with and without exact AMP, alongside experimental data~\cite{NNDC}.  
Including AMP reduces the half-lives by about 60–10\% for $\nuclide[62]{Fe}$–$\nuclide[66]{Fe}$.
In contrast, for $\nuclide[68]{Fe}$, AMP increases the half-life by roughly 10\%, as it suppresses the strength of the first Gamow–Teller peak.
Overall, the pnFAM calculations systematically underestimate the measured half-lives, and the inclusion of exact AMP further worsens the agreement with experiment.
Other extensions to the pnFAM, including two-body currents ~\cite{Ney:2022PRC} and particle-vibration coupling~\cite{Liu:2024PRC}, have also been explored recently, and each has a significant effect on half-lives~\cite{Niu:2015, Shafer:2016etk}.  
Because parameters such as the spin-density coupling constant $C^s_1$~\cite{Mustonen:2015sfa} that multiply time-odd terms in the EDF (as well as the strength of isoscalar pairing) are fitted in part to half-lives, the addition of any new physics, whether AMP or something else, requires a refitting of those parameters. 
These considerations aside, exact AMP has a clear impact on the $\beta$-decay half-lives of $\nuclide[62,64]{Fe}$ and a moderate impact on those of $\nuclide[66,68]{Fe}$.  

\section{Summary and perspectives}
\label{sec:summary}

In this work, we have extended the axially deformed pnQRPA framework by implementing exact AMP in the calculation of transition matrix elements for nuclear $\beta$ decay with a Skyrme energy density functional.  
We have investigated the impact of AMP on Fermi and GT transitions, as well as on the resulting $\beta$ decay half-lives of $\nuclide[62-68]{Fe}$, and have closely examined the effects of deformation more broadly by starting from a set of axially deformed HFB states with varying amounts of deformation.
Our results demonstrate that deformation usually reduces $\beta$ decay half-lives, and that exact AMP makes half-lives up to 60\% shorter than those computed in the needle approximation to projection.  
This result is an example of the essential role of accurate symmetry restoration in achieving reliable results for transition rates in EDF theory. 

We note, finally, that the present study is restricted to AMP within the projection-after-variation (PAV) scheme. Moreover, this work does not yet include particle-number projection. As a result, the Ikeda sum rule is not exactly fulfilled in the QRPA calculations with AMP, since the average particle number of the angular-momentum projected reference state is no longer conserved. In addition, the normalization factor appearing in the denominator of the transition matrix elements renders the associated closure relation only approximate, introducing a further source of sum-rule violation. A more complete assessment will be pursued once these refinements are incorporated.

\section*{acknowledgments} 
We thank N. Hinohara, C.F. Jiao, Q.Q. Liu, W.L. Lu, E. M. Ney,  Y.F. Niu, and P. Ring for fruitful discussions.  This work is supported in part by the National Natural Science Foundation of China (Grant Nos.  12375119, 12305129, and 12141501), and the Guangdong Basic and Applied Basic Research Foundation (2023A1515010936).  J.E. is  supported by the US Department of Energy, Office of Nuclear Physics, under grant DE-FG02-97ER41019.

 \section*{Data availability statement}
The data that support the findings of this article are openly available \cite{Data}.

\begin{appendix}

\section{Normalization factors for the wave functions of ground state and QRPA excited states}
 \label{eq:appendA}
The  normalization factor $\mathcal{N}_i$ for the wave function of the initial state (ground state) with spin-parity $J^\pi_i=0^+$  is defined as follows:
\beqn  
 \mathcal{N}_i  
&\equiv& \bra{J^{\pi}_i}    J^{\pi}_i\rangle \nonumber\\
&=&  \bra{\QRPA}  \hat P^{J_i}_{KK} \ket{\QRPA}  \nonumber\\
&\simeq& \bra{\HFB } \hat{P}^{0}_{00}  \ket{\HFB } \,,
\eeqn 
where the QRPA ground state is approximated by an axially  deformed HFB state labeled by the quantum numbers $K^\pi_i$.  The AMP operator is~\cite{Ring:1980}  
\begin{eqnarray}
    \hat{P}^{J}_{MK}= \frac{2J+1}{8\pi^2} \int d \Omega D^{J\ast}_{MK}(\Omega) \hat{R}(\Omega) \,,
\end{eqnarray} 
where $D^{J}_{MK}(\Omega)$ is the Wigner-D function,
and  $\hat{R}(\Omega) = e^{-i\alpha \hat{J}_{z}} e^{-i\beta \hat{J}_{y}}e^{-i\gamma \hat{J}_{z}}$ is the rotation operator. 
Substituting  the above expressions, one finds for the initial-state normalization factor, 
   \begin{eqnarray}
     \label{normalization:N_i}
    \mathcal{N}_i 
     & =& \frac{2J_i+1}{2} \int^{+1}_{-1} d(\cos\beta)  d^{J_i}_{K K}(\beta)         \delta_{KK_i} \mathbb{N}( \beta) \,.
    \end{eqnarray}  

For even-even nuclei, $J_i=K_i=0$. The normalization factor $ \mathcal{N}_i $ reduces to
\beq 
N_i  = \frac{1}{2} \int^{+1}_{-1} d(\cos\beta) \mathbb{N}(\beta) \,,
\eeq 
where the norm overlap is defined by 
\begin{eqnarray}
\label{eq:norm_overlap_angles}
 {\mathbb{N}}( \beta)
& \equiv &\langle {\rm HFB}  | e^{-i\beta\hat{J}_{y}}| {\rm HFB}
\rangle={\mathbb{N}}^{(N)}( \beta){\mathbb{N}}^{(P)}( \beta) \,.
\end{eqnarray}
Because HFB is time-reversal invariant, the overlap can be determined by the Onishi formula~\cite{Onishi:1966}, 
 \begin{eqnarray}
 {\mathbb{N}}^{(\tau)}( \beta)
 =  \sqrt{\rm det [\mathbb{U}^{(\tau)}( \beta)]}.
\end{eqnarray}
 \eda{More discussions can be found in Refs.~\cite{Neergard:1983,Robledo2009yd,Mizusaki2017ist,Porro2022tgc}.}  The matrix element $\mathbb{U}^{(\tau)}( \beta) $ is defined by~\cite{Yao:2022PPNP} 
 \begin{eqnarray}
 \label{eq:bbU}
   \mathbb{U}^{(\tau)}( \beta) & \equiv U^{{(\tau)}\dagger}( \beta )U^{{(\tau)}}   + V^{(\tau)\dagger} ( \beta) V^{{(\tau)}}  \,,
\end{eqnarray}
in terms of the rotated $U$ and  $V$ matrices,
\bsub\beqn
U^{(\tau)}(  \beta) 
&=& D[\hat{R}_y(\beta)]U^{(\tau)},\\
V^{(\tau)}(  \beta) 
&=&  D^\ast[\hat{R}_y(\beta)]V^{(\tau)} \,.
\eeqn 
\esub
Here, $\mathbb{U}^{(\nu)}$ and $\mathbb{U}^{(\pi)}$ denote the matrix elements (\ref{eq:bbU}) of neutrons and protons, respectively. The matrix $D[\hat{R}_y(\beta)]$ is the matrix representation of the rotation operator $\hat R_y(\beta)$.

The  normalization factor $N^{(N)}_f$ for the $N$-th final state with spin-parity $J^\pi_f (K) $ is defined as follows:
\beqn  
\mathcal{N}^{(N)}_f 
&\equiv& \bra{ N, J^\pi   MK}  N, J^\pi   MK \rangle \nonumber\\ 
&=& \eda{\bra{\QRPA } \hat{Q}_{N, K^\pi}  \hat{P}^{J_f}_{K K}      \hat{Q}^{\dagger}_{N, K^\pi}-\hat{Q}^\dagger_{N, K^\pi}  \hat{P}^{J_f}_{K K}      \hat{Q}_{N, K^\pi}  \ket{\QRPA  }}\nonumber\\
&=& \sum_{pnp'n'} \Bigg[X^{(N, K^{\pi})\ast}_{pn} X^{(N, K^{\pi})}_{p'n'}
-Y^{(N, K^\pi)\ast}_{pn}Y^{(N, K^\pi)}_{p'n'}\Bigg] \,.
\nonumber\\
&&\times \bra{\QRPA } 
 \beta_{p}\beta_{n}     \hat{P}^{J_f}_{K K}      \beta^\dagger_{n' }  \beta^\dagger_{p' }  \ket{\QRPA  }.
\eeqn
In the above derivation, we have used the property,
 \beq 
\bra{\QRPA }\hat{Q}^\dagger_{N, K^\pi} = 0,\quad 
\hat{Q}_{N, K^\pi} \ket{\QRPA }=0
\eeq 
Substituting the expressions for the  projection  operators and making the quasi-boson approximation~\cite{Ring:1980}, one finds 
    \begin{eqnarray}
    \label{app:norm_f}
    \mathcal{N}^{(N)}_f 
     & =& \frac{2J_f+1}{2} \int^{+1}_{-1} d(\cos\beta)  d^{J_f}_{KK}(\beta)     \nonumber\\
     & \times&\Bigg(\bra{\HFB } \hat{Q}_{N, K^\pi}e^{-i\beta \hat{J_{y}}}        \hat{Q}^{\dagger}_{N, K^\pi}   \ket{\HFB }  \nonumber\\
     & & -\bra{\HFB } \hat{Q}^\dagger_{N, K^\pi}e^{-i\beta \hat{J_{y}}}        \hat{Q}_{N, K^\pi}   \ket{\HFB }\Bigg) \,, 
    \end{eqnarray}  
where  the overlap takes the form 
 \beqn 
  &&   \bra{\HFB } \hat{Q}_{N, K^\pi}e^{-i\beta \hat{J_{y}} }       \hat{Q}^{\dagger}_{N, K^\pi}   \ket{\HFB } \nonumber\\
&=&\sum_{pnp'n'}X^{(N, K^{\pi})*}_{pn} X^{(N, K^{\pi})}_{p'n'}\bra{\HFB }   \beta_{n}\beta_{p}    e^{-i\beta \hat{J_{y}}}          \beta^{\dagger}_{p'}\beta^{\dagger}_{n'} \ket{\HFB } \nonumber\\ 
   &=&\sum_{pnp'n'}X^{(N, K^{\pi})*}_{pn} X^{(N, K^{\pi})}_{p'n'} \mathbb{U}^{(\pi)-1}_{pp'} (  \beta) \mathbb{U}^{(\nu)-1}_{nn'} (  \beta) 
            \mathbb{N}(  \beta) \,, 
\eeqn  
and  we have introduced the rotated HFB state,
\beq 
\ket{\HFB(\beta)}
\equiv  \frac{e^{-i\beta\hat{J}_{y}}\ket{\HFB }}{\bra{\HFB } e^{-i\beta\hat{J}_{y}}\ket{\HFB }} \,.
\eeq 
 The second term in Eq.\ (\ref{app:norm_f}) can be simplified:
 \begin{eqnarray}
     &&   \bra{\HFB} \hat{Q}^\dagger_{N, K^\pi}e^{-i\beta\hat{ J_{y}}}        \hat{Q}_{N, K^\pi}   \ket{\HFB}\nonumber\\
&=&\sum_{pnp'n'}Y^{(N, K^{\pi})}_{p'n'}Y^{(N, K^{\pi})*}_{pn}  \mathbb{U}^{(\pi)-1}_{p'p}
       (  \beta) \mathbb{U}^{(\nu)-1}_{n'n}
       (  \beta)\mathbb{N}(  \beta) \,,
       \end{eqnarray}
where we have used the relation~\cite{Yao:2010}, 
\begin{eqnarray}
 \bra{\rm HFB}    \beta_{n}\beta_{p} e^{-i\beta\hat{ J_{y}}}          \beta^{\dagger}_{p'}\beta^{\dagger}_{n'}  \ket{\rm HFB} = \mathbb{U}^{(\pi)-1}_{pp'} (  \beta) \mathbb{U}^{(\nu)-1}_{nn'} (  \beta) \mathbb{N}(  \beta) \,.\nonumber\\
\end{eqnarray}
The normalization factor for the AMP-projected excited state can be calculated as follows: 
    \begin{eqnarray}
    \mathcal{N}^{(N)}_f 
     & =& \frac{2J_f+1}{2} \int^{+1}_{-1} d(\cos\beta)  d^{J_f}_{KK}(\beta) \mathbb{N}(  \beta)\nonumber\\
     & \times&
  \eda{ {\rm Tr}\Bigg[(\mathbb{U}^{(\nu)-1})^{T}(  \beta)X^{(N,K^{\pi})\dagger} \mathbb{U}^{(\pi)-1}(  \beta)X^{(N,K^{\pi})}} \nonumber\\
   && \eda{-(\mathbb{U}^{(\nu)-1})^{T}(  \beta)Y^{(N,K^{\pi})\dagger} \mathbb{U}^{(\pi)-1}(  \beta)Y^{(N,K^{\pi})}\Bigg]} \,, \nonumber\\
    \end{eqnarray}  
  where the norm overlap is in Eq.(\ref{eq:norm_overlap_angles}).

\section{Transition matrix elements}
\label{eq:appendB}
With the wave functions of the initial state $\ket{ J^\pi_i M_i K_i }$ and the $n$-th final state $\ket{ N, J^\pi_f M_f K_f}$ projected onto states with good angular momentum, the transition matrix element of the tensor operator $\hat{T}_{\lambda\mu}$ is determined by
   \begin{eqnarray}
    \label{eq:projected-ME-appendix}
  M^{(N)}_{\rm PAV}
   &=&   \bra{  N, J^\pi_f M_f K_f}  \hat{T}_{\lambda\mu} \ket{  J^\pi_i M_i K_i  } \nonumber\\
    &=&  \frac{1} {\sqrt{N_{i} N^{(N)}_{f}}}\langle {\rm QRPA}   | \hat{Q}_{N, K^{\pi}}          \hat{P}^{J_f}_{K_{f} M_f}  \hat{T}_{\lambda\mu}           \hat{P}^{J_i}_{M_iK_{i}} | {\rm QRPA}  \rangle   \nonumber\\
       &\simeq& \frac{1}{\sqrt{N_{i} N^{(N)}_{f}}}C_{J_{i} M_{i} \lambda \mu}^{J_{f} M_{f}} \sum_{K_{i}^{\prime}} \sum_{\nu} C_{J_{i} K_{i}^{\prime} \lambda \nu}^{J_{f} K_{f}}   \nonumber\\
      &\times&
      \bra{\HFB} \hat{Q}_{N, K^{\pi}}\hat{T}_{\lambda \nu} \hat{P}_{K_{i}^{\prime} K_{i}}^{J_{i}} 
      -\hat{P}_{K_{i}^{\prime} K_{i}}^{J_{i}}       \hat{T}_{\lambda \nu} \hat{Q}_{N, K^{\pi}} \ket{\HFB}, \nonumber \\
    \end{eqnarray}  
 \eda{where the QRPA ground state is replaced by the HFB state, in close analogy to the quasi-boson approximation commonly employed in QRPA calculations without symmetry projections~\cite{Ring:1980}.} Besides, we have used the relation~\cite{Yao:2010},
\beq 
\hat{P}_{K_{f} M_{f}}^{J_{f}} \hat{T}_{\lambda \mu} \hat{P}_{M_{i} K_{i}}^{J_{i}}
=C_{J_{i} M_{i} \lambda \mu}^{J_{f} M_{f}} \sum_{K_{i}^{\prime}} \sum_{\nu} C_{J_{i} K_{i}^{\prime} \lambda \nu}^{J_{f} K_{f}} \hat{T}_{\lambda \nu} \hat{P}_{K_{i}^{\prime} K_{i}}^{J_{i}} \, \ed{,}
\eeq   
For the axial deformation, the first term in Eq.(\ref{eq:projected-ME-appendix}) can be simplified via the relation
\beqn
    && \langle {\rm HFB}   | \hat{Q}_{N, K^{\pi}}  \hat{T}_{\lambda \nu} \hat{P}^{J_i}_{K'_i, K_i}       
       | {\rm HFB}   \rangle\nonumber\\ 
         & =&  
          \delta_{K_i, 0}\delta_{K'_i+\nu-K,0} 
          \frac{2J_i+1}{2} \int^{+1}_{-1} d(\cos\beta)  d^{J_i}_{K'_i, K_i}(\beta)  \nonumber\\ 
     & &\times\langle {\rm HFB}  |\hat{Q}_{N, K^\pi}  \hat T_{\lambda\nu}   e^{-i\beta\hat{J}_{y}} | {\rm HFB}   \rangle  \,,
     \eeqn 
where we have used the fact that
\beq 
e^{i\alpha \hat J_z} \hat T_{\lambda\mu}e^{-i\alpha \hat J_z}
=\sum_{\nu} \bra{\lambda \nu} e^{i\alpha \hat J_z} \ket{\lambda \mu}\hat{T}_{\lambda\nu }
=e^{i\mu \alpha}\hat{T}_{\lambda\mu } \,.
\eeq 
For the $\beta^-$ decay, the tensor operator $\hat T_{\lambda\mu}$ changes a neutron to a proton, i.e.,
\beq 
\hat T_{\lambda\mu}
=\sum_{ p n } \bra{p}\hat{T}_{\lambda\mu}  \ket{n} c^{\dagger}_{p} c_{n} \,.
\eeq 
Substituting the expressions for the $\hat{Q}_{N, K^\pi} $ and $\hat T_{\lambda\mu}$ operators, one finds the overlap
\beqn 
&& \langle {\rm HFB} |\hat{Q}_{N, K^\pi}  \hat T_{\lambda\nu} e^{-i\beta\hat{J}_{y}} | {\rm HFB}   \rangle \nonumber\\
& =&  \sum_{p' n' pn}
X^{(N, K^{\pi})*}_{p' n'} \bra{p}\hat{T}_{\lambda\nu}\ket{n}\nonumber\\
&&\times \bra{\HFB } \beta_{n'}\beta_{p'} c^{\dagger}_{p} c_{n}  \ket{\HFB (  \beta)} \mathbb{N}(  \beta) \,, 
 \eeqn  
where
\beqn
&&\bra{\HFB } \beta_{n'}\beta_{p'} c^{\dagger}_{p} c_{n}  \ket{\HFB (  \beta)} \nonumber\\
&=&\bra{\HFB }  \beta_{p'} c^{\dagger}_{p}  \ket{\HFB (  \beta)} \bra{\HFB } \beta_{n'}  c_{n}  \ket{\HFB (  \beta)} \,,  \nonumber\\
\eeqn 
with the overlaps given by
\begin{eqnarray}
\bra{\HFB }  \beta_{p'} c^{\dagger}_{p}  \ket{\HFB (  \beta)} &=&\left[ \mathbb{U}^{(\pi)-1}(\beta ) U^{(\pi)\dagger}(\beta)\right]_{p' p}, \\
\bra{\HFB } \beta_{n'}  c_{n}  \ket{\HFB (  \beta)} &=& 
      \left[ \mathbb{U}^{(\nu)-1}(\beta ) V^{(\nu)\dagger}(\beta )\right]_{n' n}. 
      \end{eqnarray}
Finally, one finds
\beqn 
&& \langle {\HFB} |\hat{Q}_{N, K^\pi_f}  \hat T_{\lambda\nu}   e^{-i\beta\hat{J}_{y}} | {\rm HFB}   \rangle \nonumber\\
&=&  \sum_{p'n'} 
   X^{(N)\dagger}_{p' n'}   \left[ \mathbb{U}^{(\pi)-1}(  \beta) U^{(\pi)\dagger}(  \beta)  T_{\lambda\mu}   V^{(\nu)*}(  \beta) (\mathbb{U}^{(\nu)-1})^{T}(  \beta) \right] _{p' n'} \mathbb{N}(\beta),  \nonumber\\ 
\eeqn
For the second term in Eq.(\ref{eq:projected-ME-appendix}), one has the simplified formula  
\beqn
    && -\langle {\rm HFB}   |  \hat{P}^{J_i}_{K'_i, K_i}         \hat{T}_{\lambda \nu}\hat{Q}_{N, K^{\pi}}
       | {\rm HFB}  \rangle \nonumber\\ 
  & =& \sum_{p'n'} Y^{(N, K^\pi)\ast}_{p'n'}\frac{2J_i+1}{2}  \delta_{K, \nu+K_i}    \delta_{K'_{i},0}    \nonumber\\
          &&\times\int^{+1}_{-1} d(\cos\beta)  d^{J_i}_{K_{i} K'_{i}}(\beta)     \bra{\HFB}       e^{-i\beta \hat{J}_{y}}\hat{T}_{\lambda\nu}\beta^\dagger_{n'}\beta^\dagger_{p'} 
     \ket{ \HFB},     \nonumber\\  
\eeqn  
where the overlap for the $\beta^-$ decay  can be derived in the same way as the first term, 
\beqn 
&& \bra{\HFB}       e^{-i\beta \hat{J}_{y}}\hat{T}_{\lambda\nu}\beta^\dagger_{n'}\beta^\dagger_{p'} 
     | {\rm HFB}   \rangle \nonumber\\
     &=& \langle {\rm HFB}( \beta)| \hat{T}_{\lambda\nu}\beta^\dagger_{n'}\beta^\dagger_{p'} 
     \ket{\HFB} \mathbb{N}(  \beta),
\eeqn 
where
\beqn 
&&\bra{\HFB(\beta)}\hat{T}_{\lambda\nu}\beta^\dagger_{n'}\beta^\dagger_{p'} 
    \ket{\HFB}\nonumber\\ 
    &=& -\sum_{pn} \bra{p} T_{\lambda \mu}\ket{n}\langle {\rm HFB}( \beta)  |   c^{\dagger}_{p}  \beta^{\dagger}_{p'}   | {\rm HFB}  \rangle \nonumber\\
    &&\times\langle {\rm HFB}( \beta)  |   c_{n} \beta^{\dagger}_{n'}  | {\rm HFB}  \rangle \,,\nonumber
\eeqn
with the overlap
\beqn
&&\bra{\HFB( \beta)  }   c^{\dagger}_{p}  \beta^{\dagger}_{p'}  \ket{\HFB}= \Bigg[(\mathbb{U}^{(\pi)-1})^{\ast} (\beta)  V^{(\pi)T}( \beta)\Bigg]_{p'p} \,,
\eeqn 
 and
 \beqn 
 &&\bra{\HFB( \beta)  }  c_{n} \beta^{\dagger}_{n'}  \ket{\HFB}= \Bigg[(\mathbb{U}^{(\nu)-1})^{\ast} (\beta)  U^{(\nu)T}( \beta)\Bigg]_{n'n} \,.
 \eeqn 

Finally, one finds
\beqn  
&&\bra{\HFB( \beta)}  \hat{T}_{\lambda\nu}\beta^\dagger_{n'}\beta^\dagger_{p'} \ket{\HFB} \nonumber\\ 
    &=&  - 
     \Bigg[(\mathbb{U}^{(\pi)-1})^{\ast} (  \beta)  V^{(\pi)T}( \beta) 
      T_{\lambda \mu} 
     U^{(\nu)}( \beta)(\mathbb{U}^{(\nu)-1})^{\dagger}(  \beta)   \Bigg]_{p'n'} \,.
\eeqn  
Thus, for the second term in Eq.(\ref{eq:projected-ME-appendix}),
\beqn
    && -\langle {\rm HFB}    | \hat{P}^{J_i}_{K_i K^{\prime}_{i }}\hat{T}_{\lambda \nu}\hat{Q}_{N, K^{\pi}}
       | {\rm HFB}  \rangle \nonumber\\ 
  & =& \sum_{p'n'} 
  Y^{(N, K^\pi)\ast}_{p'n'} 
  \frac{2J_i+1}{2}    \int^{+1}_{-1} d(\cos\beta)   d^{J_i}_{K_{i} K'_{i}}(\beta)   \delta_{  \nu+K^{\prime}_i-K,0}    \delta_{K_{i},0}  \mathbb{N}(  \beta)  \nonumber\\ &&\Bigg[(\mathbb{U}^{(\pi)-1})^{\ast}  (  \beta)  V^{(\pi)T}( \beta) 
      T_{\lambda \mu}  
     U^{(\nu)}( \beta)[(\mathbb{U}^{(\nu)-1})^{\dagger} (  \beta)]   \Bigg]_{p'n'} \,.\nonumber\\
  \eeqn

\section{Transition matrix elements with projection}
\label{eq:appendC}

Before presenting the expression for the  transition matrix element with projection, we first prove the relation in Eq.\ (\ref{eq:FAM4XY}) when there is no projection. 
The transition matrix element of the operator $\hat{F}\equiv\hat T_{\lambda\mu}$ in the standard pnQRPA is given by
\begin{eqnarray}
 \label{MEwithoutp}
    M^{(N)}(\hat F) &=& \bra{N, K^{\pi}}   \hat F \ket{\QRPA}\nonumber    \\  
    &\approx& \bra{\HFB}\bigg [\hat{Q}_{N, K^{\pi}}, \hat F\bigg]\ket{\HFB}\nonumber    \\
&=& \sum_{pn} X^{(N, K^{\pi})\ast}_{pn}  F^{20}_{pn}+
Y^{(N, K^{\pi})\ast}_{pn}  F^{02}_{pn} \,.
\end{eqnarray}
Substituting Eq.\ (\ref{eq:X_QRPA_aprooximation_X} ) into the above equation, we obtain 
\begin{eqnarray}
      M^{(N)}(\hat F)  
     &\simeq& -\frac{ (\Omega_N -\omega ){\rm Im}(\mathcal{X}^{(K^{\pi})}_{pn})}{\bra{N, K^{\pi}}\hat{F}\ket{\QRPA}} F^{20}_{pn}-\frac{ (\Omega_N - \omega ){\rm Im}\mathcal{Y}^{(K^{\pi})}_{pn})}{\bra{N, K^{\pi}}\hat F\ket{\QRPA}} F^{02}_{pn}\nonumber\\
     &=& -\frac{ (\Omega_N -\omega )}{\bra{N, K^{\pi}}\hat F\ket{\QRPA}}\Bigg[{{\rm Im}(\mathcal{X}^{(K^{\pi})}_{pn})F^{20} _{pn}+{\rm Im}(\mathcal{Y}^{(K^{\pi})}_{pn}}) F^{02}_{pn}\Bigg]  \nonumber\\
     &=& -\frac{ (\Omega_N - \omega  )}{\bra{N, K^{\pi}}\hat F\ket{\QRPA}} {\rm Im}[S(\hat F, K^{\pi};\omega)] \,,
\end{eqnarray} 
 where on the right hand side, $\omega=\Omega+i\gamma$ with $\Omega\to\Omega_N$.  Using Eq.~(\ref{eq:strength}), we then arrive at the relation,
\beq 
|M^{(N)}(\hat F)|^2  \simeq  \pi\gamma \frac{d B(\hat{F},\omega)}{d\omega}\Bigg|_{\omega=\Omega_N+i\gamma} \,, 
\eeq   
which confirms the relation in Eq.\ (\ref{eq:FAM4XY}).
 
The transition matrix  elements $M^{(N)}_{\rm PAV}(\hat F)$ for even-even nuclei with AMP in Eq.\ (\ref{eq:projected_ME}) are determined by, 
\begin{eqnarray} 
\label{eq:NME_AMP}
  && M^{(N)}_{\rm PAV}(\hat  F) \nonumber\\ 
    &=& \frac{1}{\sqrt{N_{i} N^{(N)}_{f}}}C_{J_{i} M_{i} \lambda \mu}^{J_{f} M_{f}} \sum_{K_{i}^{\prime}} \sum_{\nu} C_{J_{i} K_{i}^{\prime} \lambda \nu}^{J_{f} K_{f}}  \nonumber \\
    &\times& \frac{2J_i+1}{2}   \int^{+1}_{-1} d(\cos\beta)  d^{J_i=0}_{00}(\beta)  \mathbb{N}(  \beta)  \nonumber  \\
  &\times&  \sum_{pn} 
  \bigg\{X^{(N, K^{\pi})*}_{pn} \left[ \mathbb{U}^{(\pi)-1}(    \beta )  U^{(\pi)\dagger} (    \beta )f V^{(\nu)*}(    \beta  )(\mathbb{U}^{(\nu)-1})^{T}(    \beta  )\right]_{pn}\nonumber\\ 
  && 
      +Y^{(N, K^{\pi})}_{pn} \left[(\mathbb{U}^{(\pi)-1})^{ *}( \beta)V^{(\pi)T}( \beta) f U^{(\nu)} ( \beta)[(\mathbb{U}^{(\nu)-1})^{\dagger} (  \beta)]  \right]_{pn} \bigg\} \,.\nonumber\\
\end{eqnarray}
   
\end{appendix}

 \bibliographystyle{apsrev4-1} 

\begin{thebibliography}{101}%
\makeatletter
\providecommand \@ifxundefined [1]{%
 \@ifx{#1\undefined}
}%
\providecommand \@ifnum [1]{%
 \ifnum #1\expandafter \@firstoftwo
 \else \expandafter \@secondoftwo
 \fi
}%
\providecommand \@ifx [1]{%
 \ifx #1\expandafter \@firstoftwo
 \else \expandafter \@secondoftwo
 \fi
}%
\providecommand \natexlab [1]{#1}%
\providecommand \enquote  [1]{``#1''}%
\providecommand \bibnamefont  [1]{#1}%
\providecommand \bibfnamefont [1]{#1}%
\providecommand \citenamefont [1]{#1}%
\providecommand \href@noop [0]{\@secondoftwo}%
\providecommand \href [0]{\begingroup \@sanitize@url \@href}%
\providecommand \@href[1]{\@@startlink{#1}\@@href}%
\providecommand \@@href[1]{\endgroup#1\@@endlink}%
\providecommand \@sanitize@url [0]{\catcode `\\12\catcode `\$12\catcode
  `\&12\catcode `\#12\catcode `\^12\catcode `\_12\catcode `\%12\relax}%
\providecommand \@@startlink[1]{}%
\providecommand \@@endlink[0]{}%
\providecommand \url  [0]{\begingroup\@sanitize@url \@url }%
\providecommand \@url [1]{\endgroup\@href {#1}{\urlprefix }}%
\providecommand \urlprefix  [0]{URL }%
\providecommand \Eprint [0]{\href }%
\providecommand \doibase [0]{http://dx.doi.org/}%
\providecommand \selectlanguage [0]{\@gobble}%
\providecommand \bibinfo  [0]{\@secondoftwo}%
\providecommand \bibfield  [0]{\@secondoftwo}%
\providecommand \translation [1]{[#1]}%
\providecommand \BibitemOpen [0]{}%
\providecommand \bibitemStop [0]{}%
\providecommand \bibitemNoStop [0]{.\EOS\space}%
\providecommand \EOS [0]{\spacefactor3000\relax}%
\providecommand \BibitemShut  [1]{\csname bibitem#1\endcsname}%
\let\auto@bib@innerbib\@empty
\bibitem [{\citenamefont {Langanke}\ and\ \citenamefont
  {Mart\'{\i}nez-Pinedo}(2003)}]{Langanke:2003RMP}%
  \BibitemOpen
  \bibfield  {author} {\bibinfo {author} {\bibfnamefont {K.}~\bibnamefont
  {Langanke}}\ and\ \bibinfo {author} {\bibfnamefont {G.}~\bibnamefont
  {Mart\'{\i}nez-Pinedo}},\ }\href {\doibase 10.1103/RevModPhys.75.819}
  {\bibfield  {journal} {\bibinfo  {journal} {Rev. Mod. Phys.}\ }\textbf
  {\bibinfo {volume} {75}},\ \bibinfo {pages} {819} (\bibinfo {year}
  {2003})}\BibitemShut {NoStop}%
\bibitem [{\citenamefont {Fischer}\ \emph {et~al.}(2024)\citenamefont
  {Fischer}, \citenamefont {Guo}, \citenamefont {Langanke}, \citenamefont
  {Martinez-Pinedo}, \citenamefont {Qian},\ and\ \citenamefont
  {Wu}}]{Fischer:2024PPNP}%
  \BibitemOpen
  \bibfield  {author} {\bibinfo {author} {\bibfnamefont {T.}~\bibnamefont
  {Fischer}}, \bibinfo {author} {\bibfnamefont {G.}~\bibnamefont {Guo}},
  \bibinfo {author} {\bibfnamefont {K.}~\bibnamefont {Langanke}}, \bibinfo
  {author} {\bibfnamefont {G.}~\bibnamefont {Martinez-Pinedo}}, \bibinfo
  {author} {\bibfnamefont {Y.-Z.}\ \bibnamefont {Qian}}, \ and\ \bibinfo
  {author} {\bibfnamefont {M.-R.}\ \bibnamefont {Wu}},\ }\href {\doibase
  10.1016/j.ppnp.2024.104107} {\bibfield  {journal} {\bibinfo  {journal} {Prog.
  Part. Nucl. Phys.}\ }\textbf {\bibinfo {volume} {137}},\ \bibinfo {pages}
  {104107} (\bibinfo {year} {2024})},\ \Eprint
  {http://arxiv.org/abs/2308.03962} {arXiv:2308.03962 [astro-ph.HE]}
  \BibitemShut {NoStop}%
\bibitem [{\citenamefont {Suzuki}(2022)}]{Suzuki:2022PPNP}%
  \BibitemOpen
  \bibfield  {author} {\bibinfo {author} {\bibfnamefont {T.}~\bibnamefont
  {Suzuki}},\ }\href {\doibase 10.1016/j.ppnp.2022.103974} {\bibfield
  {journal} {\bibinfo  {journal} {Prog. Part. Nucl. Phys.}\ }\textbf {\bibinfo
  {volume} {126}},\ \bibinfo {pages} {103974} (\bibinfo {year} {2022})},\
  \Eprint {http://arxiv.org/abs/2205.09262} {arXiv:2205.09262 [nucl-th]}
  \BibitemShut {NoStop}%
\bibitem [{\citenamefont {Severijns}\ \emph
  {et~al.}(2006{\natexlab{a}})\citenamefont {Severijns}, \citenamefont {Beck},\
  and\ \citenamefont {Naviliat-Cuncic}}]{Severijns:2006RMP}%
  \BibitemOpen
  \bibfield  {author} {\bibinfo {author} {\bibfnamefont {N.}~\bibnamefont
  {Severijns}}, \bibinfo {author} {\bibfnamefont {M.}~\bibnamefont {Beck}}, \
  and\ \bibinfo {author} {\bibfnamefont {O.}~\bibnamefont {Naviliat-Cuncic}},\
  }\href {\doibase 10.1103/RevModPhys.78.991} {\bibfield  {journal} {\bibinfo
  {journal} {Rev. Mod. Phys.}\ }\textbf {\bibinfo {volume} {78}},\ \bibinfo
  {pages} {991} (\bibinfo {year} {2006}{\natexlab{a}})},\ \Eprint
  {http://arxiv.org/abs/nucl-ex/0605029} {arXiv:nucl-ex/0605029} \BibitemShut
  {NoStop}%
\bibitem [{\citenamefont {Towner}\ and\ \citenamefont
  {Hardy}(2010)}]{Towner_2010}%
  \BibitemOpen
  \bibfield  {author} {\bibinfo {author} {\bibfnamefont {I.~S.}\ \bibnamefont
  {Towner}}\ and\ \bibinfo {author} {\bibfnamefont {J.~C.}\ \bibnamefont
  {Hardy}},\ }\href {\doibase 10.1088/0034-4885/73/4/046301} {\bibfield
  {journal} {\bibinfo  {journal} {Reports on Progress in Physics}\ }\textbf
  {\bibinfo {volume} {73}},\ \bibinfo {pages} {046301} (\bibinfo {year}
  {2010})}\BibitemShut {NoStop}%
\bibitem [{\citenamefont {Hayen}(2024)}]{Hayen:2024}%
  \BibitemOpen
  \bibfield  {author} {\bibinfo {author} {\bibfnamefont {L.}~\bibnamefont
  {Hayen}},\ }\href {\doibase 10.1146/annurev-nucl-121423-100730} {\bibfield
  {journal} {\bibinfo  {journal} {Ann. Rev. Nucl. Part. Sci.}\ }\textbf
  {\bibinfo {volume} {74}},\ \bibinfo {pages} {497} (\bibinfo {year} {2024})},\
  \Eprint {http://arxiv.org/abs/2403.08485} {arXiv:2403.08485 [nucl-th]}
  \BibitemShut {NoStop}%
\bibitem [{\citenamefont {Herczeg}(2001)}]{Herczeg:2001PPNP}%
  \BibitemOpen
  \bibfield  {author} {\bibinfo {author} {\bibfnamefont {P.}~\bibnamefont
  {Herczeg}},\ }\href {\doibase https://doi.org/10.1016/S0146-6410(01)00149-1}
  {\bibfield  {journal} {\bibinfo  {journal} {Prog. Part. Nucl. Phys.}\
  }\textbf {\bibinfo {volume} {46}},\ \bibinfo {pages} {413 } (\bibinfo {year}
  {2001})}\BibitemShut {NoStop}%
\bibitem [{\citenamefont {Severijns}\ \emph
  {et~al.}(2006{\natexlab{b}})\citenamefont {Severijns}, \citenamefont {Beck},\
  and\ \citenamefont {Naviliat-Cuncic}}]{Severijns:2006}%
  \BibitemOpen
  \bibfield  {author} {\bibinfo {author} {\bibfnamefont {N.}~\bibnamefont
  {Severijns}}, \bibinfo {author} {\bibfnamefont {M.}~\bibnamefont {Beck}}, \
  and\ \bibinfo {author} {\bibfnamefont {O.}~\bibnamefont {Naviliat-Cuncic}},\
  }\href {\doibase 10.1103/RevModPhys.78.991} {\bibfield  {journal} {\bibinfo
  {journal} {Rev. Mod. Phys.}\ }\textbf {\bibinfo {volume} {78}},\ \bibinfo
  {pages} {991} (\bibinfo {year} {2006}{\natexlab{b}})}\BibitemShut {NoStop}%
\bibitem [{\citenamefont {Otten}\ and\ \citenamefont
  {Weinheimer}(2008)}]{Otten:2008RPP}%
  \BibitemOpen
  \bibfield  {author} {\bibinfo {author} {\bibfnamefont {E.~W.}\ \bibnamefont
  {Otten}}\ and\ \bibinfo {author} {\bibfnamefont {C.}~\bibnamefont
  {Weinheimer}},\ }\href {\doibase 10.1088/0034-4885/71/8/086201} {\bibfield
  {journal} {\bibinfo  {journal} {Rep. Prog. Phys.}\ }\textbf {\bibinfo
  {volume} {71}},\ \bibinfo {pages} {086201} (\bibinfo {year}
  {2008})}\BibitemShut {NoStop}%
\bibitem [{\citenamefont {Falkowski}\ \emph {et~al.}(2021)\citenamefont
  {Falkowski}, \citenamefont {Gonz\'alez-Alonso},\ and\ \citenamefont
  {Naviliat-Cuncic}}]{Falkowski:2021JHEP}%
  \BibitemOpen
  \bibfield  {author} {\bibinfo {author} {\bibfnamefont {A.}~\bibnamefont
  {Falkowski}}, \bibinfo {author} {\bibfnamefont {M.}~\bibnamefont
  {Gonz\'alez-Alonso}}, \ and\ \bibinfo {author} {\bibfnamefont
  {O.}~\bibnamefont {Naviliat-Cuncic}},\ }\href {\doibase
  10.1007/JHEP04(2021)126} {\bibfield  {journal} {\bibinfo  {journal} {JHEP}\
  }\textbf {\bibinfo {volume} {04}},\ \bibinfo {pages} {126} (\bibinfo {year}
  {2021})},\ \Eprint {http://arxiv.org/abs/2010.13797} {arXiv:2010.13797
  [hep-ph]} \BibitemShut {NoStop}%
\bibitem [{\citenamefont {Burbidge}\ \emph {et~al.}(1957)\citenamefont
  {Burbidge}, \citenamefont {Burbidge}, \citenamefont {Fowler},\ and\
  \citenamefont {Hoyle}}]{Burbidge:1957RMP}%
  \BibitemOpen
  \bibfield  {author} {\bibinfo {author} {\bibfnamefont {M.~E.}\ \bibnamefont
  {Burbidge}}, \bibinfo {author} {\bibfnamefont {G.~R.}\ \bibnamefont
  {Burbidge}}, \bibinfo {author} {\bibfnamefont {W.~A.}\ \bibnamefont
  {Fowler}}, \ and\ \bibinfo {author} {\bibfnamefont {F.}~\bibnamefont
  {Hoyle}},\ }\href {\doibase 10.1103/RevModPhys.29.547} {\bibfield  {journal}
  {\bibinfo  {journal} {Rev. Mod. Phys.}\ }\textbf {\bibinfo {volume} {29}},\
  \bibinfo {pages} {547} (\bibinfo {year} {1957})}\BibitemShut {NoStop}%
\bibitem [{\citenamefont {Cowan}\ \emph {et~al.}(1991)\citenamefont {Cowan},
  \citenamefont {Thielemann},\ and\ \citenamefont {Truran}}]{Cowan:1991PR}%
  \BibitemOpen
  \bibfield  {author} {\bibinfo {author} {\bibfnamefont {J.~J.}\ \bibnamefont
  {Cowan}}, \bibinfo {author} {\bibfnamefont {F.-K.}\ \bibnamefont
  {Thielemann}}, \ and\ \bibinfo {author} {\bibfnamefont {J.~W.}\ \bibnamefont
  {Truran}},\ }\href {\doibase 10.1016/0370-1573(91)90070-3} {\bibfield
  {journal} {\bibinfo  {journal} {Phys. Rept.}\ }\textbf {\bibinfo {volume}
  {208}},\ \bibinfo {pages} {267} (\bibinfo {year} {1991})}\BibitemShut
  {NoStop}%
\bibitem [{\citenamefont {Qian}\ and\ \citenamefont
  {Wasserburg}(2007)}]{Qian:2007PR}%
  \BibitemOpen
  \bibfield  {author} {\bibinfo {author} {\bibfnamefont {Y.~Z.}\ \bibnamefont
  {Qian}}\ and\ \bibinfo {author} {\bibfnamefont {G.~J.}\ \bibnamefont
  {Wasserburg}},\ }\href {\doibase 10.1016/j.physrep.2007.02.006} {\bibfield
  {journal} {\bibinfo  {journal} {Phys. Rept.}\ }\textbf {\bibinfo {volume}
  {442}},\ \bibinfo {pages} {237} (\bibinfo {year} {2007})},\ \Eprint
  {http://arxiv.org/abs/0708.1767} {arXiv:0708.1767 [astro-ph]} \BibitemShut
  {NoStop}%
\bibitem [{\citenamefont {Moller}\ \emph {et~al.}(2003)\citenamefont {Moller},
  \citenamefont {Pfeiffer},\ and\ \citenamefont {Kratz}}]{Moller:2003PRC}%
  \BibitemOpen
  \bibfield  {author} {\bibinfo {author} {\bibfnamefont {P.}~\bibnamefont
  {Moller}}, \bibinfo {author} {\bibfnamefont {B.}~\bibnamefont {Pfeiffer}}, \
  and\ \bibinfo {author} {\bibfnamefont {K.-L.}\ \bibnamefont {Kratz}},\ }\href
  {\doibase 10.1103/PhysRevC.67.055802} {\bibfield  {journal} {\bibinfo
  {journal} {Phys. Rev. C}\ }\textbf {\bibinfo {volume} {67}},\ \bibinfo
  {pages} {055802} (\bibinfo {year} {2003})}\BibitemShut {NoStop}%
\bibitem [{\citenamefont {Pereira}\ \emph {et~al.}(2009)\citenamefont {Pereira}
  \emph {et~al.}}]{Pereira:2009PRC}%
  \BibitemOpen
  \bibfield  {author} {\bibinfo {author} {\bibfnamefont {J.}~\bibnamefont
  {Pereira}} \emph {et~al.},\ }\href {\doibase 10.1103/PhysRevC.79.035806}
  {\bibfield  {journal} {\bibinfo  {journal} {Phys. Rev. C}\ }\textbf {\bibinfo
  {volume} {79}},\ \bibinfo {pages} {035806} (\bibinfo {year} {2009})},\
  \Eprint {http://arxiv.org/abs/0902.1705} {arXiv:0902.1705 [nucl-ex]}
  \BibitemShut {NoStop}%
\bibitem [{\citenamefont {Nishimura}\ \emph {et~al.}(2011)\citenamefont
  {Nishimura}, \citenamefont {Li}, \citenamefont {Watanabe}, \citenamefont
  {Yoshinaga} \emph {et~al.}}]{Nishimura:2011PRL}%
  \BibitemOpen
  \bibfield  {author} {\bibinfo {author} {\bibfnamefont {S.}~\bibnamefont
  {Nishimura}}, \bibinfo {author} {\bibfnamefont {Z.}~\bibnamefont {Li}},
  \bibinfo {author} {\bibfnamefont {H.}~\bibnamefont {Watanabe}}, \bibinfo
  {author} {\bibnamefont {Yoshinaga}},  \emph {et~al.},\ }\href {\doibase
  10.1103/PhysRevLett.106.052502} {\bibfield  {journal} {\bibinfo  {journal}
  {Phys. Rev. Lett.}\ }\textbf {\bibinfo {volume} {106}},\ \bibinfo {pages}
  {052502} (\bibinfo {year} {2011})}\BibitemShut {NoStop}%
\bibitem [{\citenamefont {Moller}\ \emph {et~al.}(1997)\citenamefont {Moller},
  \citenamefont {Nix},\ and\ \citenamefont {Kratz}}]{Moller:1997}%
  \BibitemOpen
  \bibfield  {author} {\bibinfo {author} {\bibfnamefont {P.}~\bibnamefont
  {Moller}}, \bibinfo {author} {\bibfnamefont {J.~R.}\ \bibnamefont {Nix}}, \
  and\ \bibinfo {author} {\bibfnamefont {K.~L.}\ \bibnamefont {Kratz}},\ }\href
  {\doibase 10.1006/adnd.1997.0746} {\bibfield  {journal} {\bibinfo  {journal}
  {Atom. Data Nucl. Data Tabl.}\ }\textbf {\bibinfo {volume} {66}},\ \bibinfo
  {pages} {131} (\bibinfo {year} {1997})},\ \Eprint
  {http://arxiv.org/abs/nucl-th/9601043} {arXiv:nucl-th/9601043} \BibitemShut
  {NoStop}%
\bibitem [{\citenamefont {Weinberg}(1991)}]{Weinberg:1991}%
  \BibitemOpen
  \bibfield  {author} {\bibinfo {author} {\bibfnamefont {S.}~\bibnamefont
  {Weinberg}},\ }\href {\doibase 10.1016/0550-3213(91)90231-L} {\bibfield
  {journal} {\bibinfo  {journal} {Nucl. Phys. B}\ }\textbf {\bibinfo {volume}
  {363}},\ \bibinfo {pages} {3} (\bibinfo {year} {1991})}\BibitemShut {NoStop}%
\bibitem [{\citenamefont {Ekstr\"om}\ \emph {et~al.}(2014)\citenamefont
  {Ekstr\"om}, \citenamefont {Jansen}, \citenamefont {Wendt}, \citenamefont
  {Hagen}, \citenamefont {Papenbrock}, \citenamefont {Bacca}, \citenamefont
  {Carlsson},\ and\ \citenamefont {Gazit}}]{Ekstrom:2014PRL}%
  \BibitemOpen
  \bibfield  {author} {\bibinfo {author} {\bibfnamefont {A.}~\bibnamefont
  {Ekstr\"om}}, \bibinfo {author} {\bibfnamefont {G.~R.}\ \bibnamefont
  {Jansen}}, \bibinfo {author} {\bibfnamefont {K.~A.}\ \bibnamefont {Wendt}},
  \bibinfo {author} {\bibfnamefont {G.}~\bibnamefont {Hagen}}, \bibinfo
  {author} {\bibfnamefont {T.}~\bibnamefont {Papenbrock}}, \bibinfo {author}
  {\bibfnamefont {S.}~\bibnamefont {Bacca}}, \bibinfo {author} {\bibfnamefont
  {B.}~\bibnamefont {Carlsson}}, \ and\ \bibinfo {author} {\bibfnamefont
  {D.}~\bibnamefont {Gazit}},\ }\href {\doibase 10.1103/PhysRevLett.113.262504}
  {\bibfield  {journal} {\bibinfo  {journal} {Phys. Rev. Lett.}\ }\textbf
  {\bibinfo {volume} {113}},\ \bibinfo {pages} {262504} (\bibinfo {year}
  {2014})}\BibitemShut {NoStop}%
\bibitem [{\citenamefont {Gysbers}\ \emph {et~al.}(2019)\citenamefont {Gysbers}
  \emph {et~al.}}]{Gysbers:2019}%
  \BibitemOpen
  \bibfield  {author} {\bibinfo {author} {\bibfnamefont {P.}~\bibnamefont
  {Gysbers}} \emph {et~al.},\ }\href {\doibase 10.1038/s41567-019-0450-7}
  {\bibfield  {journal} {\bibinfo  {journal} {Nature Phys.}\ }\textbf {\bibinfo
  {volume} {15}},\ \bibinfo {pages} {428} (\bibinfo {year} {2019})},\ \Eprint
  {http://arxiv.org/abs/1903.00047} {arXiv:1903.00047 [nucl-th]} \BibitemShut
  {NoStop}%
\bibitem [{\citenamefont {Stroberg}(2021)}]{Stroberg:2021}%
  \BibitemOpen
  \bibfield  {author} {\bibinfo {author} {\bibfnamefont {S.~R.}\ \bibnamefont
  {Stroberg}},\ }\href {\doibase 10.3390/particles4040038} {\bibfield
  {journal} {\bibinfo  {journal} {Particles}\ }\textbf {\bibinfo {volume}
  {4}},\ \bibinfo {pages} {521} (\bibinfo {year} {2021})},\ \Eprint
  {http://arxiv.org/abs/2109.13462} {arXiv:2109.13462 [nucl-th]} \BibitemShut
  {NoStop}%
\bibitem [{\citenamefont {King}\ \emph {et~al.}(2023)\citenamefont {King},
  \citenamefont {Baroni}, \citenamefont {Cirigliano}, \citenamefont {Gandolfi},
  \citenamefont {Hayen}, \citenamefont {Mereghetti}, \citenamefont {Pastore},\
  and\ \citenamefont {Piarulli}}]{King:2023}%
  \BibitemOpen
  \bibfield  {author} {\bibinfo {author} {\bibfnamefont {G.~B.}\ \bibnamefont
  {King}}, \bibinfo {author} {\bibfnamefont {A.}~\bibnamefont {Baroni}},
  \bibinfo {author} {\bibfnamefont {V.}~\bibnamefont {Cirigliano}}, \bibinfo
  {author} {\bibfnamefont {S.}~\bibnamefont {Gandolfi}}, \bibinfo {author}
  {\bibfnamefont {L.}~\bibnamefont {Hayen}}, \bibinfo {author} {\bibfnamefont
  {E.}~\bibnamefont {Mereghetti}}, \bibinfo {author} {\bibfnamefont
  {S.}~\bibnamefont {Pastore}}, \ and\ \bibinfo {author} {\bibfnamefont
  {M.}~\bibnamefont {Piarulli}},\ }\href {\doibase 10.1103/PhysRevC.107.015503}
  {\bibfield  {journal} {\bibinfo  {journal} {Phys. Rev. C}\ }\textbf {\bibinfo
  {volume} {107}},\ \bibinfo {pages} {015503} (\bibinfo {year} {2023})},\
  \Eprint {http://arxiv.org/abs/2207.11179} {arXiv:2207.11179 [nucl-th]}
  \BibitemShut {NoStop}%
\bibitem [{\citenamefont {King}\ and\ \citenamefont
  {Pastore}(2024)}]{King:2024}%
  \BibitemOpen
  \bibfield  {author} {\bibinfo {author} {\bibfnamefont {G.~B.}\ \bibnamefont
  {King}}\ and\ \bibinfo {author} {\bibfnamefont {S.}~\bibnamefont {Pastore}},\
  }\href {\doibase 10.1146/annurev-nucl-101920-021401} {\bibfield  {journal}
  {\bibinfo  {journal} {Ann. Rev. Nucl. Part. Sci.}\ }\textbf {\bibinfo
  {volume} {74}},\ \bibinfo {pages} {343} (\bibinfo {year} {2024})},\ \Eprint
  {http://arxiv.org/abs/2402.06602} {arXiv:2402.06602 [nucl-th]} \BibitemShut
  {NoStop}%
\bibitem [{\citenamefont {Li}\ \emph {et~al.}(2025)\citenamefont {Li},
  \citenamefont {Miyagi},\ and\ \citenamefont {Schwenk}}]{Li:2025}%
  \BibitemOpen
  \bibfield  {author} {\bibinfo {author} {\bibfnamefont {Z.}~\bibnamefont
  {Li}}, \bibinfo {author} {\bibfnamefont {T.}~\bibnamefont {Miyagi}}, \ and\
  \bibinfo {author} {\bibfnamefont {A.}~\bibnamefont {Schwenk}},\ }\href@noop
  {} {\  (\bibinfo {year} {2025})},\ \Eprint {http://arxiv.org/abs/2509.06812}
  {arXiv:2509.06812 [nucl-th]} \BibitemShut {NoStop}%
\bibitem [{\citenamefont {Beaujeault-Taudi{\`e}re}\ \emph
  {et~al.}(2023)\citenamefont {Beaujeault-Taudi{\`e}re}, \citenamefont
  {Frosini}, \citenamefont {Ebran}, \citenamefont {Duguet}, \citenamefont
  {Roth},\ and\ \citenamefont {Som{\`a}}}]{Beaujeault:2022ayi}%
  \BibitemOpen
  \bibfield  {author} {\bibinfo {author} {\bibfnamefont {Y.}~\bibnamefont
  {Beaujeault-Taudi{\`e}re}}, \bibinfo {author} {\bibfnamefont
  {M.}~\bibnamefont {Frosini}}, \bibinfo {author} {\bibfnamefont {J.~P.}\
  \bibnamefont {Ebran}}, \bibinfo {author} {\bibfnamefont {T.}~\bibnamefont
  {Duguet}}, \bibinfo {author} {\bibfnamefont {R.}~\bibnamefont {Roth}}, \ and\
  \bibinfo {author} {\bibfnamefont {V.}~\bibnamefont {Som{\`a}}},\ }\href
  {\doibase 10.1103/PhysRevC.107.L021302} {\bibfield  {journal} {\bibinfo
  {journal} {Phys. Rev. C}\ }\textbf {\bibinfo {volume} {107}},\ \bibinfo
  {pages} {L021302} (\bibinfo {year} {2023})},\ \Eprint
  {http://arxiv.org/abs/2203.13513} {arXiv:2203.13513 [nucl-th]} \BibitemShut
  {NoStop}%
\bibitem [{\citenamefont {Zaragoza}\ \emph {et~al.}(2024)\citenamefont
  {Zaragoza}, \citenamefont {Ebran}, \citenamefont {Hilaire}, \citenamefont
  {P{\'e}ru}, \citenamefont {Frosini},\ and\ \citenamefont
  {Duguet}}]{Zaragoza:2024jvk}%
  \BibitemOpen
  \bibfield  {author} {\bibinfo {author} {\bibfnamefont {L.~G.-M.}\
  \bibnamefont {Zaragoza}}, \bibinfo {author} {\bibfnamefont {J.-P.}\
  \bibnamefont {Ebran}}, \bibinfo {author} {\bibfnamefont {S.}~\bibnamefont
  {Hilaire}}, \bibinfo {author} {\bibfnamefont {S.}~\bibnamefont {P{\'e}ru}},
  \bibinfo {author} {\bibfnamefont {M.}~\bibnamefont {Frosini}}, \ and\
  \bibinfo {author} {\bibfnamefont {T.}~\bibnamefont {Duguet}},\ }\href
  {\doibase 10.1051/epjconf/202429403003} {\bibfield  {journal} {\bibinfo
  {journal} {EPJ Web Conf.}\ }\textbf {\bibinfo {volume} {294}},\ \bibinfo
  {pages} {03003} (\bibinfo {year} {2024})}\BibitemShut {NoStop}%
\bibitem [{\citenamefont {Schuck}\ \emph {et~al.}(2021)\citenamefont {Schuck},
  \citenamefont {Delion}, \citenamefont {Dukelsky}, \citenamefont {Jemai},
  \citenamefont {Litvinova}, \citenamefont {Roepke},\ and\ \citenamefont
  {Tohyama}}]{Schuck:2021}%
  \BibitemOpen
  \bibfield  {author} {\bibinfo {author} {\bibfnamefont {P.}~\bibnamefont
  {Schuck}}, \bibinfo {author} {\bibfnamefont {D.~S.}\ \bibnamefont {Delion}},
  \bibinfo {author} {\bibfnamefont {J.}~\bibnamefont {Dukelsky}}, \bibinfo
  {author} {\bibfnamefont {M.}~\bibnamefont {Jemai}}, \bibinfo {author}
  {\bibfnamefont {E.}~\bibnamefont {Litvinova}}, \bibinfo {author}
  {\bibfnamefont {G.}~\bibnamefont {Roepke}}, \ and\ \bibinfo {author}
  {\bibfnamefont {M.}~\bibnamefont {Tohyama}},\ }\href {\doibase
  10.1016/j.physrep.2021.06.001} {\bibfield  {journal} {\bibinfo  {journal}
  {Phys. Rept.}\ }\textbf {\bibinfo {volume} {929}},\ \bibinfo {pages} {1}
  (\bibinfo {year} {2021})},\ \Eprint {http://arxiv.org/abs/2009.00591}
  {arXiv:2009.00591 [nucl-th]} \BibitemShut {NoStop}%
\bibitem [{\citenamefont {Delion}\ \emph {et~al.}(1997)\citenamefont {Delion},
  \citenamefont {Dukelsky},\ and\ \citenamefont {Schuck}}]{Delion:1997vr}%
  \BibitemOpen
  \bibfield  {author} {\bibinfo {author} {\bibfnamefont {D.~S.}\ \bibnamefont
  {Delion}}, \bibinfo {author} {\bibfnamefont {J.}~\bibnamefont {Dukelsky}}, \
  and\ \bibinfo {author} {\bibfnamefont {P.}~\bibnamefont {Schuck}},\ }\href
  {\doibase 10.1103/PhysRevC.55.2340} {\bibfield  {journal} {\bibinfo
  {journal} {Phys. Rev. C}\ }\textbf {\bibinfo {volume} {55}},\ \bibinfo
  {pages} {2340} (\bibinfo {year} {1997})}\BibitemShut {NoStop}%
\bibitem [{\citenamefont {Caurier}\ \emph
  {et~al.}(2005{\natexlab{a}})\citenamefont {Caurier}, \citenamefont
  {Martinez-Pinedo}, \citenamefont {Nowacki}, \citenamefont {Poves},\ and\
  \citenamefont {Zuker}}]{Caurier:2004gf}%
  \BibitemOpen
  \bibfield  {author} {\bibinfo {author} {\bibfnamefont {E.}~\bibnamefont
  {Caurier}}, \bibinfo {author} {\bibfnamefont {G.}~\bibnamefont
  {Martinez-Pinedo}}, \bibinfo {author} {\bibfnamefont {F.}~\bibnamefont
  {Nowacki}}, \bibinfo {author} {\bibfnamefont {A.}~\bibnamefont {Poves}}, \
  and\ \bibinfo {author} {\bibfnamefont {A.~P.}\ \bibnamefont {Zuker}},\ }\href
  {\doibase 10.1103/RevModPhys.77.427} {\bibfield  {journal} {\bibinfo
  {journal} {Rev. Mod. Phys.}\ }\textbf {\bibinfo {volume} {77}},\ \bibinfo
  {pages} {427} (\bibinfo {year} {2005}{\natexlab{a}})},\ \Eprint
  {http://arxiv.org/abs/nucl-th/0402046} {arXiv:nucl-th/0402046} \BibitemShut
  {NoStop}%
\bibitem [{\citenamefont {Martinez-Pinedo}\ and\ \citenamefont
  {Langanke}(1999)}]{Martinez-Pinedo:1999:PRL}%
  \BibitemOpen
  \bibfield  {author} {\bibinfo {author} {\bibfnamefont {G.}~\bibnamefont
  {Martinez-Pinedo}}\ and\ \bibinfo {author} {\bibfnamefont {K.}~\bibnamefont
  {Langanke}},\ }\href {\doibase 10.1103/PhysRevLett.83.4502} {\bibfield
  {journal} {\bibinfo  {journal} {Phys. Rev. Lett.}\ }\textbf {\bibinfo
  {volume} {83}},\ \bibinfo {pages} {4502} (\bibinfo {year} {1999})},\ \Eprint
  {http://arxiv.org/abs/astro-ph/9907274} {arXiv:astro-ph/9907274} \BibitemShut
  {NoStop}%
\bibitem [{\citenamefont {Zhi}\ \emph {et~al.}(2013)\citenamefont {Zhi},
  \citenamefont {Caurier}, \citenamefont {Cuenca-Garcia}, \citenamefont
  {Langanke}, \citenamefont {Martinez-Pinedo},\ and\ \citenamefont
  {Sieja}}]{Zhi:2013hg}%
  \BibitemOpen
  \bibfield  {author} {\bibinfo {author} {\bibfnamefont {Q.}~\bibnamefont
  {Zhi}}, \bibinfo {author} {\bibfnamefont {E.}~\bibnamefont {Caurier}},
  \bibinfo {author} {\bibfnamefont {J.~J.}\ \bibnamefont {Cuenca-Garcia}},
  \bibinfo {author} {\bibfnamefont {K.}~\bibnamefont {Langanke}}, \bibinfo
  {author} {\bibfnamefont {G.}~\bibnamefont {Martinez-Pinedo}}, \ and\ \bibinfo
  {author} {\bibfnamefont {K.}~\bibnamefont {Sieja}},\ }\href {\doibase
  10.1103/PhysRevC.87.025803} {\bibfield  {journal} {\bibinfo  {journal} {Phys.
  Rev. C}\ }\textbf {\bibinfo {volume} {87}},\ \bibinfo {pages} {025803}
  (\bibinfo {year} {2013})},\ \Eprint {http://arxiv.org/abs/1301.5225}
  {arXiv:1301.5225 [nucl-th]} \BibitemShut {NoStop}%
\bibitem [{\citenamefont {Caurier}\ \emph
  {et~al.}(2005{\natexlab{b}})\citenamefont {Caurier}, \citenamefont
  {Mart\'{\i}nez-Pinedo}, \citenamefont {Nowacki}, \citenamefont {Poves},\ and\
  \citenamefont {Zuker}}]{Caurier:2005RMP}%
  \BibitemOpen
  \bibfield  {author} {\bibinfo {author} {\bibfnamefont {E.}~\bibnamefont
  {Caurier}}, \bibinfo {author} {\bibfnamefont {G.}~\bibnamefont
  {Mart\'{\i}nez-Pinedo}}, \bibinfo {author} {\bibfnamefont {F.}~\bibnamefont
  {Nowacki}}, \bibinfo {author} {\bibfnamefont {A.}~\bibnamefont {Poves}}, \
  and\ \bibinfo {author} {\bibfnamefont {A.~P.}\ \bibnamefont {Zuker}},\ }\href
  {\doibase 10.1103/RevModPhys.77.427} {\bibfield  {journal} {\bibinfo
  {journal} {Rev. Mod. Phys.}\ }\textbf {\bibinfo {volume} {77}},\ \bibinfo
  {pages} {427} (\bibinfo {year} {2005}{\natexlab{b}})}\BibitemShut {NoStop}%
\bibitem [{\citenamefont {Fang}\ \emph {et~al.}(2013)\citenamefont {Fang},
  \citenamefont {Brown},\ and\ \citenamefont {Suzuki}}]{Fang:2013}%
  \BibitemOpen
  \bibfield  {author} {\bibinfo {author} {\bibfnamefont {D.-L.}\ \bibnamefont
  {Fang}}, \bibinfo {author} {\bibfnamefont {B.~A.}\ \bibnamefont {Brown}}, \
  and\ \bibinfo {author} {\bibfnamefont {T.}~\bibnamefont {Suzuki}},\ }\href
  {\doibase 10.1103/PhysRevC.88.024314} {\bibfield  {journal} {\bibinfo
  {journal} {Phys. Rev. C}\ }\textbf {\bibinfo {volume} {88}},\ \bibinfo
  {pages} {024314} (\bibinfo {year} {2013})}\BibitemShut {NoStop}%
\bibitem [{\citenamefont {Ni}\ and\ \citenamefont {Ren}(2014)}]{Ni:2014}%
  \BibitemOpen
  \bibfield  {author} {\bibinfo {author} {\bibfnamefont {D.}~\bibnamefont
  {Ni}}\ and\ \bibinfo {author} {\bibfnamefont {Z.}~\bibnamefont {Ren}},\
  }\href {\doibase 10.1103/PhysRevC.89.064320} {\bibfield  {journal} {\bibinfo
  {journal} {Phys. Rev. C}\ }\textbf {\bibinfo {volume} {89}},\ \bibinfo
  {pages} {064320} (\bibinfo {year} {2014})}\BibitemShut {NoStop}%
\bibitem [{\citenamefont {Engel}\ \emph {et~al.}(1999)\citenamefont {Engel},
  \citenamefont {Bender}, \citenamefont {Dobaczewski}, \citenamefont
  {Nazarewicz},\ and\ \citenamefont {Surman}}]{Engel:1999}%
  \BibitemOpen
  \bibfield  {author} {\bibinfo {author} {\bibfnamefont {J.}~\bibnamefont
  {Engel}}, \bibinfo {author} {\bibfnamefont {M.}~\bibnamefont {Bender}},
  \bibinfo {author} {\bibfnamefont {J.}~\bibnamefont {Dobaczewski}}, \bibinfo
  {author} {\bibfnamefont {W.}~\bibnamefont {Nazarewicz}}, \ and\ \bibinfo
  {author} {\bibfnamefont {R.}~\bibnamefont {Surman}},\ }\href {\doibase
  10.1103/PhysRevC.60.014302} {\bibfield  {journal} {\bibinfo  {journal} {Phys.
  Rev. C}\ }\textbf {\bibinfo {volume} {60}},\ \bibinfo {pages} {014302}
  (\bibinfo {year} {1999})},\ \Eprint {http://arxiv.org/abs/nucl-th/9902059}
  {arXiv:nucl-th/9902059} \BibitemShut {NoStop}%
\bibitem [{\citenamefont {Niu}\ \emph {et~al.}(2015)\citenamefont {Niu},
  \citenamefont {Niu}, \citenamefont {Colo},\ and\ \citenamefont
  {Vigezzi}}]{Niu:2015}%
  \BibitemOpen
  \bibfield  {author} {\bibinfo {author} {\bibfnamefont {Y.}~\bibnamefont
  {Niu}}, \bibinfo {author} {\bibfnamefont {Z.}~\bibnamefont {Niu}}, \bibinfo
  {author} {\bibfnamefont {G.}~\bibnamefont {Colo}}, \ and\ \bibinfo {author}
  {\bibfnamefont {E.}~\bibnamefont {Vigezzi}},\ }\href {\doibase
  10.1103/PhysRevLett.114.142501} {\bibfield  {journal} {\bibinfo  {journal}
  {Phys. Rev. Lett.}\ }\textbf {\bibinfo {volume} {114}},\ \bibinfo {pages}
  {142501} (\bibinfo {year} {2015})},\ \Eprint
  {http://arxiv.org/abs/1502.04830} {arXiv:1502.04830 [nucl-th]} \BibitemShut
  {NoStop}%
\bibitem [{\citenamefont {Gambacurta}\ \emph {et~al.}(2020)\citenamefont
  {Gambacurta}, \citenamefont {Grasso},\ and\ \citenamefont
  {Engel}}]{Gambacurta:2020}%
  \BibitemOpen
  \bibfield  {author} {\bibinfo {author} {\bibfnamefont {D.}~\bibnamefont
  {Gambacurta}}, \bibinfo {author} {\bibfnamefont {M.}~\bibnamefont {Grasso}},
  \ and\ \bibinfo {author} {\bibfnamefont {J.}~\bibnamefont {Engel}},\ }\href
  {\doibase 10.1103/PhysRevLett.125.212501} {\bibfield  {journal} {\bibinfo
  {journal} {Phys. Rev. Lett.}\ }\textbf {\bibinfo {volume} {125}},\ \bibinfo
  {pages} {212501} (\bibinfo {year} {2020})}\BibitemShut {NoStop}%
\bibitem [{\citenamefont {Bai}\ \emph {et~al.}(2022)\citenamefont {Bai},
  \citenamefont {Fang},\ and\ \citenamefont {Zhang}}]{Bai:2022pys}%
  \BibitemOpen
  \bibfield  {author} {\bibinfo {author} {\bibfnamefont {C.~L.}\ \bibnamefont
  {Bai}}, \bibinfo {author} {\bibfnamefont {D.~L.}\ \bibnamefont {Fang}}, \
  and\ \bibinfo {author} {\bibfnamefont {H.~Q.}\ \bibnamefont {Zhang}},\ }\href
  {\doibase 10.1088/1674-1137/ac80ee} {\bibfield  {journal} {\bibinfo
  {journal} {Chin. Phys. C}\ }\textbf {\bibinfo {volume} {46}},\ \bibinfo
  {pages} {114104} (\bibinfo {year} {2022})}\BibitemShut {NoStop}%
\bibitem [{\citenamefont {Suhonen}\ \emph {et~al.}(1988)\citenamefont
  {Suhonen}, \citenamefont {Taigel},\ and\ \citenamefont
  {Faessler}}]{Suhonen:1988umy}%
  \BibitemOpen
  \bibfield  {author} {\bibinfo {author} {\bibfnamefont {J.}~\bibnamefont
  {Suhonen}}, \bibinfo {author} {\bibfnamefont {T.}~\bibnamefont {Taigel}}, \
  and\ \bibinfo {author} {\bibfnamefont {A.}~\bibnamefont {Faessler}},\ }\href
  {\doibase 10.1016/0375-9474(88)90041-3} {\bibfield  {journal} {\bibinfo
  {journal} {Nucl. Phys. A}\ }\textbf {\bibinfo {volume} {486}},\ \bibinfo
  {pages} {91} (\bibinfo {year} {1988})}\BibitemShut {NoStop}%
\bibitem [{\citenamefont {Paar}\ \emph {et~al.}(2004)\citenamefont {Paar},
  \citenamefont {Niksic}, \citenamefont {Vretenar},\ and\ \citenamefont
  {Ring}}]{Paar:2004}%
  \BibitemOpen
  \bibfield  {author} {\bibinfo {author} {\bibfnamefont {N.}~\bibnamefont
  {Paar}}, \bibinfo {author} {\bibfnamefont {T.}~\bibnamefont {Niksic}},
  \bibinfo {author} {\bibfnamefont {D.}~\bibnamefont {Vretenar}}, \ and\
  \bibinfo {author} {\bibfnamefont {P.}~\bibnamefont {Ring}},\ }\href {\doibase
  10.1103/PhysRevC.69.054303} {\bibfield  {journal} {\bibinfo  {journal} {Phys.
  Rev. C}\ }\textbf {\bibinfo {volume} {69}},\ \bibinfo {pages} {054303}
  (\bibinfo {year} {2004})},\ \Eprint {http://arxiv.org/abs/nucl-th/0402094}
  {arXiv:nucl-th/0402094} \BibitemShut {NoStop}%
\bibitem [{\citenamefont {Niksic}\ \emph {et~al.}(2005)\citenamefont {Niksic},
  \citenamefont {Marketin}, \citenamefont {Vretenar}, \citenamefont {Paar},\
  and\ \citenamefont {Ring}}]{Niksic:2005}%
  \BibitemOpen
  \bibfield  {author} {\bibinfo {author} {\bibfnamefont {T.}~\bibnamefont
  {Niksic}}, \bibinfo {author} {\bibfnamefont {T.}~\bibnamefont {Marketin}},
  \bibinfo {author} {\bibfnamefont {D.}~\bibnamefont {Vretenar}}, \bibinfo
  {author} {\bibfnamefont {N.}~\bibnamefont {Paar}}, \ and\ \bibinfo {author}
  {\bibfnamefont {P.}~\bibnamefont {Ring}},\ }\href {\doibase
  10.1103/PhysRevC.71.014308} {\bibfield  {journal} {\bibinfo  {journal} {Phys.
  Rev. C}\ }\textbf {\bibinfo {volume} {71}},\ \bibinfo {pages} {014308}
  (\bibinfo {year} {2005})},\ \Eprint {http://arxiv.org/abs/nucl-th/0412028}
  {arXiv:nucl-th/0412028} \BibitemShut {NoStop}%
\bibitem [{\citenamefont {Marketin}\ \emph {et~al.}(2007)\citenamefont
  {Marketin}, \citenamefont {Vretenar},\ and\ \citenamefont
  {Ring}}]{Marketin:2007}%
  \BibitemOpen
  \bibfield  {author} {\bibinfo {author} {\bibfnamefont {T.}~\bibnamefont
  {Marketin}}, \bibinfo {author} {\bibfnamefont {D.}~\bibnamefont {Vretenar}},
  \ and\ \bibinfo {author} {\bibfnamefont {P.}~\bibnamefont {Ring}},\ }\href
  {\doibase 10.1103/PhysRevC.75.024304} {\bibfield  {journal} {\bibinfo
  {journal} {Phys. Rev. C}\ }\textbf {\bibinfo {volume} {75}},\ \bibinfo
  {pages} {024304} (\bibinfo {year} {2007})},\ \Eprint
  {http://arxiv.org/abs/nucl-th/0701025} {arXiv:nucl-th/0701025} \BibitemShut
  {NoStop}%
\bibitem [{\citenamefont {Niu}\ \emph {et~al.}(2013)\citenamefont {Niu},
  \citenamefont {Niu}, \citenamefont {Liang}, \citenamefont {Long},
  \citenamefont {Nik\v{s}i\'{c}}, \citenamefont {Vretenar},\ and\ \citenamefont
  {Meng}}]{Niu:2013PLB}%
  \BibitemOpen
  \bibfield  {author} {\bibinfo {author} {\bibfnamefont {Z.~M.}\ \bibnamefont
  {Niu}}, \bibinfo {author} {\bibfnamefont {Y.~F.}\ \bibnamefont {Niu}},
  \bibinfo {author} {\bibfnamefont {H.~Z.}\ \bibnamefont {Liang}}, \bibinfo
  {author} {\bibfnamefont {W.~H.}\ \bibnamefont {Long}}, \bibinfo {author}
  {\bibfnamefont {T.}~\bibnamefont {Nik\v{s}i\'{c}}}, \bibinfo {author}
  {\bibfnamefont {D.}~\bibnamefont {Vretenar}}, \ and\ \bibinfo {author}
  {\bibfnamefont {J.}~\bibnamefont {Meng}},\ }\href {\doibase
  http://dx.doi.org/10.1016/j.physletb.2013.04.048} {\bibfield  {journal}
  {\bibinfo  {journal} {Phys. Lett. B}\ }\textbf {\bibinfo {volume} {723}},\
  \bibinfo {pages} {172 } (\bibinfo {year} {2013})}\BibitemShut {NoStop}%
\bibitem [{\citenamefont {Liang}\ \emph {et~al.}(2011)\citenamefont {Liang},
  \citenamefont {Zhao}, \citenamefont {Li},\ and\ \citenamefont
  {Meng}}]{Liang:2010dy}%
  \BibitemOpen
  \bibfield  {author} {\bibinfo {author} {\bibfnamefont {H.}~\bibnamefont
  {Liang}}, \bibinfo {author} {\bibfnamefont {P.}~\bibnamefont {Zhao}},
  \bibinfo {author} {\bibfnamefont {L.}~\bibnamefont {Li}}, \ and\ \bibinfo
  {author} {\bibfnamefont {J.}~\bibnamefont {Meng}},\ }\href {\doibase
  10.1103/PhysRevC.83.011302} {\bibfield  {journal} {\bibinfo  {journal} {Phys.
  Rev. C}\ }\textbf {\bibinfo {volume} {83}},\ \bibinfo {pages} {011302}
  (\bibinfo {year} {2011})},\ \Eprint {http://arxiv.org/abs/1012.5865}
  {arXiv:1012.5865 [nucl-th]} \BibitemShut {NoStop}%
\bibitem [{\citenamefont {Marketin}\ \emph {et~al.}(2016)\citenamefont
  {Marketin}, \citenamefont {Huther},\ and\ \citenamefont
  {Mart\'\i{}nez-Pinedo}}]{Marketin:2015gya}%
  \BibitemOpen
  \bibfield  {author} {\bibinfo {author} {\bibfnamefont {T.}~\bibnamefont
  {Marketin}}, \bibinfo {author} {\bibfnamefont {L.}~\bibnamefont {Huther}}, \
  and\ \bibinfo {author} {\bibfnamefont {G.}~\bibnamefont
  {Mart\'\i{}nez-Pinedo}},\ }\href {\doibase 10.1103/PhysRevC.93.025805}
  {\bibfield  {journal} {\bibinfo  {journal} {Phys. Rev. C}\ }\textbf {\bibinfo
  {volume} {93}},\ \bibinfo {pages} {025805} (\bibinfo {year} {2016})},\
  \Eprint {http://arxiv.org/abs/1507.07442} {arXiv:1507.07442 [nucl-th]}
  \BibitemShut {NoStop}%
\bibitem [{\citenamefont {Liu}\ \emph {et~al.}(2024)\citenamefont {Liu},
  \citenamefont {Engel}, \citenamefont {Hinohara},\ and\ \citenamefont
  {Kortelainen}}]{Liu:2024PRC}%
  \BibitemOpen
  \bibfield  {author} {\bibinfo {author} {\bibfnamefont {Q.}~\bibnamefont
  {Liu}}, \bibinfo {author} {\bibfnamefont {J.}~\bibnamefont {Engel}}, \bibinfo
  {author} {\bibfnamefont {N.}~\bibnamefont {Hinohara}}, \ and\ \bibinfo
  {author} {\bibfnamefont {M.}~\bibnamefont {Kortelainen}},\ }\href {\doibase
  10.1103/PhysRevC.109.044308} {\bibfield  {journal} {\bibinfo  {journal}
  {Phys. Rev. C}\ }\textbf {\bibinfo {volume} {109}},\ \bibinfo {pages}
  {044308} (\bibinfo {year} {2024})},\ \Eprint
  {http://arxiv.org/abs/2308.11802} {arXiv:2308.11802 [nucl-th]} \BibitemShut
  {NoStop}%
\bibitem [{\citenamefont {Arteaga}\ and\ \citenamefont
  {Ring}(2008)}]{Arteaga:2008}%
  \BibitemOpen
  \bibfield  {author} {\bibinfo {author} {\bibfnamefont {D.~P.}\ \bibnamefont
  {Arteaga}}\ and\ \bibinfo {author} {\bibfnamefont {P.}~\bibnamefont {Ring}},\
  }\href {\doibase 10.1103/PhysRevC.77.034317} {\bibfield  {journal} {\bibinfo
  {journal} {Phys. Rev. C}\ }\textbf {\bibinfo {volume} {77}},\ \bibinfo
  {pages} {034317} (\bibinfo {year} {2008})}\BibitemShut {NoStop}%
\bibitem [{\citenamefont {Peru}\ and\ \citenamefont
  {Goutte}(2008)}]{Peru:2008}%
  \BibitemOpen
  \bibfield  {author} {\bibinfo {author} {\bibfnamefont {S.}~\bibnamefont
  {Peru}}\ and\ \bibinfo {author} {\bibfnamefont {H.}~\bibnamefont {Goutte}},\
  }\href {\doibase 10.1103/PhysRevC.77.044313} {\bibfield  {journal} {\bibinfo
  {journal} {Phys. Rev. C}\ }\textbf {\bibinfo {volume} {77}},\ \bibinfo
  {pages} {044313} (\bibinfo {year} {2008})},\ \Eprint
  {http://arxiv.org/abs/0804.0130} {arXiv:0804.0130 [nucl-th]} \BibitemShut
  {NoStop}%
\bibitem [{\citenamefont {Yoshida}\ and\ \citenamefont
  {Van~Giai}(2008)}]{Yoshida:2008}%
  \BibitemOpen
  \bibfield  {author} {\bibinfo {author} {\bibfnamefont {K.}~\bibnamefont
  {Yoshida}}\ and\ \bibinfo {author} {\bibfnamefont {N.}~\bibnamefont
  {Van~Giai}},\ }\href {\doibase 10.1103/PhysRevC.78.064316} {\bibfield
  {journal} {\bibinfo  {journal} {Phys. Rev. C}\ }\textbf {\bibinfo {volume}
  {78}},\ \bibinfo {pages} {064316} (\bibinfo {year} {2008})},\ \Eprint
  {http://arxiv.org/abs/0809.0169} {arXiv:0809.0169 [nucl-th]} \BibitemShut
  {NoStop}%
\bibitem [{\citenamefont {Losa}\ \emph {et~al.}(2010)\citenamefont {Losa},
  \citenamefont {Pastore}, \citenamefont {Dossing}, \citenamefont {Vigezzi},\
  and\ \citenamefont {Broglia}}]{Losa:2010}%
  \BibitemOpen
  \bibfield  {author} {\bibinfo {author} {\bibfnamefont {C.}~\bibnamefont
  {Losa}}, \bibinfo {author} {\bibfnamefont {A.}~\bibnamefont {Pastore}},
  \bibinfo {author} {\bibfnamefont {T.}~\bibnamefont {Dossing}}, \bibinfo
  {author} {\bibfnamefont {E.}~\bibnamefont {Vigezzi}}, \ and\ \bibinfo
  {author} {\bibfnamefont {R.~A.}\ \bibnamefont {Broglia}},\ }\href {\doibase
  10.1103/PhysRevC.81.064307} {\bibfield  {journal} {\bibinfo  {journal} {Phys.
  Rev. C}\ }\textbf {\bibinfo {volume} {81}},\ \bibinfo {pages} {064307}
  (\bibinfo {year} {2010})},\ \Eprint {http://arxiv.org/abs/1002.4351}
  {arXiv:1002.4351 [nucl-th]} \BibitemShut {NoStop}%
\bibitem [{\citenamefont {Terasaki}\ and\ \citenamefont
  {Engel}(2010)}]{Terasaki:2010}%
  \BibitemOpen
  \bibfield  {author} {\bibinfo {author} {\bibfnamefont {J.}~\bibnamefont
  {Terasaki}}\ and\ \bibinfo {author} {\bibfnamefont {J.}~\bibnamefont
  {Engel}},\ }\href {\doibase 10.1103/PhysRevC.82.034326} {\bibfield  {journal}
  {\bibinfo  {journal} {Phys. Rev. C}\ }\textbf {\bibinfo {volume} {82}},\
  \bibinfo {pages} {034326} (\bibinfo {year} {2010})},\ \Eprint
  {http://arxiv.org/abs/1006.0010} {arXiv:1006.0010 [nucl-th]} \BibitemShut
  {NoStop}%
\bibitem [{\citenamefont {Repko}\ \emph {et~al.}(2019)\citenamefont {Repko},
  \citenamefont {Kvasil},\ and\ \citenamefont {Nesterenko}}]{Repko:2018gcn}%
  \BibitemOpen
  \bibfield  {author} {\bibinfo {author} {\bibfnamefont {A.}~\bibnamefont
  {Repko}}, \bibinfo {author} {\bibfnamefont {J.}~\bibnamefont {Kvasil}}, \
  and\ \bibinfo {author} {\bibfnamefont {V.~O.}\ \bibnamefont {Nesterenko}},\
  }\href {\doibase 10.1103/PhysRevC.99.044307} {\bibfield  {journal} {\bibinfo
  {journal} {Phys. Rev. C}\ }\textbf {\bibinfo {volume} {99}},\ \bibinfo
  {pages} {044307} (\bibinfo {year} {2019})},\ \Eprint
  {http://arxiv.org/abs/1809.01911} {arXiv:1809.01911 [nucl-th]} \BibitemShut
  {NoStop}%
\bibitem [{\citenamefont {Kvasil}\ \emph {et~al.}(2019)\citenamefont {Kvasil},
  \citenamefont {Repko},\ and\ \citenamefont {Nesterenko}}]{Kvasil:2019giv}%
  \BibitemOpen
  \bibfield  {author} {\bibinfo {author} {\bibfnamefont {J.}~\bibnamefont
  {Kvasil}}, \bibinfo {author} {\bibfnamefont {A.}~\bibnamefont {Repko}}, \
  and\ \bibinfo {author} {\bibfnamefont {V.~O.}\ \bibnamefont {Nesterenko}},\
  }\href {\doibase 10.1140/epja/i2019-12898-7} {\bibfield  {journal} {\bibinfo
  {journal} {Eur. Phys. J. A}\ }\textbf {\bibinfo {volume} {55}},\ \bibinfo
  {pages} {213} (\bibinfo {year} {2019})}\BibitemShut {NoStop}%
\bibitem [{\citenamefont {Terasaki}(2016)}]{Terasaki:2016}%
  \BibitemOpen
  \bibfield  {author} {\bibinfo {author} {\bibfnamefont {J.}~\bibnamefont
  {Terasaki}},\ }\href {\doibase 10.1103/PhysRevC.93.024317} {\bibfield
  {journal} {\bibinfo  {journal} {Phys. Rev. C}\ }\textbf {\bibinfo {volume}
  {93}},\ \bibinfo {pages} {024317} (\bibinfo {year} {2016})}\BibitemShut
  {NoStop}%
\bibitem [{\citenamefont {Nakatsukasa}\ \emph {et~al.}(2007)\citenamefont
  {Nakatsukasa}, \citenamefont {Inakura},\ and\ \citenamefont
  {Yabana}}]{Nakatsukasa:2007qj}%
  \BibitemOpen
  \bibfield  {author} {\bibinfo {author} {\bibfnamefont {T.}~\bibnamefont
  {Nakatsukasa}}, \bibinfo {author} {\bibfnamefont {T.}~\bibnamefont
  {Inakura}}, \ and\ \bibinfo {author} {\bibfnamefont {K.}~\bibnamefont
  {Yabana}},\ }\href {\doibase 10.1103/PhysRevC.76.024318} {\bibfield
  {journal} {\bibinfo  {journal} {Phys. Rev. C}\ }\textbf {\bibinfo {volume}
  {76}},\ \bibinfo {pages} {024318} (\bibinfo {year} {2007})},\ \Eprint
  {http://arxiv.org/abs/nucl-th/0703100} {arXiv:nucl-th/0703100} \BibitemShut
  {NoStop}%
\bibitem [{\citenamefont {Inakura}\ \emph
  {et~al.}(2009{\natexlab{a}})\citenamefont {Inakura}, \citenamefont
  {Nakatsukasa},\ and\ \citenamefont {Yabana}}]{Inakura:2009PRC}%
  \BibitemOpen
  \bibfield  {author} {\bibinfo {author} {\bibfnamefont {T.}~\bibnamefont
  {Inakura}}, \bibinfo {author} {\bibfnamefont {T.}~\bibnamefont
  {Nakatsukasa}}, \ and\ \bibinfo {author} {\bibfnamefont {K.}~\bibnamefont
  {Yabana}},\ }\href {\doibase 10.1103/PhysRevC.80.044301} {\bibfield
  {journal} {\bibinfo  {journal} {Phys. Rev. C}\ }\textbf {\bibinfo {volume}
  {80}},\ \bibinfo {pages} {044301} (\bibinfo {year} {2009}{\natexlab{a}})},\
  \Eprint {http://arxiv.org/abs/0906.5239} {arXiv:0906.5239 [nucl-th]}
  \BibitemShut {NoStop}%
\bibitem [{\citenamefont {Inakura}\ \emph {et~al.}(2010)\citenamefont
  {Inakura}, \citenamefont {Nakatsukasa},\ and\ \citenamefont
  {Yabana}}]{Inakura:2010zz}%
  \BibitemOpen
  \bibfield  {author} {\bibinfo {author} {\bibfnamefont {T.}~\bibnamefont
  {Inakura}}, \bibinfo {author} {\bibfnamefont {T.}~\bibnamefont
  {Nakatsukasa}}, \ and\ \bibinfo {author} {\bibfnamefont {K.}~\bibnamefont
  {Yabana}},\ }\href {\doibase 10.1142/S0217732310000678} {\bibfield  {journal}
  {\bibinfo  {journal} {Mod. Phys. Lett. A}\ }\textbf {\bibinfo {volume}
  {25}},\ \bibinfo {pages} {1931} (\bibinfo {year} {2010})}\BibitemShut
  {NoStop}%
\bibitem [{\citenamefont {Stoitsov}\ \emph {et~al.}(2011)\citenamefont
  {Stoitsov}, \citenamefont {Kortelainen}, \citenamefont {Nakatsukasa},
  \citenamefont {Losa},\ and\ \citenamefont {Nazarewicz}}]{Stoitsov:2011zz}%
  \BibitemOpen
  \bibfield  {author} {\bibinfo {author} {\bibfnamefont {M.}~\bibnamefont
  {Stoitsov}}, \bibinfo {author} {\bibfnamefont {M.}~\bibnamefont
  {Kortelainen}}, \bibinfo {author} {\bibfnamefont {T.}~\bibnamefont
  {Nakatsukasa}}, \bibinfo {author} {\bibfnamefont {C.}~\bibnamefont {Losa}}, \
  and\ \bibinfo {author} {\bibfnamefont {W.}~\bibnamefont {Nazarewicz}},\
  }\href {\doibase 10.1103/PhysRevC.84.041305} {\bibfield  {journal} {\bibinfo
  {journal} {Phys. Rev. C}\ }\textbf {\bibinfo {volume} {84}},\ \bibinfo
  {pages} {041305} (\bibinfo {year} {2011})},\ \Eprint
  {http://arxiv.org/abs/1107.3530} {arXiv:1107.3530 [nucl-th]} \BibitemShut
  {NoStop}%
\bibitem [{\citenamefont {Avogadro}\ and\ \citenamefont
  {Nakatsukasa}(2011)}]{Avogadro:2011PRC}%
  \BibitemOpen
  \bibfield  {author} {\bibinfo {author} {\bibfnamefont {P.}~\bibnamefont
  {Avogadro}}\ and\ \bibinfo {author} {\bibfnamefont {T.}~\bibnamefont
  {Nakatsukasa}},\ }\href {\doibase 10.1103/PhysRevC.84.014314} {\bibfield
  {journal} {\bibinfo  {journal} {Phys. Rev. C}\ }\textbf {\bibinfo {volume}
  {84}},\ \bibinfo {pages} {014314} (\bibinfo {year} {2011})},\ \Eprint
  {http://arxiv.org/abs/1104.3692} {arXiv:1104.3692 [nucl-th]} \BibitemShut
  {NoStop}%
\bibitem [{\citenamefont {Niksic}\ \emph {et~al.}(2013)\citenamefont {Niksic},
  \citenamefont {Kralj}, \citenamefont {Tutis}, \citenamefont {Vretenar},\ and\
  \citenamefont {Ring}}]{Niksic:2013ega}%
  \BibitemOpen
  \bibfield  {author} {\bibinfo {author} {\bibfnamefont {T.}~\bibnamefont
  {Niksic}}, \bibinfo {author} {\bibfnamefont {N.}~\bibnamefont {Kralj}},
  \bibinfo {author} {\bibfnamefont {T.}~\bibnamefont {Tutis}}, \bibinfo
  {author} {\bibfnamefont {D.}~\bibnamefont {Vretenar}}, \ and\ \bibinfo
  {author} {\bibfnamefont {P.}~\bibnamefont {Ring}},\ }\href {\doibase
  10.1103/PhysRevC.88.044327} {\bibfield  {journal} {\bibinfo  {journal} {Phys.
  Rev. C}\ }\textbf {\bibinfo {volume} {88}},\ \bibinfo {pages} {044327}
  (\bibinfo {year} {2013})},\ \Eprint {http://arxiv.org/abs/1311.6581}
  {arXiv:1311.6581 [nucl-th]} \BibitemShut {NoStop}%
\bibitem [{\citenamefont {Liang}\ \emph {et~al.}(2013)\citenamefont {Liang},
  \citenamefont {Nakatsukasa}, \citenamefont {Niu},\ and\ \citenamefont
  {Meng}}]{Liang:2013pda}%
  \BibitemOpen
  \bibfield  {author} {\bibinfo {author} {\bibfnamefont {H.}~\bibnamefont
  {Liang}}, \bibinfo {author} {\bibfnamefont {T.}~\bibnamefont {Nakatsukasa}},
  \bibinfo {author} {\bibfnamefont {Z.}~\bibnamefont {Niu}}, \ and\ \bibinfo
  {author} {\bibfnamefont {J.}~\bibnamefont {Meng}},\ }\href {\doibase
  10.1103/PhysRevC.87.054310} {\bibfield  {journal} {\bibinfo  {journal} {Phys.
  Rev. C}\ }\textbf {\bibinfo {volume} {87}},\ \bibinfo {pages} {054310}
  (\bibinfo {year} {2013})},\ \Eprint {http://arxiv.org/abs/1304.3953}
  {arXiv:1304.3953 [nucl-th]} \BibitemShut {NoStop}%
\bibitem [{\citenamefont {Hinohara}\ \emph {et~al.}(2013)\citenamefont
  {Hinohara}, \citenamefont {Kortelainen},\ and\ \citenamefont
  {Nazarewicz}}]{Hinohara:2013qda}%
  \BibitemOpen
  \bibfield  {author} {\bibinfo {author} {\bibfnamefont {N.}~\bibnamefont
  {Hinohara}}, \bibinfo {author} {\bibfnamefont {M.}~\bibnamefont
  {Kortelainen}}, \ and\ \bibinfo {author} {\bibfnamefont {W.}~\bibnamefont
  {Nazarewicz}},\ }\href {\doibase 10.1103/PhysRevC.87.064309} {\bibfield
  {journal} {\bibinfo  {journal} {Phys. Rev. C}\ }\textbf {\bibinfo {volume}
  {87}},\ \bibinfo {pages} {064309} (\bibinfo {year} {2013})},\ \Eprint
  {http://arxiv.org/abs/1304.4008} {arXiv:1304.4008 [nucl-th]} \BibitemShut
  {NoStop}%
\bibitem [{\citenamefont {Liang}\ \emph {et~al.}(2014)\citenamefont {Liang},
  \citenamefont {Nakatsukasa}, \citenamefont {Niu},\ and\ \citenamefont
  {Meng}}]{Liang_2014}%
  \BibitemOpen
  \bibfield  {author} {\bibinfo {author} {\bibfnamefont {H.}~\bibnamefont
  {Liang}}, \bibinfo {author} {\bibfnamefont {T.}~\bibnamefont {Nakatsukasa}},
  \bibinfo {author} {\bibfnamefont {Z.}~\bibnamefont {Niu}}, \ and\ \bibinfo
  {author} {\bibfnamefont {J.}~\bibnamefont {Meng}},\ }\href {\doibase
  10.1088/0031-8949/89/5/054018} {\bibfield  {journal} {\bibinfo  {journal}
  {Physica Scripta}\ }\textbf {\bibinfo {volume} {89}},\ \bibinfo {pages}
  {054018} (\bibinfo {year} {2014})}\BibitemShut {NoStop}%
\bibitem [{\citenamefont {Oishi}\ \emph {et~al.}(2016)\citenamefont {Oishi},
  \citenamefont {Kortelainen},\ and\ \citenamefont {Hinohara}}]{Oishi:2015lph}%
  \BibitemOpen
  \bibfield  {author} {\bibinfo {author} {\bibfnamefont {T.}~\bibnamefont
  {Oishi}}, \bibinfo {author} {\bibfnamefont {M.}~\bibnamefont {Kortelainen}},
  \ and\ \bibinfo {author} {\bibfnamefont {N.}~\bibnamefont {Hinohara}},\
  }\href {\doibase 10.1103/PhysRevC.93.034329} {\bibfield  {journal} {\bibinfo
  {journal} {Phys. Rev. C}\ }\textbf {\bibinfo {volume} {93}},\ \bibinfo
  {pages} {034329} (\bibinfo {year} {2016})},\ \Eprint
  {http://arxiv.org/abs/1512.09146} {arXiv:1512.09146 [nucl-th]} \BibitemShut
  {NoStop}%
\bibitem [{\citenamefont {Inakura}\ \emph
  {et~al.}(2009{\natexlab{b}})\citenamefont {Inakura}, \citenamefont
  {Nakatsukasa},\ and\ \citenamefont {Yabana}}]{Inakura:2009tqn}%
  \BibitemOpen
  \bibfield  {author} {\bibinfo {author} {\bibfnamefont {T.}~\bibnamefont
  {Inakura}}, \bibinfo {author} {\bibfnamefont {T.}~\bibnamefont
  {Nakatsukasa}}, \ and\ \bibinfo {author} {\bibfnamefont {K.}~\bibnamefont
  {Yabana}},\ }\href {\doibase 10.1063/1.3141657} {\bibfield  {journal}
  {\bibinfo  {journal} {AIP Conf. Proc.}\ }\textbf {\bibinfo {volume} {1120}},\
  \bibinfo {pages} {260} (\bibinfo {year} {2009}{\natexlab{b}})}\BibitemShut
  {NoStop}%
\bibitem [{\citenamefont {Shafer}\ \emph {et~al.}(2016)\citenamefont {Shafer},
  \citenamefont {Engel}, \citenamefont {Fr\"ohlich}, \citenamefont
  {McLaughlin}, \citenamefont {Mumpower},\ and\ \citenamefont
  {Surman}}]{Shafer:2016etk}%
  \BibitemOpen
  \bibfield  {author} {\bibinfo {author} {\bibfnamefont {T.}~\bibnamefont
  {Shafer}}, \bibinfo {author} {\bibfnamefont {J.}~\bibnamefont {Engel}},
  \bibinfo {author} {\bibfnamefont {C.}~\bibnamefont {Fr\"ohlich}}, \bibinfo
  {author} {\bibfnamefont {G.~C.}\ \bibnamefont {McLaughlin}}, \bibinfo
  {author} {\bibfnamefont {M.}~\bibnamefont {Mumpower}}, \ and\ \bibinfo
  {author} {\bibfnamefont {R.}~\bibnamefont {Surman}},\ }\href {\doibase
  10.1103/PhysRevC.94.055802} {\bibfield  {journal} {\bibinfo  {journal} {Phys.
  Rev. C}\ }\textbf {\bibinfo {volume} {94}},\ \bibinfo {pages} {055802}
  (\bibinfo {year} {2016})},\ \Eprint {http://arxiv.org/abs/1606.05909}
  {arXiv:1606.05909 [nucl-th]} \BibitemShut {NoStop}%
\bibitem [{\citenamefont {Mustonen}\ \emph {et~al.}(2014)\citenamefont
  {Mustonen}, \citenamefont {Shafer}, \citenamefont {Zenginerler},\ and\
  \citenamefont {Engel}}]{Mustonen:2014bya}%
  \BibitemOpen
  \bibfield  {author} {\bibinfo {author} {\bibfnamefont {M.~T.}\ \bibnamefont
  {Mustonen}}, \bibinfo {author} {\bibfnamefont {T.}~\bibnamefont {Shafer}},
  \bibinfo {author} {\bibfnamefont {Z.}~\bibnamefont {Zenginerler}}, \ and\
  \bibinfo {author} {\bibfnamefont {J.}~\bibnamefont {Engel}},\ }\href
  {\doibase 10.1103/PhysRevC.90.024308} {\bibfield  {journal} {\bibinfo
  {journal} {Phys. Rev. C}\ }\textbf {\bibinfo {volume} {90}},\ \bibinfo
  {pages} {024308} (\bibinfo {year} {2014})},\ \Eprint
  {http://arxiv.org/abs/1405.0254} {arXiv:1405.0254 [nucl-th]} \BibitemShut
  {NoStop}%
\bibitem [{\citenamefont {Mustonen}\ and\ \citenamefont
  {Engel}(2016)}]{Mustonen:2015sfa}%
  \BibitemOpen
  \bibfield  {author} {\bibinfo {author} {\bibfnamefont {M.~T.}\ \bibnamefont
  {Mustonen}}\ and\ \bibinfo {author} {\bibfnamefont {J.}~\bibnamefont
  {Engel}},\ }\href {\doibase 10.1103/PhysRevC.93.014304} {\bibfield  {journal}
  {\bibinfo  {journal} {Phys. Rev. C}\ }\textbf {\bibinfo {volume} {93}},\
  \bibinfo {pages} {014304} (\bibinfo {year} {2016})},\ \Eprint
  {http://arxiv.org/abs/1510.02136} {arXiv:1510.02136 [nucl-th]} \BibitemShut
  {NoStop}%
\bibitem [{\citenamefont {Ney}\ \emph {et~al.}(2020)\citenamefont {Ney},
  \citenamefont {Engel}, \citenamefont {Schunck},\ and\ \citenamefont
  {Li}}]{Ney:2020mnx}%
  \BibitemOpen
  \bibfield  {author} {\bibinfo {author} {\bibfnamefont {E.~M.}\ \bibnamefont
  {Ney}}, \bibinfo {author} {\bibfnamefont {J.}~\bibnamefont {Engel}}, \bibinfo
  {author} {\bibfnamefont {N.}~\bibnamefont {Schunck}}, \ and\ \bibinfo
  {author} {\bibfnamefont {T.}~\bibnamefont {Li}},\ }\href {\doibase
  10.1103/PhysRevC.102.034326} {\bibfield  {journal} {\bibinfo  {journal}
  {Phys. Rev. C}\ }\textbf {\bibinfo {volume} {102}},\ \bibinfo {pages}
  {034326} (\bibinfo {year} {2020})},\ \Eprint
  {http://arxiv.org/abs/2005.12883} {arXiv:2005.12883 [nucl-th]} \BibitemShut
  {NoStop}%
\bibitem [{\citenamefont {Federschmidt}\ and\ \citenamefont
  {Ring}(1985)}]{Federschmidt:1985}%
  \BibitemOpen
  \bibfield  {author} {\bibinfo {author} {\bibfnamefont {C.}~\bibnamefont
  {Federschmidt}}\ and\ \bibinfo {author} {\bibfnamefont {P.}~\bibnamefont
  {Ring}},\ }\href {\doibase 10.1016/0375-9474(85)90307-0} {\bibfield
  {journal} {\bibinfo  {journal} {Nucl. Phys. A}\ }\textbf {\bibinfo {volume}
  {435}},\ \bibinfo {pages} {110} (\bibinfo {year} {1985})}\BibitemShut
  {NoStop}%
\bibitem [{\citenamefont {Krumlinde}\ and\ \citenamefont
  {M\"oller}(1984)}]{Krumlinde:1984hbw}%
  \BibitemOpen
  \bibfield  {author} {\bibinfo {author} {\bibfnamefont {J.}~\bibnamefont
  {Krumlinde}}\ and\ \bibinfo {author} {\bibfnamefont {P.}~\bibnamefont
  {M\"oller}},\ }\href {\doibase 10.1016/0375-9474(84)90406-8} {\bibfield
  {journal} {\bibinfo  {journal} {Nucl. Phys. A}\ }\textbf {\bibinfo {volume}
  {417}},\ \bibinfo {pages} {419} (\bibinfo {year} {1984})}\BibitemShut
  {NoStop}%
\bibitem [{\citenamefont {Yousef}\ \emph {et~al.}(2009)\citenamefont {Yousef},
  \citenamefont {Rodin}, \citenamefont {Faessler},\ and\ \citenamefont
  {Simkovic}}]{Yousef:2009}%
  \BibitemOpen
  \bibfield  {author} {\bibinfo {author} {\bibfnamefont {M.~S.}\ \bibnamefont
  {Yousef}}, \bibinfo {author} {\bibfnamefont {V.}~\bibnamefont {Rodin}},
  \bibinfo {author} {\bibfnamefont {A.}~\bibnamefont {Faessler}}, \ and\
  \bibinfo {author} {\bibfnamefont {F.}~\bibnamefont {Simkovic}},\ }\href
  {\doibase 10.1103/PhysRevC.79.014314} {\bibfield  {journal} {\bibinfo
  {journal} {Phys. Rev. C}\ }\textbf {\bibinfo {volume} {79}},\ \bibinfo
  {pages} {014314} (\bibinfo {year} {2009})},\ \Eprint
  {http://arxiv.org/abs/0806.0964} {arXiv:0806.0964 [nucl-th]} \BibitemShut
  {NoStop}%
\bibitem [{\citenamefont {Ravli\'c}\ \emph {et~al.}(2024)\citenamefont
  {Ravli\'c}, \citenamefont {Nik\v{s}i\'c}, \citenamefont {Niu}, \citenamefont
  {Ring},\ and\ \citenamefont {Paar}}]{Ravlic:2024hpi}%
  \BibitemOpen
  \bibfield  {author} {\bibinfo {author} {\bibfnamefont {A.}~\bibnamefont
  {Ravli\'c}}, \bibinfo {author} {\bibfnamefont {T.}~\bibnamefont
  {Nik\v{s}i\'c}}, \bibinfo {author} {\bibfnamefont {Y.~F.}\ \bibnamefont
  {Niu}}, \bibinfo {author} {\bibfnamefont {P.}~\bibnamefont {Ring}}, \ and\
  \bibinfo {author} {\bibfnamefont {N.}~\bibnamefont {Paar}},\ }\href {\doibase
  10.1103/PhysRevC.110.024323} {\bibfield  {journal} {\bibinfo  {journal}
  {Phys. Rev. C}\ }\textbf {\bibinfo {volume} {110}},\ \bibinfo {pages}
  {024323} (\bibinfo {year} {2024})},\ \Eprint
  {http://arxiv.org/abs/2404.13266} {arXiv:2404.13266 [nucl-th]} \BibitemShut
  {NoStop}%
\bibitem [{\citenamefont {Ring}\ and\ \citenamefont
  {Schuck}(1980)}]{Ring:1980}%
  \BibitemOpen
  \bibfield  {author} {\bibinfo {author} {\bibfnamefont {P.}~\bibnamefont
  {Ring}}\ and\ \bibinfo {author} {\bibfnamefont {P.}~\bibnamefont {Schuck}},\
  }\href@noop {} {\emph {\bibinfo {title} {The nuclear many-body problem}}}\
  (\bibinfo  {publisher} {Springer-Verlag},\ \bibinfo {address} {New York},\
  \bibinfo {year} {1980})\BibitemShut {NoStop}%
\bibitem [{\citenamefont {Erler}\ and\ \citenamefont
  {Roth}(2014)}]{Erler:2014}%
  \BibitemOpen
  \bibfield  {author} {\bibinfo {author} {\bibfnamefont {B.}~\bibnamefont
  {Erler}}\ and\ \bibinfo {author} {\bibfnamefont {R.}~\bibnamefont {Roth}},\
  }\href@noop {} {\  (\bibinfo {year} {2014})},\ \Eprint
  {http://arxiv.org/abs/1409.0826} {arXiv:1409.0826 [nucl-th]} \BibitemShut
  {NoStop}%
\bibitem [{\citenamefont {Porro}\ \emph {et~al.}(2024)\citenamefont {Porro},
  \citenamefont {Col\`o}, \citenamefont {Duguet}, \citenamefont {Gambacurta},\
  and\ \citenamefont {Som\`a}}]{Porro:2024}%
  \BibitemOpen
  \bibfield  {author} {\bibinfo {author} {\bibfnamefont {A.}~\bibnamefont
  {Porro}}, \bibinfo {author} {\bibfnamefont {G.}~\bibnamefont {Col\`o}},
  \bibinfo {author} {\bibfnamefont {T.}~\bibnamefont {Duguet}}, \bibinfo
  {author} {\bibfnamefont {D.}~\bibnamefont {Gambacurta}}, \ and\ \bibinfo
  {author} {\bibfnamefont {V.}~\bibnamefont {Som\`a}},\ }\href {\doibase
  10.1103/PhysRevC.109.044315} {\bibfield  {journal} {\bibinfo  {journal}
  {Phys. Rev. C}\ }\textbf {\bibinfo {volume} {109}},\ \bibinfo {pages}
  {044315} (\bibinfo {year} {2024})},\ \Eprint
  {http://arxiv.org/abs/2312.10410} {arXiv:2312.10410 [nucl-th]} \BibitemShut
  {NoStop}%
\bibitem [{\citenamefont {Martini}\ \emph {et~al.}(2014)\citenamefont
  {Martini}, \citenamefont {Peru},\ and\ \citenamefont
  {Goriely}}]{Martini:2014}%
  \BibitemOpen
  \bibfield  {author} {\bibinfo {author} {\bibfnamefont {M.}~\bibnamefont
  {Martini}}, \bibinfo {author} {\bibfnamefont {S.}~\bibnamefont {Peru}}, \
  and\ \bibinfo {author} {\bibfnamefont {S.}~\bibnamefont {Goriely}},\ }\href
  {\doibase 10.1103/PhysRevC.89.044306} {\bibfield  {journal} {\bibinfo
  {journal} {Phys. Rev. C}\ }\textbf {\bibinfo {volume} {89}},\ \bibinfo
  {pages} {044306} (\bibinfo {year} {2014})},\ \Eprint
  {http://arxiv.org/abs/1404.1493} {arXiv:1404.1493 [nucl-th]} \BibitemShut
  {NoStop}%
\bibitem [{\citenamefont {Hinohara}\ and\ \citenamefont
  {Engel}(2022)}]{Hinohara:2022uip}%
  \BibitemOpen
  \bibfield  {author} {\bibinfo {author} {\bibfnamefont {N.}~\bibnamefont
  {Hinohara}}\ and\ \bibinfo {author} {\bibfnamefont {J.}~\bibnamefont
  {Engel}},\ }\href {\doibase 10.1103/PhysRevC.105.044314} {\bibfield
  {journal} {\bibinfo  {journal} {Phys. Rev. C}\ }\textbf {\bibinfo {volume}
  {105}},\ \bibinfo {pages} {044314} (\bibinfo {year} {2022})},\ \Eprint
  {http://arxiv.org/abs/2201.12983} {arXiv:2201.12983 [nucl-th]} \BibitemShut
  {NoStop}%
\bibitem [{\citenamefont {Litvinova}\ \emph {et~al.}(2007)\citenamefont
  {Litvinova}, \citenamefont {Ring},\ and\ \citenamefont
  {Tselyaev}}]{Litvinova:2007gg}%
  \BibitemOpen
  \bibfield  {author} {\bibinfo {author} {\bibfnamefont {E.}~\bibnamefont
  {Litvinova}}, \bibinfo {author} {\bibfnamefont {P.}~\bibnamefont {Ring}}, \
  and\ \bibinfo {author} {\bibfnamefont {V.}~\bibnamefont {Tselyaev}},\ }\href
  {\doibase 10.1103/PhysRevC.75.064308} {\bibfield  {journal} {\bibinfo
  {journal} {Phys. Rev. C}\ }\textbf {\bibinfo {volume} {75}},\ \bibinfo
  {pages} {064308} (\bibinfo {year} {2007})},\ \Eprint
  {http://arxiv.org/abs/0705.1044} {arXiv:0705.1044 [nucl-th]} \BibitemShut
  {NoStop}%
\bibitem [{\citenamefont {Bahcall}(1966)}]{Bahcal:1966}%
  \BibitemOpen
  \bibfield  {author} {\bibinfo {author} {\bibfnamefont {J.~N.}\ \bibnamefont
  {Bahcall}},\ }\href {\doibase https://doi.org/10.1016/0029-5582(66)90745-0}
  {\bibfield  {journal} {\bibinfo  {journal} {Nuclear Physics}\ }\textbf
  {\bibinfo {volume} {75}},\ \bibinfo {pages} {10} (\bibinfo {year}
  {1966})}\BibitemShut {NoStop}%
\bibitem [{\citenamefont {Doi}\ \emph {et~al.}(1985)\citenamefont {Doi},
  \citenamefont {Kotani},\ and\ \citenamefont {Takasugi}}]{Doi:1985}%
  \BibitemOpen
  \bibfield  {author} {\bibinfo {author} {\bibfnamefont {M.}~\bibnamefont
  {Doi}}, \bibinfo {author} {\bibfnamefont {T.}~\bibnamefont {Kotani}}, \ and\
  \bibinfo {author} {\bibfnamefont {E.}~\bibnamefont {Takasugi}},\ }\href
  {\doibase 10.1143/PTPS.83.1} {\bibfield  {journal} {\bibinfo  {journal}
  {Prog. Theor. Phys. Suppl.}\ }\textbf {\bibinfo {volume} {83}},\ \bibinfo
  {pages} {1} (\bibinfo {year} {1985})}\BibitemShut {NoStop}%
\bibitem [{\citenamefont {Stoitsov}\ \emph {et~al.}(2005)\citenamefont
  {Stoitsov}, \citenamefont {Dobaczewski}, \citenamefont {Nazarewicz},\ and\
  \citenamefont {Ring}}]{STOITSOV200543}%
  \BibitemOpen
  \bibfield  {author} {\bibinfo {author} {\bibfnamefont {M.}~\bibnamefont
  {Stoitsov}}, \bibinfo {author} {\bibfnamefont {J.}~\bibnamefont
  {Dobaczewski}}, \bibinfo {author} {\bibfnamefont {W.}~\bibnamefont
  {Nazarewicz}}, \ and\ \bibinfo {author} {\bibfnamefont {P.}~\bibnamefont
  {Ring}},\ }\href {\doibase https://doi.org/10.1016/j.cpc.2005.01.001}
  {\bibfield  {journal} {\bibinfo  {journal} {Computer Physics Communications}\
  }\textbf {\bibinfo {volume} {167}},\ \bibinfo {pages} {43} (\bibinfo {year}
  {2005})}\BibitemShut {NoStop}%
\bibitem [{\citenamefont {Stoitsov}\ \emph {et~al.}(2013)\citenamefont
  {Stoitsov}, \citenamefont {Schunck}, \citenamefont {Kortelainen},
  \citenamefont {Michel}, \citenamefont {Nam}, \citenamefont {Olsen},
  \citenamefont {Sarich},\ and\ \citenamefont {Wild}}]{STOITSOV20131592}%
  \BibitemOpen
  \bibfield  {author} {\bibinfo {author} {\bibfnamefont {M.}~\bibnamefont
  {Stoitsov}}, \bibinfo {author} {\bibfnamefont {N.}~\bibnamefont {Schunck}},
  \bibinfo {author} {\bibfnamefont {M.}~\bibnamefont {Kortelainen}}, \bibinfo
  {author} {\bibfnamefont {N.}~\bibnamefont {Michel}}, \bibinfo {author}
  {\bibfnamefont {H.}~\bibnamefont {Nam}}, \bibinfo {author} {\bibfnamefont
  {E.}~\bibnamefont {Olsen}}, \bibinfo {author} {\bibfnamefont
  {J.}~\bibnamefont {Sarich}}, \ and\ \bibinfo {author} {\bibfnamefont
  {S.}~\bibnamefont {Wild}},\ }\href {\doibase
  https://doi.org/10.1016/j.cpc.2013.01.013} {\bibfield  {journal} {\bibinfo
  {journal} {Computer Physics Communications}\ }\textbf {\bibinfo {volume}
  {184}},\ \bibinfo {pages} {1592} (\bibinfo {year} {2013})}\BibitemShut
  {NoStop}%
\bibitem [{\citenamefont {Perez}\ \emph {et~al.}(2017)\citenamefont {Perez},
  \citenamefont {Schunck}, \citenamefont {Lasseri}, \citenamefont {Zhang},\
  and\ \citenamefont {Sarich}}]{PEREZ2017363}%
  \BibitemOpen
  \bibfield  {author} {\bibinfo {author} {\bibfnamefont {R.~N.}\ \bibnamefont
  {Perez}}, \bibinfo {author} {\bibfnamefont {N.}~\bibnamefont {Schunck}},
  \bibinfo {author} {\bibfnamefont {R.-D.}\ \bibnamefont {Lasseri}}, \bibinfo
  {author} {\bibfnamefont {C.}~\bibnamefont {Zhang}}, \ and\ \bibinfo {author}
  {\bibfnamefont {J.}~\bibnamefont {Sarich}},\ }\href {\doibase
  https://doi.org/10.1016/j.cpc.2017.06.022} {\bibfield  {journal} {\bibinfo
  {journal} {Computer Physics Communications}\ }\textbf {\bibinfo {volume}
  {220}},\ \bibinfo {pages} {363} (\bibinfo {year} {2017})}\BibitemShut
  {NoStop}%
\bibitem [{\citenamefont {Reinhard}\ \emph {et~al.}(1999)\citenamefont
  {Reinhard}, \citenamefont {Dean}, \citenamefont {Nazarewicz}, \citenamefont
  {Dobaczewski}, \citenamefont {Maruhn},\ and\ \citenamefont
  {Strayer}}]{Reinhard:1999ut}%
  \BibitemOpen
  \bibfield  {author} {\bibinfo {author} {\bibfnamefont {P.~G.}\ \bibnamefont
  {Reinhard}}, \bibinfo {author} {\bibfnamefont {D.~J.}\ \bibnamefont {Dean}},
  \bibinfo {author} {\bibfnamefont {W.}~\bibnamefont {Nazarewicz}}, \bibinfo
  {author} {\bibfnamefont {J.}~\bibnamefont {Dobaczewski}}, \bibinfo {author}
  {\bibfnamefont {J.~A.}\ \bibnamefont {Maruhn}}, \ and\ \bibinfo {author}
  {\bibfnamefont {M.~R.}\ \bibnamefont {Strayer}},\ }\href {\doibase
  10.1103/PhysRevC.60.014316} {\bibfield  {journal} {\bibinfo  {journal} {Phys.
  Rev. C}\ }\textbf {\bibinfo {volume} {60}},\ \bibinfo {pages} {014316}
  (\bibinfo {year} {1999})},\ \Eprint {http://arxiv.org/abs/nucl-th/9903037}
  {arXiv:nucl-th/9903037} \BibitemShut {NoStop}%
\bibitem [{\citenamefont {Bally}\ and\ \citenamefont
  {Bender}(2021)}]{Bally:2020wkb}%
  \BibitemOpen
  \bibfield  {author} {\bibinfo {author} {\bibfnamefont {B.}~\bibnamefont
  {Bally}}\ and\ \bibinfo {author} {\bibfnamefont {M.}~\bibnamefont {Bender}},\
  }\href {\doibase 10.1103/PhysRevC.103.024315} {\bibfield  {journal} {\bibinfo
   {journal} {Phys. Rev. C}\ }\textbf {\bibinfo {volume} {103}},\ \bibinfo
  {pages} {024315} (\bibinfo {year} {2021})},\ \Eprint
  {http://arxiv.org/abs/2010.15224} {arXiv:2010.15224 [nucl-th]} \BibitemShut
  {NoStop}%
\bibitem [{\citenamefont {{National Nuclear Data Center}}(2020)}]{NNDC}%
  \BibitemOpen
  \bibfield  {author} {\bibinfo {author} {\bibnamefont {{National Nuclear Data
  Center}}},\ }\href {https://www.nndc.bnl.gov/nudat2} {\enquote {\bibinfo
  {title} {{NuDat 2 Database}},}\ } (\bibinfo {year} {2020}),\ \bibinfo {note}
  {\url{https://www.nndc.bnl.gov/nudat2}}\BibitemShut {NoStop}%
\bibitem [{\citenamefont {Mercier}\ \emph {et~al.}(2022)\citenamefont
  {Mercier}, \citenamefont {Ebran},\ and\ \citenamefont {Khan}}]{Mercier:2022}%
  \BibitemOpen
  \bibfield  {author} {\bibinfo {author} {\bibfnamefont {F.}~\bibnamefont
  {Mercier}}, \bibinfo {author} {\bibfnamefont {J.~P.}\ \bibnamefont {Ebran}},
  \ and\ \bibinfo {author} {\bibfnamefont {E.}~\bibnamefont {Khan}},\ }\href
  {\doibase 10.1103/PhysRevC.105.034343} {\bibfield  {journal} {\bibinfo
  {journal} {Phys. Rev. C}\ }\textbf {\bibinfo {volume} {105}},\ \bibinfo
  {pages} {034343} (\bibinfo {year} {2022})},\ \Eprint
  {http://arxiv.org/abs/2109.02498} {arXiv:2109.02498 [nucl-th]} \BibitemShut
  {NoStop}%
\bibitem [{\citenamefont {Ikeda}(1964)}]{Ikeda64}%
  \BibitemOpen
  \bibfield  {author} {\bibinfo {author} {\bibfnamefont {K.}~\bibnamefont
  {Ikeda}},\ }\href {\doibase 10.1143/PTP.31.434} {\bibfield  {journal}
  {\bibinfo  {journal} {Progress of Theoretical Physics}\ }\textbf {\bibinfo
  {volume} {31}},\ \bibinfo {pages} {434} (\bibinfo {year} {1964})},\ \Eprint
  {http://arxiv.org/abs/https://academic.oup.com/ptp/article-pdf/31/3/434/5364608/31-3-434.pdf}
  {https://academic.oup.com/ptp/article-pdf/31/3/434/5364608/31-3-434.pdf}
  \BibitemShut {NoStop}%
\bibitem [{\citenamefont {Boillos}\ and\ \citenamefont
  {Sarriguren}(2015)}]{Boillos:2015}%
  \BibitemOpen
  \bibfield  {author} {\bibinfo {author} {\bibfnamefont {J.~M.}\ \bibnamefont
  {Boillos}}\ and\ \bibinfo {author} {\bibfnamefont {P.}~\bibnamefont
  {Sarriguren}},\ }\href {\doibase 10.1103/PhysRevC.91.034311} {\bibfield
  {journal} {\bibinfo  {journal} {Phys. Rev. C}\ }\textbf {\bibinfo {volume}
  {91}},\ \bibinfo {pages} {034311} (\bibinfo {year} {2015})},\ \Eprint
  {http://arxiv.org/abs/1503.02944} {arXiv:1503.02944 [nucl-th]} \BibitemShut
  {NoStop}%
\bibitem [{\citenamefont {Sarriguren}(2017)}]{Sarriguren:2017}%
  \BibitemOpen
  \bibfield  {author} {\bibinfo {author} {\bibfnamefont {P.}~\bibnamefont
  {Sarriguren}},\ }\href {\doibase 10.1103/PhysRevC.95.014304} {\bibfield
  {journal} {\bibinfo  {journal} {Phys. Rev. C}\ }\textbf {\bibinfo {volume}
  {95}},\ \bibinfo {pages} {014304} (\bibinfo {year} {2017})},\ \Eprint
  {http://arxiv.org/abs/1612.04084} {arXiv:1612.04084 [nucl-th]} \BibitemShut
  {NoStop}%
\bibitem [{\citenamefont {Yoshida}\ \emph {et~al.}(2023)\citenamefont
  {Yoshida}, \citenamefont {Niu},\ and\ \citenamefont {Minato}}]{Yoshida:2023}%
  \BibitemOpen
  \bibfield  {author} {\bibinfo {author} {\bibfnamefont {K.}~\bibnamefont
  {Yoshida}}, \bibinfo {author} {\bibfnamefont {Y.}~\bibnamefont {Niu}}, \ and\
  \bibinfo {author} {\bibfnamefont {F.}~\bibnamefont {Minato}},\ }\href
  {\doibase 10.1103/PhysRevC.108.034305} {\bibfield  {journal} {\bibinfo
  {journal} {Phys. Rev. C}\ }\textbf {\bibinfo {volume} {108}},\ \bibinfo
  {pages} {034305} (\bibinfo {year} {2023})}\BibitemShut {NoStop}%
\bibitem [{\citenamefont {Ney}\ \emph {et~al.}(2022)\citenamefont {Ney},
  \citenamefont {Engel},\ and\ \citenamefont {Schunck}}]{Ney:2022PRC}%
  \BibitemOpen
  \bibfield  {author} {\bibinfo {author} {\bibfnamefont {E.~M.}\ \bibnamefont
  {Ney}}, \bibinfo {author} {\bibfnamefont {J.}~\bibnamefont {Engel}}, \ and\
  \bibinfo {author} {\bibfnamefont {N.}~\bibnamefont {Schunck}},\ }\href
  {\doibase 10.1103/PhysRevC.105.034349} {\bibfield  {journal} {\bibinfo
  {journal} {Phys. Rev. C}\ }\textbf {\bibinfo {volume} {105}},\ \bibinfo
  {pages} {034349} (\bibinfo {year} {2022})}\BibitemShut {NoStop}%
\bibitem [{\citenamefont {Chen}\ \emph {et~al.}(2026)\citenamefont {Chen},
  \citenamefont {Zhang}, \citenamefont {Yao},\ and\ \citenamefont
  {Engel}}]{Data}%
  \BibitemOpen
  \bibfield  {author} {\bibinfo {author} {\bibfnamefont {R.~N.}\ \bibnamefont
  {Chen}}, \bibinfo {author} {\bibfnamefont {Y.~N.}\ \bibnamefont {Zhang}},
  \bibinfo {author} {\bibfnamefont {J.~M.}\ \bibnamefont {Yao}}, \ and\
  \bibinfo {author} {\bibfnamefont {J.}~\bibnamefont {Engel}},\ }\href
  {\doibase 10.5281/zenodo.18703406} {\enquote {\bibinfo {title}
  {10.5281/zenodo.18703406},}\ } (\bibinfo {year} {2026})\BibitemShut {NoStop}%
\bibitem [{\citenamefont {Onishi}\ and\ \citenamefont
  {Yoshida}(1966)}]{Onishi:1966}%
  \BibitemOpen
  \bibfield  {author} {\bibinfo {author} {\bibfnamefont {N.}~\bibnamefont
  {Onishi}}\ and\ \bibinfo {author} {\bibfnamefont {S.}~\bibnamefont
  {Yoshida}},\ }\href {\doibase https://doi.org/10.1016/0029-5582(66)90096-4}
  {\bibfield  {journal} {\bibinfo  {journal} {Nuclear Physics}\ }\textbf
  {\bibinfo {volume} {80}},\ \bibinfo {pages} {367} (\bibinfo {year}
  {1966})}\BibitemShut {NoStop}%
\bibitem [{\citenamefont {Neerg\r{a}rd}\ and\ \citenamefont
  {W\"ust}(1983)}]{Neergard:1983}%
  \BibitemOpen
  \bibfield  {author} {\bibinfo {author} {\bibfnamefont {K.}~\bibnamefont
  {Neerg\r{a}rd}}\ and\ \bibinfo {author} {\bibfnamefont {E.}~\bibnamefont
  {W\"ust}},\ }\href {\doibase 10.1016/0375-9474(83)90501-8} {\bibfield
  {journal} {\bibinfo  {journal} {Nucl. Phys. A}\ }\textbf {\bibinfo {volume}
  {402}},\ \bibinfo {pages} {311} (\bibinfo {year} {1983})}\BibitemShut
  {NoStop}%
\bibitem [{\citenamefont {Robledo}(2009)}]{Robledo2009yd}%
  \BibitemOpen
  \bibfield  {author} {\bibinfo {author} {\bibfnamefont {L.~M.}\ \bibnamefont
  {Robledo}},\ }\href {\doibase 10.1103/PhysRevC.79.021302} {\bibfield
  {journal} {\bibinfo  {journal} {Phys. Rev. C}\ }\textbf {\bibinfo {volume}
  {79}},\ \bibinfo {pages} {021302} (\bibinfo {year} {2009})},\ \Eprint
  {http://arxiv.org/abs/0901.3213} {arXiv:0901.3213 [nucl-th]} \BibitemShut
  {NoStop}%
\bibitem [{\citenamefont {Mizusaki}\ \emph {et~al.}(2018)\citenamefont
  {Mizusaki}, \citenamefont {Oi},\ and\ \citenamefont
  {Shimizu}}]{Mizusaki2017ist}%
  \BibitemOpen
  \bibfield  {author} {\bibinfo {author} {\bibfnamefont {T.}~\bibnamefont
  {Mizusaki}}, \bibinfo {author} {\bibfnamefont {M.}~\bibnamefont {Oi}}, \ and\
  \bibinfo {author} {\bibfnamefont {N.}~\bibnamefont {Shimizu}},\ }\href
  {\doibase 10.1016/j.physletb.2018.02.012} {\bibfield  {journal} {\bibinfo
  {journal} {Phys. Lett. B}\ }\textbf {\bibinfo {volume} {779}},\ \bibinfo
  {pages} {237} (\bibinfo {year} {2018})},\ \Eprint
  {http://arxiv.org/abs/1711.00369} {arXiv:1711.00369 [nucl-th]} \BibitemShut
  {NoStop}%
\bibitem [{\citenamefont {Porro}\ and\ \citenamefont
  {Duguet}(2022)}]{Porro2022tgc}%
  \BibitemOpen
  \bibfield  {author} {\bibinfo {author} {\bibfnamefont {A.}~\bibnamefont
  {Porro}}\ and\ \bibinfo {author} {\bibfnamefont {T.}~\bibnamefont {Duguet}},\
  }\href {\doibase 10.1140/epja/s10050-022-00843-2} {\bibfield  {journal}
  {\bibinfo  {journal} {Eur. Phys. J. A}\ }\textbf {\bibinfo {volume} {58}},\
  \bibinfo {pages} {197} (\bibinfo {year} {2022})},\ \Eprint
  {http://arxiv.org/abs/2206.03781} {arXiv:2206.03781 [nucl-th]} \BibitemShut
  {NoStop}%
\bibitem [{\citenamefont {Yao}\ \emph {et~al.}(2022)\citenamefont {Yao},
  \citenamefont {Meng}, \citenamefont {Niu},\ and\ \citenamefont
  {Ring}}]{Yao:2022PPNP}%
  \BibitemOpen
  \bibfield  {author} {\bibinfo {author} {\bibfnamefont {J.~M.}\ \bibnamefont
  {Yao}}, \bibinfo {author} {\bibfnamefont {J.}~\bibnamefont {Meng}}, \bibinfo
  {author} {\bibfnamefont {Y.~F.}\ \bibnamefont {Niu}}, \ and\ \bibinfo
  {author} {\bibfnamefont {P.}~\bibnamefont {Ring}},\ }\href {\doibase
  10.1016/j.ppnp.2022.103965} {\bibfield  {journal} {\bibinfo  {journal} {Prog.
  Part. Nucl. Phys.}\ }\textbf {\bibinfo {volume} {126}},\ \bibinfo {pages}
  {103965} (\bibinfo {year} {2022})},\ \Eprint
  {http://arxiv.org/abs/2111.15543} {arXiv:2111.15543 [nucl-th]} \BibitemShut
  {NoStop}%
\bibitem [{\citenamefont {Yao}\ \emph {et~al.}(2010)\citenamefont {Yao},
  \citenamefont {Meng}, \citenamefont {Ring},\ and\ \citenamefont
  {Vretenar}}]{Yao:2010}%
  \BibitemOpen
  \bibfield  {author} {\bibinfo {author} {\bibfnamefont {J.~M.}\ \bibnamefont
  {Yao}}, \bibinfo {author} {\bibfnamefont {J.}~\bibnamefont {Meng}}, \bibinfo
  {author} {\bibfnamefont {P.}~\bibnamefont {Ring}}, \ and\ \bibinfo {author}
  {\bibfnamefont {D.}~\bibnamefont {Vretenar}},\ }\href {\doibase
  10.1103/PhysRevC.81.044311} {\bibfield  {journal} {\bibinfo  {journal} {Phys.
  Rev. C}\ }\textbf {\bibinfo {volume} {81}},\ \bibinfo {pages} {044311}
  (\bibinfo {year} {2010})}\BibitemShut {NoStop}%
\end{thebibliography}

%

\end{document}